\documentclass[12pt,preprint]{aastex52/aastex} 

\slugcomment{Cha~II Synthesis: \today}

\shorttitle{The Spitzer c2d Survey in Chamaeleon II}
\shortauthors{Alcal\'a et al.}

\begin{document}


\title{The Spitzer c2d Survey of Large, Nearby, Interstellar Clouds.~X. \\
The Chamaeleon~II Pre-Main Sequence Population as Observed With IRAC and MIPS}


\author{
Juan~M.~Alcal\'a\altaffilmark{1}, 
Loredana~Spezzi\altaffilmark{1,4},  
Nicholas~Chapman\altaffilmark{2}, 
Neal~J.~Evans~II\altaffilmark{3}, 
Tracy~L.~Huard\altaffilmark{5}, 
Jes~K.~J{\o}rgensen\altaffilmark{6},
Bruno~Mer\'{i}n\altaffilmark{7,10}, 
Karl~R.~Stapelfeldt\altaffilmark{9},
Elvira~Covino\altaffilmark{1}, 
Antonio~Frasca\altaffilmark{4}, 
Davide~Gandolfi\altaffilmark{4}, 
Isa~Oliveira\altaffilmark{8,10} 
}

  
 
\altaffiltext{1}{INAF-OA Capodimonte, via Moiariello 16, 80131, Naples, Italy; alcala@oacn.inaf.it}
\altaffiltext{2}{Astronomy Department, University of Maryland, College Park, MD~20742; chapman@astro.umd.edu}
\altaffiltext{3}{Astronomy Department, University of Texas at Austin, 1 University Station C1400, Austin, TX~78712-0259; nje@astro.as.utexas.edu}
\altaffiltext{4}{INAF-OA Catania, via S. Sofia, 78, 95123 Catania, Italy; afr@oact.inaf.it}
\altaffiltext{5}{Smithsonian Astrophysical Observatory, 60 Garden Street, MS42, Cambridge, MA~02138; thuard@cfa.harvard.edu}
\altaffiltext{6}{Argelander-Institut f\"ur Astronomie, University of Bonn, Auf dem H\"ugel 71, 53121 Bonn, Germany; jes@astro.uni-bonn.de}
\altaffiltext{7}{Research and Scientific Support Dept. (ESTEC), European Space Agency, Keplerlaan, 1,  PO Box 299, 2200 AG Noordwijk, The Netherlands; bmerin@rssd.esa.int}
\altaffiltext{8}{Division of Geological and Planetary Sciences, California Institute of Technology, Pasadena, CA 91125, USA; isa@gps.caltech.edu}
\altaffiltext{9}{Jet Propulsion Laboratory, California Institute of Technology, Pasadena, CA; krs@exoplanet.jpl.nasa.gov}
\altaffiltext{10}{Leiden Observatory, Leiden University, P.O. Box 9513, 2300 RA Leiden, The Netherlands}

 
\begin{abstract}
We discuss the results from the combined IRAC and MIPS c2d Spitzer Legacy
survey observations and complementary optical and near infrared data of 
the Chamaeleon~II (Cha~II) dark cloud. We perform a census of the young 
population of Cha~II, in a mapped area of $\sim$1.75~deg$^2$, and study 
the spatial distribution and properties of the cloud members and candidate 
pre-main sequence (PMS) objects and their circumstellar matter. 
Our census of PMS objects and candidates in ChaII is complete down to the 
sub-stellar regime ($M \approx 0.03~M_{\odot}$), at the assumed cloud 
distance of 178~pc. The population consists of 51 certified and 
11 candidate PMS objects, most of them located in the Eastern part of 
the cloud, but approximately following the dust emission lanes of the 
c2d extinction map. From the analysis of the volume density of the PMS 
objects and candidates we find two tight groups of objects with volume 
densities higher than 25~M$_{\odot}$~pc$^{-3}$ and 5-10 members each. 
These groups correlate well in space with the regions of high extinction.
A multiplicity fraction of about 13$\pm$3\% is observed for objects 
with separations 0.8" $< \theta <$ 6.0" (142 - 1065 AU). No evidence for
variability in the IRAC bands between the two epochs of the c2d data set,
$\Delta t \sim$6 hours, is detected. 
Using the results of masses and ages from a companion paper, we estimate 
the star formation efficiency to be 1-4\%, consistent with the estimates 
for Taurus and  Lupus, but significantly lower than for Cha~I. This might 
mean that different star-formation activities in the Chamaeleon clouds 
reflect a different history of star formation.
We also find that the Cha~II cloud is turning some 6-7 $M_{\odot}$ into 
stars every Myr, which is low in comparison with the star formation 
rate in other c2d clouds. On  the other hand, the disk fraction of 
70-80\% that we estimate in Cha~II is much higher than in other star 
forming regions and indicates that the population in this cloud is dominated 
by objects with active accretion, with only a minority being systems with 
passive and debris disks. The circumstellar envelope/disk properties of 
the PMS objects and candidates are also investigated. Finally, the Cha~II 
outflows are discussed, with particular regard to the discovery of a new 
Herbig-Haro outflow, HH~939, driven by the classical T~Tauri star Sz~50.

\end{abstract}
 
 
 
 

\newpage
\section{Introduction}
 
The Chamaeleon cloud complex is one of the major star-forming 
regions  within 200~pc from the Sun. It consists of three 
main dark clouds, namely Cha~I, Cha~II and Cha~III \citep{Sch91}.
In terms of star formation, the most active of the three is Cha~I, 
while Cha~III is the least active one \citep{Sch77, Het88}. 

At a distance of 178~pc \citep{Whi97}, the Cha~II dark cloud 
(R.A.~$\approx 13^h 00^m$ and Dec.~$\approx -77^\circ 06'$) is a 
well-defined structure that covers about 2~deg$^2$ in the sky.
The presence of pre-main sequence (PMS) objects in this cloud has 
been traced by a variety of observations, including objective prism 
surveys \citep{Sch77, Har93}, X-ray imaging with ROSAT 
\citep[][and references therein]{Alc00} and infrared (IR) surveys 
with IRAS \citep{Whi91, Pru92, Lar98}, and ISO \citep{Per03}. 
Some 20 classical T~Tauri stars (CTTSs) and a number of weak-line T~Tauri 
star (WTTS) candidates have been identified in the cloud so far from 		
ground-based optical and near-IR photometry, and subsequent spectroscopic 
follow-up \citep{Hug92, Alc95, Cov97a, Lar98, Vuo01, Bar04a, All07}. 
In addition to the low-mass young stars, one intermediate-mass 
AeBe star, \emph{IRAS~12496$-$7650}, and the well-studied Herbig-Haro 
object \emph{HH 54} \citep{Kne92, Lis96, Nis96, Neu98, Neu06} are also 
associated with Cha~II. 

The Cha~II dark cloud is one of the five star formation regions selected by 
the Legacy program \emph{From Molecular Cores to Planet-forming Disks} 
\citep[c2d,][]{Eva03}. 
Cha~II was chosen for the c2d program as an example of a cloud with modest 
star formation activity, based on previous data. One of the major goals 
was to investigate whether the modest star formation activity in the 
cloud is the result of observational biases and the more sensitive data 
from Spitzer would discover a more vigorous star formation, or whether the 
modest star formation is real. For instance, previous ground-based studies 
\citep{Bar04a} failed to find young sub-stellar objects. Hence, an important 
goal of the c2d program has been to confirm whether the non-detection of 
brown dwarfs (BDs) in Cha~II is due to incompleteness of the previous 
surveys, both in sky area and flux, or to a real lack of BDs in the 
region. Recent follow-up of the c2d observations have now confirmed 
the sub-stellar nature of some objects in Cha~II 
\citep{All06a, Alc06, Jay06, All07, Spe07b}.

The Spitzer observations in Cha~II are reported in previous basic data 
papers of this series: the MIPS (Multi-band Imaging Photometer for Spitzer) 
observations are presented in \citet{You05}, while those with IRAC 
(InfraRed Array Camera) are reported in \citet{Por07}. \citet{You05} also 
presented a 1.2 square degrees millimeter map of Cha~II obtained with 
SIMBA on the Swedish-ESO Submillimetre Telescope (SEST); they show that 
only a small fraction of the gas is in compact structures with high 
column density. Complementary to the c2d survey, data in Cha~II in 
optical and near-infrared (IR) wavelengths are presented in 
\citet[][hereafter S07a]{Spe07a} and \citet{All06a}, respectively. 

The main goal of this paper is to combine the c2d IRAC and MIPS data 
with the complementary data, and with other existing in the literature, 
to make a synthesis of the young population of Cha~II. Our intention 
is to deliver the most reliable census to date of the young population 
that permits a thorough study of the star formation in this cloud. 
To this aim, we report the available photometry in the range of 
0.36\,$\mu$m to 1300\,$\mu$m for as many Cha~II members as possible. 
The IRAC observations reported in \citet{Por07} were performed with 
the criterion to cover the area with visual extinction greater 
than 2~mag. The consequence of this is that only the central and 
''densest" part of the Cha~II cloud was investigated with IRAC. 
Thus, to perform a complete census we also include the PMS objects 
and candidates that lie off the cloud, in regions where no IRAC 
observations were performed. The data presented here are used 
to investigate the spectral energy distribution (SED) of the 
Cha~II certified and candidate pre-main sequence (PMS) objects 
and study their circumstellar disk/envelope properties.

Throughout this paper, we will use the terms young stellar object (YSO)
and PMS object. The c2d convention is to define the former as those 
PMS objects with detectable IR excess in at least one of the Spitzer 
bands and that satisfy the multi-color criteria described in 
\citet[][hereafter H07]{Har07} and \citet{Eva07}, while the latter considers 
a broader class which includes not only YSOs, but also PMS objects 
discovered using other observational techniques like H$\alpha$ and X-ray 
surveys or chromospheric activity indicators. Thus, by PMS objects we 
mean objects not yet on the main sequence that may or may not have 
IR excess. That is, all YSOs are PMS objects, but not vice versa. 
Therefore, in this paper the term YSO includes all the envelope and 
disk stages as a sub-set of all PMS phases. 

In \S~\ref{data} we present an overview on the observational photometric
data set used, while in \S~\ref{sample} the census of the PMS population 
in Cha~II is performed by merging the c2d data with those from optical 
and IR surveys. In \S~\ref{variability}, a search for variability in the 
c2d data set is described. In \S~\ref{disk_prop} the photometric data set 
is used to investigate the global properties of the circumstellar material, 
including the comparison with accretion models in color-color diagrams, 
as well as from fits of both accretion and reprocessing models to the 
SEDs. The disk frequency and disk properties versus stellar parameters 
are also discussed in this section. In \S~\ref{overallsatarformation} the 
results on the efficiency and rate of star formation in the cloud are 
discussed, making use of the extinction maps so far derived in Cha~II 
to estimate the cloud mass. Finally, in \S~\ref{outflows} we give an 
overview of the outflows in Cha~II, and report the discovery of new 
Herbig-Haro objects. We summarize our results in \S~\ref{sum}.

\section{Photometric data set}
\label{data}

The basic data set used for this paper comes from the Spitzer (IRAC+MIPS)
c2d observations of Cha~II, complemented with optical and near-IR imaging
surveys. 
In addition, data from the Spitzer Wide-area InfraRed Extragalactic 
\citep[SWIRE;][]{Sur04} survey are used for statistical analysis and 
corrections for background contaminants.

\subsection{Spitzer data}  
\label{spitzerdata} 

The observations, data reduction, source statistics and the results from 
the IRAC observations are reported in \citet{Por07}. An area of 1.04 square 
degrees in Cha~II was mapped in the four IRAC bands (3.6, 4.5, 5.8 and 8 $\mu$m) 
with the criterion that $A_V>$2 mag in the extinction map reported by \citet{Cam99}. 
In order to make statistical comparisons of source counts and determine the 
contamination by background objects, six off-cloud fields were also observed 
with IRAC adopting the criterion of $A_V<$0.5~mag in the \citet{Cam99} map. 
Based on the 2005 catalog, some 40 sources were classified as YSO candidates 
from the IRAC observations by \citet{Por07}. However, as pointed out in that 
paper, a significant fraction of these candidates may correspond to 
extragalactic contaminants. This issue will be discussed in more detail in \S~3.

The results of the MIPS survey in Cha~II are presented in \citet{You05}.
The surveyed area of 1.5 square degrees entirely covers the one mapped 
with IRAC. By combining the MIPS data with those of the Two-Micron All-Sky 
Survey (2MASS) catalog \citep{Cut03}, and using color-color and color-magnitude 
diagrams, \citet{You05} classified more than 40 sources as potential  
YSO candidates. Considering the completeness in area and flux, and despite 
the low number of potential YSO candidates, \citet{You05} concluded that the 
population of YSO candidates found in previous surveys is nearly complete.

An important step beyond the 2005 c2d catalogs was the creation of 
new "band-filled" catalogs. Such band-filling was performed by the 
c2d catalog infrastructure in order to obtain upper limits 
or low S/N detections of objects that were not detected in the original 
source extraction processing. Details on the band-filling process, 
which is applied only to sources clearly detected in at least one of the 
Spitzer bands, are described in the final c2d data delivery document 
\citep{Eva07}. For a short description we refer the reader to H07.

\subsection{Complementary Optical to Millimeter data}

In the following sections we describe the complementary data used 
for this work. 

The optical imaging survey was performed 
using the Wide-Field Imager (WFI) attached to the ESO 2.2~m telescope, 
at La Silla, Chile. The observational strategy, data reduction and 
source extraction are described in S07a.
An area of about 1.75 square degrees in Cha~II was observed in the 
$R_C$, $I_C$, $z$ and H$\alpha$ bands and in two intermediate-band 
filters. The optical observations cover the overlap areas observed 
with IRAC and MIPS. By combining the optical data with those 
of 2MASS and using color-magnitude and color-color diagrams, 
S07a selected 10 new candidate PMS objects not revealed by previous 
surveys. These photometric data also confirmed the H$\alpha$
emission from these objects.
They concluded that these objects are most likely Class~III sources 
belonging to the WTTS class, as expected by the complementary 
character of the optical survey with respect to the IR  surveys. 
In line with \citet{You05}, S07a concluded that the optical surveys 
performed so far in Cha~II may have already provided a population 
of PMS objects which is basically complete. 

In addition to the optical survey, a deep near-IR survey  was 
performed by \citet{All06a} over an area of about 0.6 square degrees
in Cha~II. Based on color-magnitude and color-color diagrams, 
they selected seven candidate very low-mass PMS objects in Cha~II. 
Three of them were confirmed to be young sub-stellar 
objects based on follow-up spectroscopy \citep{Alc06, Jay06, All07},
among these the lowest mass objects in Cha~II \citep{All07}.

Photometric data from the ``Naval Observatory Merged Astrometric 
Dataset'' \citep[NOMAD,][]{Zac05} in the optical and from the ISO 
observations  by \citet{Per03} in the IR, and by \citet{Hen93} in 
mm wavelengths, are used as ancillary data for this work. 
In particular, these data are incorporated in the SEDs presented 
in \S~\ref{sec_seds}.

Finally, the spectral types and physical parameters of the PMS objects 
and candidates in Cha~II derived by \citet[][hereafter S07b]{Spe07b} on 
the basis of optical spectroscopic follow-ups, are used here for the 
determination and analysis of the SEDs, as well as for the analysis of 
the overall star formation in Cha~II.

\subsection{SWIRE data}

In order to perform statistical comparisons and correct for extragalactic 
contaminants, we also use in this work the much larger and deeper catalog 
coming from the SWIRE observations \citep{Sur04} of the ELAIS N1 field 
\citep{Row04}. 
The SWIRE catalog was extincted and re-sampled as accurately as possible 
to match the c2d sensitivity limits, as well as extinction in Cha~II. 
For the details on how the trimmed re-sampled SWIRE comparison catalog 
was created, we refer the reader to the final c2d data delivery document
\citep{Eva07} and for a short description to H07.

\section{The sample of PMS objects and candidates in Cha~II}
\label{sample}

The IRAC observations in Cha~II cover only the central and ''densest" 
part of the cloud. However, some previously known members and candidate 
PMS objects lie to the East of the Cha~II cloud (Figure~\ref{spa_distr}). 
Thus, in order to include these objects, our census is performed by 
merging the c2d data with those from previous optical and near-IR surveys. 
In the following we describe how our list of certified and 
candidate members of Cha~II was constructed. The IRAC and MIPS data are 
used first to select new YSO candidates and to estimate their number 
in regions where no IRAC observations were performed. The certified 
and candidate PMS objects without IRAC observations are then discussed.

The samples of PMS objects and candidates presented in the previous
surveys plus the sources selected from new c2d criteria (see \S~\ref{new_crit}) 
in this work were merged into a single list which is given in 
Table~\ref{tab:yso}. A spectroscopic follow-up of many of the candidate PMS 
objects has been performed with FLAMES at the ESO VLT, but the details and 
results of these observations are published in a companion paper \citep{Spe07b}. 
From such investigation we conclude that there are 51 certified PMS objects 
plus 11 candidates that still need spectroscopic investigation in order 
to firmly assess their nature. Of these 62 objects, 26 are YSOs selected 
from the c2d criteria; 2 of these 26 were previously unknown, while 
the remaining 24 coincide with Cha~II members or candidate PMS objects 
identified in early surveys. 
The sample is flagged in column~5 of Table~\ref{tab:yso} with "PMS" indicating 
a certified PMS object, with "CND" a candidate PMS object and with "YSO" if 
the object passed the new c2d selection criteria described in \S~\ref{sel_ysos}. 
If the young nature of the 11 candidates will be confirmed, the 
population of PMS objects in Cha~II will increase to more than 60 members.

In Table~\ref{tab:yso} we also report the sources HH~54 (or IRAS~12522$-$7640)
and IRAS~13036$-$7644. The former corresponds to the head of the HH~54 
outflow (see \S~\ref{outflows}), while the latter, also known as BHR~86 
\citep{Bou95}, has been reported by \citet{Leh05} to be a transition object 
between Class~0 and I. BHR~86 is located to the East, somewhat detached 
from the main body of the Cha~II cloud. BHR~86 is matter of focused 
c2d observations and will be studied in more detail in a future paper 
(Huard et al., in preparation).
 
The Spitzer fluxes were extracted from the new band-filled catalog 
described in \S~\ref{spitzerdata}. These, in combination with the optical 
magnitudes reported by S07a and those at longer wavelengths 
available in the literature for some objects, constitute the main data 
set for our subsequent analysis. In Tables~\ref{tab:flux_opt}, 
\ref{tab:flux_J_IR3}, \ref{tab:flux_iso6.7_iras60} and \ref{tab:flux_70_1300} 
we report the available photometric data for the sources reported in 
Table~\ref{tab:yso}. 

We remind the reader that the IRAC observations were performed in such 
a way that the coverage at 3.6$~\mu$m and 5.8$~\mu$m is shifted by about 
7~arc-min to the east of the area covered at 4.5$~\mu$m and 8.0$~\mu$m 
\citep[see Figure~1 by][]{Por07}. Thus, some sources possessing fluxes at 
3.6$~\mu$m and 5.8$~\mu$m may lack data at 4.5$~\mu$m and 8.0$~\mu$m, 
and vice versa. In column 8 of Table~\ref{tab:yso}, the flag "in"  is 
used to indicate that the object lies on the 
IRAC~3.6-8\,$\mu$m/MIPS~24\,$\mu$m overlap area, the  flag "off" is 
used otherwise.

\subsection{Selection of YSOs} 
\label{sel_ysos}

One of the main goals of the c2d observations was the selection
of new YSO candidates based on their IR emission. However, the IR 
colors of many galaxies are very similar to those of YSOs. Thus, 
a sample of YSO candidates must be corrected for the contamination 
of background extragalactic objects. This and the next subsection 
explain how the 26 YSOs were selected. 

The first criteria for the selection of YSO candidates from the 
Spitzer data are described in the third delivery document of the 
c2d program \citep{Eva05}. 
These criteria, applied to the IRAC data of the 2005 catalog, 
provided  $\sim$40 YSO candidates in Cha~II with a density 
of 38.5 candidates per square degree \citep{Por07}. 
Interestingly, many Class~I or flat-spectrum sources 
were identified among these candidates indicating a much more 
active star formation in Cha~II than previously found. It is 
also interesting to note that, despite the larger area covered 
with MIPS, \citet{You05} also selected 44 YSO candidates, 
but with the smaller density of 29.3 objects per square degree.
This cast doubts on the YSO nature of the new Class~I or 
flat-spectrum sources.
As pointed out by  \citet{Por07}, a substantial fraction of the 
low-luminosity objects, classified as Class~I or flat-spectrum 
sources from the IRAC data, may correspond to extragalactic 
contaminants. We now provide further arguments supporting this 
conclusion. 

\begin{figure} 
\epsscale{0.8}
\plotone{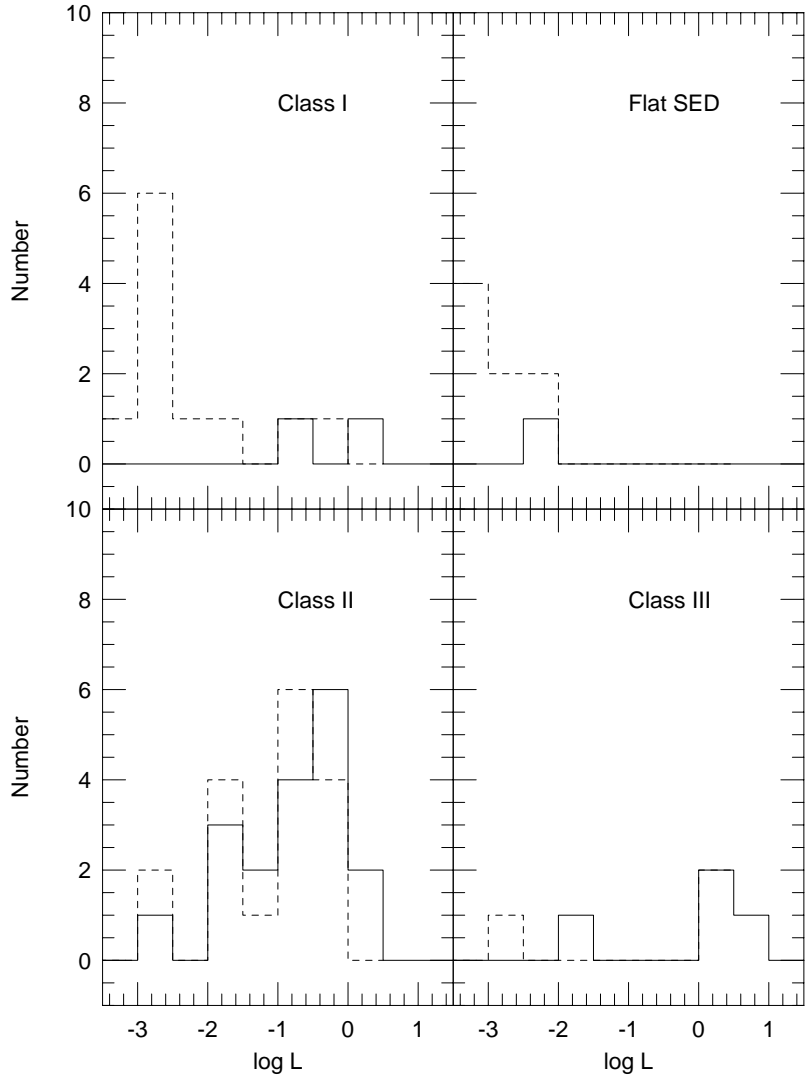}
\caption{Luminosity histograms, in solar units, of the YSO candidates
resulting from the selection criteria applied to the IRAC data alone 
(dashed histograms). Each panel represents a Lada Class, as determined 
by \citet{Por07}. Luminosities were derived by integration of the 
observed SEDs, assuming isotropic radiation and adopting the Cha~II 
distance of 178~pc. The solid histograms represent the total sample 
of YSO candidates selected from the new criteria described in the 
text.
\label{lumhistclass} }
\end{figure}

Figure~\ref{lumhistclass} shows the four luminosity histograms 
corresponding to the Lada classes as determined by \citet{Por07}. 
The luminosities were computed by integrating the SEDs, from 
1.2 to 70 micron, using the available fluxes for each source. 
The dashed histograms represent the total sample of YSO candidates,
while the solid ones represent only those selected from new
criteria described below.
In the dashed histograms, low luminosity is much more common 
for earlier Lada Classes. The few more luminous sources in this 
sample are mostly Class~III, well known stars with enough circumstellar 
dust to be luminous, but not very embedded. Essentially all the 
Class~I and flat-spectrum sources have a luminosity of less than 
0.01~L$_{\odot}$. All these ultra-low luminosity sources, with not 
much to be embedded in given the SIMBA millimeter map reported 
in \citet{You05}, are suspected to be background galaxies.

\begin{figure}[!h]
\epsscale{0.8}
\plotone{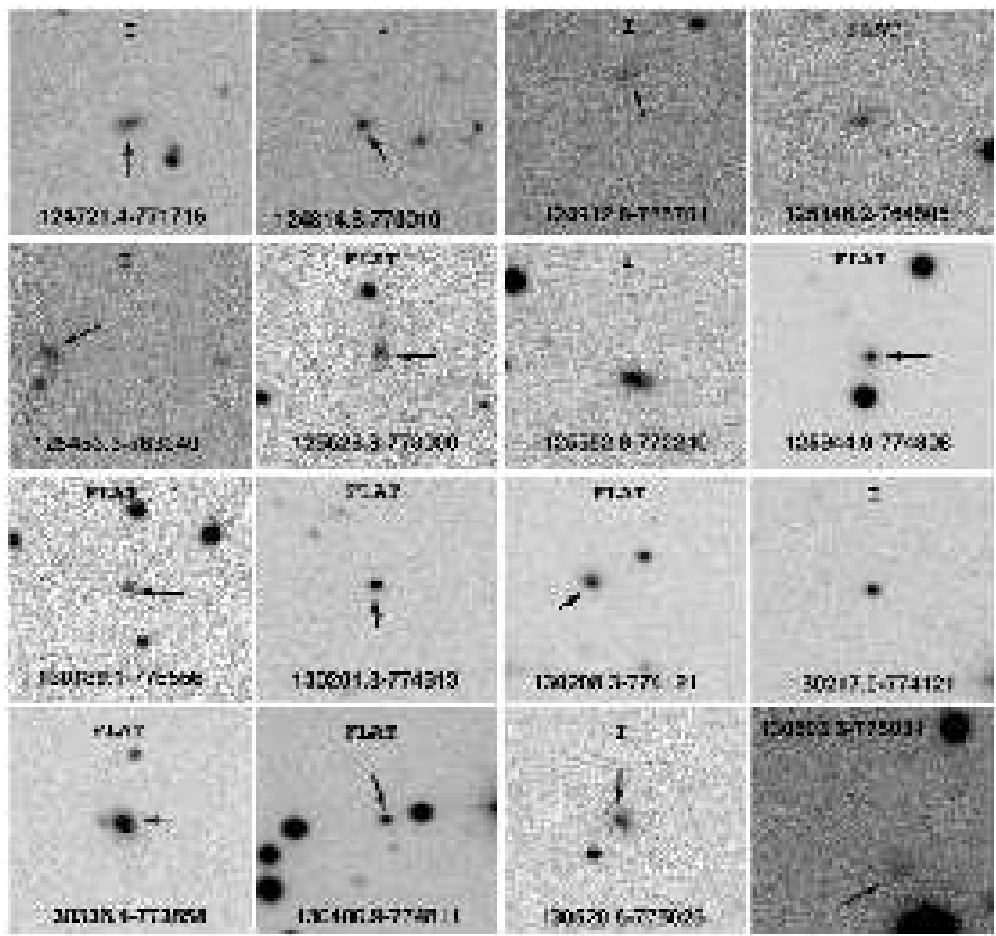}
\caption{The I-band images of objects classified as Class~I or 
flat-spectrum sources by \citet{Por07}. Each stamp covers 
an area of about 21$\times$21~arcsec$^2$. North is up and  
East to the left. The name of the sources and their classification
are indicated. In case of ambiguity, the source is indicated with 
an arrow. Two sources (SSTc2d~J125843.7$-$774728 and 125906.5$-$770740)
are not shown because nothing is seen in the I-band, nor in any other 
optical image. 
\label{Iband_images} }
\end{figure}

Additional arguments indicating that the low-luminosity sources are 
extra-galactic contaminants come from their appearance in the optical 
images. Figure~\ref{Iband_images} shows I-band images of such 
objects collected in the optical survey by S07a. Although some sources 
are barely seen, the vast majority of them clearly appear as extended, most 
likely corresponding to extra-galactic contaminants. Four sources, namely 
SSTc2d~J125944.9$-$774808, 130201.3$-$774813, 130217.0$-$774121 and 130400.9$-$774811 
seem to present a more point-like PSF than the others, but none of them has 
been selected as a candidate PMS object in previous surveys nor appears as H$\alpha$ 
emitter (see S07a). In addition, the IRAC fluxes of two of them
(SSTc2d~125944.9$-$774808 and SSTc2d~130400.9$-$774811) are partially contaminated 
by the flux of neighbor sources, most likely mimicking a flat-spectrum. 
Therefore, none of these four sources can be considered as a YSO candidate.
As we will see below, this is further confirmed by applying new c2d 
selection criteria. 

\subsection{New c2d selection criteria} 
\label{new_crit}
 
In order to improve the selection of the YSO candidates and clean the 
c2d samples as best as possible from extragalactic contamination, a new 
selection procedure was proposed by H07, using both IRAC and MIPS color 
and magnitude criteria, as well as a comparison with the re-sampled SWIRE 
catalogs matching the c2d sensitivity limits.
To be able to use these criteria, the sources must be well detected 
(S/N $\geq$ 3) in all four IRAC bands and in MIPS~24~$\mu$m. 
This restricts the identification of the YSO candidates in the 
overlapping area surveyed in all the four IRAC bands and MIPS~24~$\mu$m. 
A detailed description of the selection procedure can be 
found in the c2d final delivery document \citep{Eva07}. 

\begin{figure} 
\epsscale{1.0}
\plotone{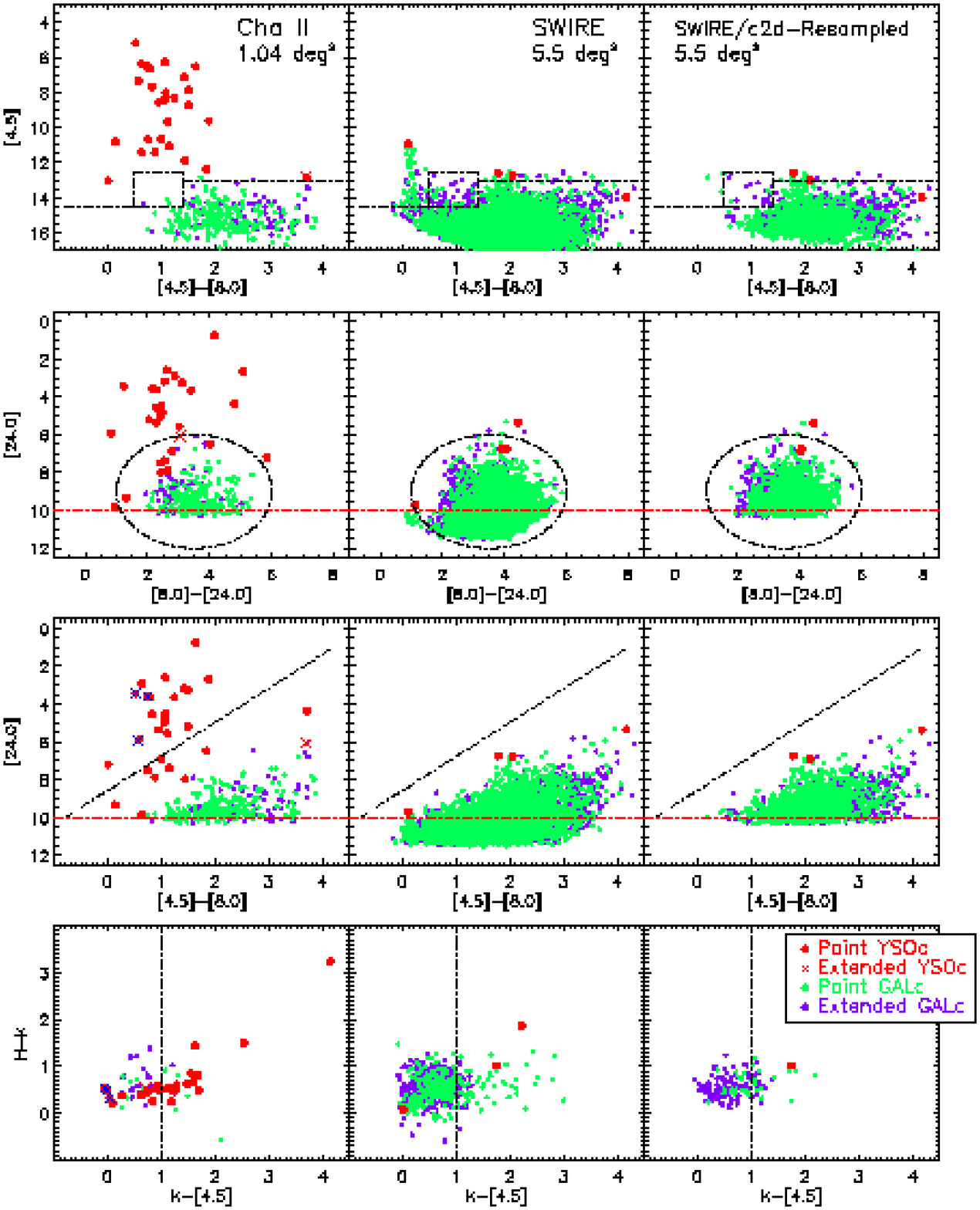}
\caption{Color-magnitude and color-color diagrams for the Cha~II cloud
(left panels), the full SWIRE catalog (central panels) and the re-sampled 
SWIRE catalog (right panels). The dot-dashed lines 
represent the color-magnitude cuts defining the YSO candidate criteria 
according to  H07.
 \label{sel_yso} }
\end{figure}

As a result of the multi-color procedure, applied to the band-filled 
catalog in Cha~II, 30 YSO candidates were selected in the 
IRAC~3.6-8~$\mu$m/MIPS~24~$\mu$m overlap area of 
1.038 square-degree\footnote{The actual number of selected objects is 29, 
but we stress that the well known young star IRAS~12496$-$7650 (DK~Cha) 
is saturated in IRAC~3.6\,$\mu$m and IRAC~4.5\,$\mu$m. By interpolating
its spectral energy distribution in these bands, the object is definitely 
selected with the new c2d criteria. In fact, its $L$ and $M$ magnitudes 
from the literature \citep{Hug92} are typical of a YSO.}. 
Figure~\ref{sel_yso} shows the  multi-color selection for Cha~II; that 
figure is the equivalent to Figure~3 by  H07. Five of the 30 objects 
are newly selected, while 25 coincide with already known  certified or 
candidate YSOs reported in previous surveys \citep[][S07a and references therein]{You05, All06b}. 
However, the latter reduce to 24 because 2MASS~12560549-7654106, originally 
proposed as a YSO candidate by \citet{You05}, has been spectroscopically 
confirmed to be a field star, unrelated to Cha~II \citep[see paper by][]{Spe07b}. 

The recovery of all the 24 previously known objects in the 
IRAC~3.6-8~$\mu$m/MIPS~24~$\mu$m overlapping area shows that the new criteria 
work quite well in selecting sources with IR excess. More important is the 
fact that, except for the previously known Class~I and flat-spectrum 
sources, all the other objects classified as Class~I or flat-spectrum sources 
in \citet{Por07} are rejected, further supporting their extragalactic nature. 
Moreover, the previously confirmed Class~I or flat-spectrum sources selected
with the new criteria possess the highest luminosities in their 
Lada class (cf. Figure~\ref{lumhistclass}).
We thus conclude that the low-luminosity YSO candidates, selected 
using the previous criteria applied to IRAC data alone and classified 
as Class~I or flat-spectrum sources in \citet{Por07}, are indeed 
extra-galactic contaminants. 

\begin{figure}[!h]
\epsscale{0.9}
\plotone{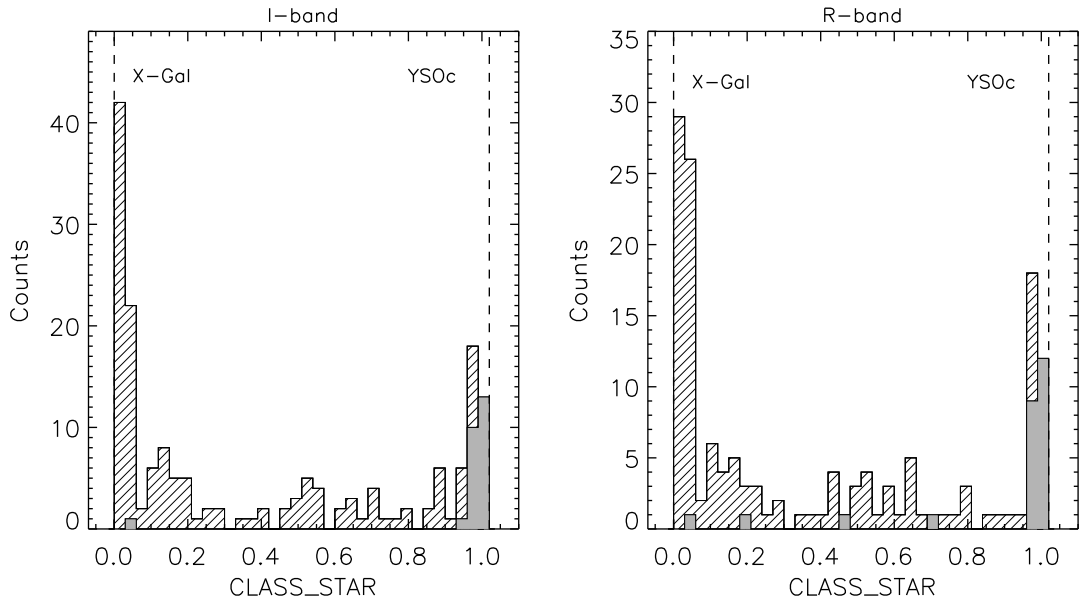}
\caption{Histograms of SExtractor {\em CLASS\_STAR} indices of the
sources extracted from the R (right) and I (left) band images by 
S07a. The hashed histograms correspond to sources having high 
probability of being extented from the new multi-color criteria, 
while the grey ones correspond to the point-like sources. 
The dashed lines at {\em CLASS\_STAR} 0 and 1 represent the limits 
for extended and point-like objects respectively. The only YSO 
candidate with {\em CLASS\_STAR} index close to zero in both bands 
corresponds to the interacting galaxies discussed in the text.
\label{class_star}}
\end{figure}

A further test for the reliability of the new selection criteria
can be performed by using the information available in the optical 
catalogs. An output parameter resulting from the source extraction
with the SExtractor tool \citep{Ber96} is the {\em CLASS\_STAR} index.
Point-like objects have {\em CLASS\_STAR} index very close to 1, while
for extended sources the index is very close to zero. Although 
some YSOs might appear as extended in some bands and some galaxies 
as point-like, statistically one expects that most YSOs will show up
as point-like and galaxies as extended.
In Figure~\ref{class_star} we show the histograms of the {\em CLASS\_STAR}
index resulting from the $R$ and $I$ imaging for the objects 
classified as YSO candidates and galaxies after applying the new 
selection criteria. For the histograms we used the optical catalogs
by S07a.  
The samples are well separated: sources classified as galaxies with
the new multi-color criteria normally have {\em CLASS\_STAR} index 
close to zero, while the index is typically 1 for the YSO candidates. 
Some galaxies may have {\em CLASS\_STAR} index close to 1, but the index 
is 1 for the vast majority of the YSO candidates. We thus conclude that 
the new selection criteria provide results which are consistent with 
the SExtractor morphological classification, giving further support 
for the new selection criteria to work well in cleaning the sample 
of YSO candidates from background extragalactic contaminants. 
Note that, while the morphological classification does separate 
extended from point-like objects, some extended YSOs would escape 
selection if we rely on such selection alone; the c2d multi-color
criteria would select these as YSO candidates.

We now briefly discuss the 5 newly selected c2d sources: 
SSTc2d~J124858.6$-$764154 and SST\-c2d\-~J125701.6\-$-$464835 are galaxies; 
the former coincides with the well know galaxy LEDA~239204, while the 
latter is clearly identified with an interacting galaxy pair in our I-band 
images. Therefore, these two are rejected as YSO candiades. The remaining 
three sources show up as follows: SSTc2d~J130521.7$-$773810 is a point-like 
source and SSTc2d\-~J130452.6$-$773947 is clearly a visual binary star in 
the I-band, while SSTc2d~J130529.0$-$77\-4140 was not detected in optical 
or near-IR bands. 
SSTc2d~J130521.7$-$773810 and both visual components of SSTc2d~J130452.6$-$773947 
are rejected as YSO candidates by the optical selection criteria by S07a. 
However, the former displays strong emission, as indicated by its photometric 
H$\alpha$ index. Thus, its optical colors might be affected by strong veiling. 
Note that such YSOs may escape selection with the S07a criteria.
Indeed, this object has been spectroscopically confirmed to show very 
strong H$\alpha$ emission \citep{Spe07b}.
On the other hand, SSTc2d~J130452.6$-$773947 is strongly contaminated 
by the MIPS~24\,$\mu$m emission of Hn~24 (see \S~\ref{multiples}), 
a certified YSO, explaining why its colors may apparently be consistent 
with those of YSOs. 
Considering that the higher resolution optical data reject this object 
and that no trace of H$\alpha$ emission is found, we removed it from the 
list of YSO candidates. Finally, not being detected in optical and near-IR 
bands, SSTc2d~J130529.0$-$774140 cannot be ruled out as a YSO candidate.
In conclusion, the three newly selected sources SSTc2d~J124858.6-764154, 
SSTc2d~J125701.6$-$464835 and SSTc2d~J130452.6$-$773947 are rejected as 
YSO candidates, while SSTc2d~J130521.7$-$773810 and SSTc2d~J130529.0$-$774140
can be considered as YSO candidates. 

To summarize, in addition to the 24 previously known YSOs and candidates 
in the IRAC~3.6-8~$\mu$m/MIPS~24\,$\mu$m overlapping area, there are at most 
two new potential YSOs. The final list of 26 certified and candidate YSOs 
selected with the new multi-color criteria by H07 are marked with a flag 
("YSO" or "YSO CND") in Table~\ref{tab:yso}. 
  
The selection criteria applied to the comparison re-sampled SWIRE data 
lead to the conclusion that 2$\pm$1 of the selected sources in Cha~II
may be galaxies (cf. Figure~\ref{sel_yso}). Indeed, two of the five 
newly selected sources in Cha~II, namely SSTc2d~J124858.6$-$764154 and 
SSTc2d~J125701.6$-$464835, were identified with galaxies. This result 
is similar to what  H07 found for the Serpens molecular cloud.

\subsection{PMS objects off the IRAC area}
\label{off_irac}

The areas mapped by the MIPS, optical and near-IR observations by 
\citet{You05}, S07a and \citet{All06a}, respectively, are larger than 
the one covered with IRAC. The different selection criteria by these 
three surveys, besides providing new candidate PMS objects, coherently 
recovered the previously known certified and candidate PMS objects. 
Moreover, since the multi-color criteria select sources with IR excess 
emission, PMS objects not possessing substantial IR excess, at least 
somewhere in the range between 3.6 and 70 $\mu$m, escape selection. 
Thus, in Table~\ref{tab:yso} we include also the PMS objects found in 
the aforementioned MIPS, optical and near-IR surveys. As reported in 
Section~\ref{sample}, the complete sample includes 62 certified and 
candidate PMS objects.

The luminosity function of the selected YSO candidates (cf. \S~\ref{lum})
demonstrates that the c2d observations are basically complete down to 
$log L/L_{\odot} \approx -2.5$\footnote{This luminosity limit corresponds
to a mass $M \approx 0.03~M_{\odot}$, assuming the \citet{Cha00} PMS 
evolutionary tracks for objects younger than 10 Myr.} in the 
IRAC~3.6-8\,$\mu$m/MIPS~24\,$\mu$m overlap area.
In the case of Serpens,  H07  extrapolate the number of YSO candidates 
selected in the IRAC~3.6-8\,$\mu$m/MIPS~24\,$\mu$m overlap area to the 
larger area surveyed only with MIPS using the percentages of sources 
classified as YSOs from different criteria. Following a similar reasoning, 
we can investigate the expected number of YSOs if IRAC observations had 
been performed in the entire area covered by MIPS~24\,$\mu$m in Cha~II. 
\citet{You05} selected 44 YSOs based on the combination of MIPS and 2MASS 
data; 20 of these fall outside the area observed with IRAC\footnote{We 
stress, however, that 3 objects (Sz~48, Sz~60 and IRAS~F13052$-$7653) 
out of these 20 are multiple visual systems that are blended in the 
MIPS data by \citet{You05}.}. Of the 24 objects selected by \citet{You05} 
in the IRAC~3.6-8\,$\mu$m/MIPS~24\,$\mu$m overlap area in Cha~II, 23 
satisfy the H07 multi-color criteria presented in this work, i.e. 96\%. 
This means that 4\% of the 20 objects ($\sim$1) outside the 
IRAC~3.6-8\,$\mu$m/MIPS~24\,$\mu$m overlap area would have 
been rejected by the new multi-color criteria. Now, 24 of the 
26 YSOs selected in the IRAC~3.6-8\,$\mu$m/MIPS~24\,$\mu$m overlap area 
in this work are recovered by the \citet{You05} criteria, which means that 
about 8\% of the 20 objects (1-2) outside the IRAC~3.6-8\,$\mu$m/MIPS~24\,$\mu$m 
overlap are missed by the \citet{You05} criteria. All this leads to an 
expected number of about 21 YSOs in the area observed only with 
MIPS~24\,$\mu$m. Therefore, the expected number of YSOs if IRAC observations 
had been performed in the entire area covered by MIPS~24\,$\mu$m would be 
something like $26+21=47$.
This means that about 24\% of the 62 certified and candidate PMS objects 
mentioned above should not display significant IR excess in any of the 
Spitzer bands and thus, are expected to be preferentially Class~III 
sources. This conclusion is in agreement with the results of our 
census (see \S~\ref{sel_ysos} and \S~\ref{alpha}).
Therefore, besides having a complete sample also outside the 
IRAC~3.6-8\,$\mu$m/MIPS~24\,$\mu$m overlap area, and despite the 
low-number statistics, our results provide support to the arguments 
by H07 for the extrapolation of the number of YSOs to the areas 
observed only with MIPS.

\subsection{Multiple visual systems}
\label{multiples}
 
In order to avoid spurious selections due to crowding, we have visually
inspected all the confirmed and candidate PMS objects in all available 
images, from optical to MIPS~24\,$\mu$m. The result is that seven objects are 
found to have neighbors within less than 10~arcsec in our optical 
images and may be blended in the IRAC or MIPS images. 
In some cases they appear as single sources, in particular in the MIPS 
images, where the spatial resolution is worse ($\sim$6~arcsec at 24$\mu$m
and $\sim$17~arcsec at 70$\mu$m) with respect to IRAC images ($<$2~arcsec). 
We now discuss these objects individually:

SSTc2d~J130452.6$-$773947: in the R-band images this source coincides with 
a visual binary system with a separation of 1.6~arcsec. The PMS star 
Hn24 is located at about 10~arcsec to the East of it. Hn24 itself
is a visual binary with a separation of about 1.8~arcsec.
In the MIPS~24\,$\mu$m  image Hn24 and SSTc2d~J130452.6$-$773947 appear as 
a single object, hence the MIPS~24\,$\mu$m  emission of SSTc2d~J130452.6$-$773947 
is strongly contaminated by the 24$\mu$m emission from Hn24. The quality 
flags from the new band-filled c2d catalog indicate that, although 
SSTc2d~J130452.6$-$773947 
was well detected in the four IRAC bands, only an upper flux limit could 
be determined in MIPS~24\,$\mu$m. Therefore, it is most likely that Hn24 is 
the main source of the 24$\mu$m emission. In fact, the optical magnitudes 
and colors of both components of SSTc2d~J130452.6$-$773947 do not satisfy 
the selection criteria by S07a, nor do their H$\alpha$ photometric 
indices indicate emission. Therefore, SSTc2d~J130452.6$-$773947 is most 
likely a spurious selection.

IRAS~F13052$-$7653: in the optical this object is revealed as a triple visual 
system (see S07a). The North and South-West components are the brightest, 
while the North-West component is the faintest one. The separation of 
the bright components is about 3~arcsec, while the separation of 
IRAS~F13052$-$7653N and IRAS~F13052$-$7653NW is about 4~arcsec. The latter
two, while being potential candidate PMS objects based also on optical 
criteria by S07a, have been spectroscopically confirmed to be PMS stars 
(S07b). The optical magnitudes and colors of the South component are 
inconsistent with those of Cha~II PMS stars \citep{Spe07a}.
The visual triple was not observed with IRAC, but in MIPS~24\,$\mu$m the 
three visual components are blended in a single source.
 
Hn22 and Hn23: in optical images these PMS stars appear as a visual pair 
with a separation of about 6~arcsec in the North-East--South-West direction. 
In the R-band the North-East component, Hn23, is brighter than the South-west 
component Hn22. The visual pair was not observed with IRAC, but an elongated 
structure in the MIPS~24\,$\mu$m  images can be appreciated.  
 
Sz48: this object appears as a visual pair with a separation of about 
1.5~arcsec in the North/East--South/West direction in the R-band image. 
The visual binary appears as a single object in the Spitzer data. The  
new band-filled c2d catalog lists a single detection in IRAC~3.6\,$\mu$m, 
IRAC~5.8\,$\mu$m, MIPS~24\,$\mu$m and MIPS~70\,$\mu$m. In the MIPS~70\,$\mu$m 
image the object falls quite near the border. This pair was also studied 
by \citet{Bra97}.

Sz59: this object has been detected by \citet{Bra97} to be a visual binary
with a separation of 0.78~arcsec. The binary appears as a single object
in optical and IR images.

Sz60:  is a visual pair with a separation of about 3~arcsec in the 
East--West direction in optical images. Both components were detected 
in IRAC~3.6\,$\mu$m, IRAC~5.8\,$\mu$m and MIPS~24\,$\mu$m, 
but their S/N ratio is too low to perform shape analysis. 
The West component was well detected in MIPS~70\,$\mu$m  but only an 
upper limit in this band could be determined for the East component. 

Sz62: \citet{Bra97} report this object as a binary with a separation of
1.1~arcsec. The object also appears as a visual binary in our optical 
images and shows up as a single object in all IR images.

Because of the difficulty to deblend the components of these objects 
in the Spitzer images, we consider their c2d fluxes as upper limits 
in Tables~\ref{tab:flux_opt}, \ref{tab:flux_J_IR3}, \ref{tab:flux_iso6.7_iras60} 
and \ref{tab:flux_70_1300}. The range of separation of the certified 
and candidate PMS objects is from about 0.8 to 6~arc-sec, the former 
being about the limiting separation we can resolve with the WFI images, 
mainly due to the seeing (see S07a). This range of separations means 
about 140 to 1060 AU at the distance of Cha~II. Considering that among 
the above objects there are 6 legitimate PMS objects and taking into 
account uncertainties on the final number of true cloud members, we 
find a multiplicity fraction of the order of 13$\pm$3\%, which is 
consistent with the observed fraction in PMS clusters in the same 
range of separations \citep[see][]{Pad97}. The result is also consistent 
with the binary fraction of 14.0$\pm$4.3\% reported by \citet{Koh01} 
for stars in the Chamaeleon complex, with separations of 0.13 to 
6~arc-sec. We do not see any dependence of binarity on spectral 
type, but the statistics are poor.

\section{Search for variability in the Spitzer bands}
\label{variability}

Since the Spitzer data were gathered in two epochs separated by about 
6 hours, we can search for variability on that timescale, similarly
as has been done in other c2d papers \citep{Har07, Reb07}. In order to 
be able to assess variability, the investigated sources need first to 
be detected with a good confidence level in both epochs and the 
artifacts due to cosmic rays, detections in image borders, etc, need 
to be cleaned up. 
Thus, in each band, we have selected all the sources detected in both 
epochs with a S/N ratio greater or equal than 3 and checked for cosmetics 
in connection with objects close to the image border or bright stars.

We did not find significant variability on the timescale of 6 hours 
for the Cha~II sources in any of the IRAC bands and in MIPS~24\,$\mu$m. 
This implies that none of the PMS objects or candidates show variability 
in the Spitzer bands on that timescale. This is in line with the c2d 
findings in other SFRs like Serpens \citep{Har07} and Perseus \citep{Reb07}.

\section{Circumstellar dust properties in Cha~II}
\label{disk_prop}

\subsection{Color-Color and Color-Magnitude diagrams}
\label{ccdiags}

Infrared color-color (CC) and color-magnitude (CM) diagrams are good 
diagnostic tools for the investigation of circumstellar matter around 
YSOs \citep[][and references therein]{Hart05, Lad06}. 
The traditional \citet{Lad87} classification of classes based on the 
slope of the SED alone may sometimes lead to misclassification. 
For instance, an embedded Class~I source seen pole-on may display a 
flat spectrum, while a normal disk seen edge-on can be confused with 
a Class~I object. In order to avoid confusion between the Lada Classes 
and their evolutionary status, \citet[][hereafter R06]{Rob06} introduced 
the term "Stages", which refers to the true evolutionary stage of the 
sources as follows: Stage~0-I have significant envelopes and possibly 
disks, Stage~II objects have optically thick disks, while sources with 
optically thin disks are classified as Stage~III sources.

\begin{figure}[!h]
\epsscale{1.0}
\plotone{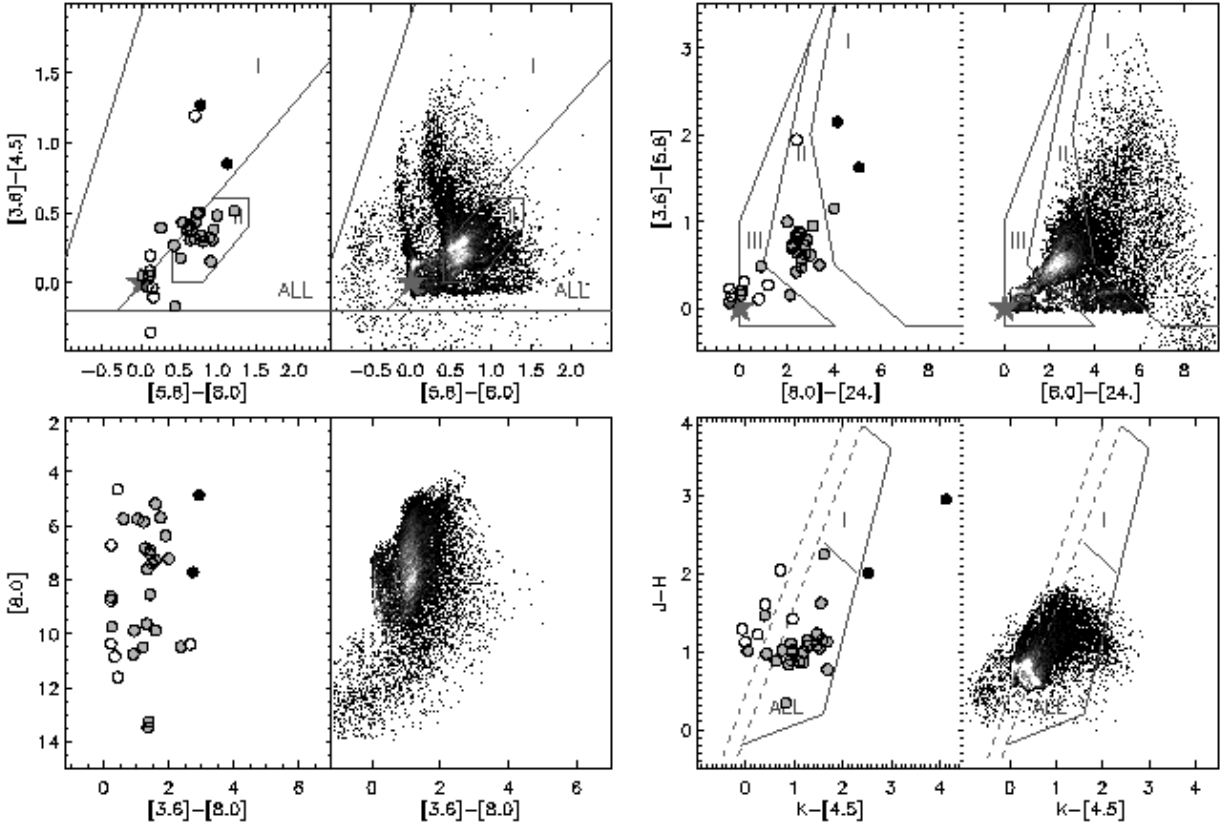}
\caption{The Color-color and color-magnitude diagrams for the spectroscopically
 confirmed (gray filled circles) and candidate (open circles) PMS objects 
 with Spitzer and 2MASS data are plotted in the left panels of each diagram. 
 The objects classified as Class~I sources are represented with black 
 filled circles. The colors derived from the SED models by R06 are plotted 
 (in gray-scale intensity representing the density of points) in the right 
 panels of each diagram. 
 The areas corresponding to the Stages I, II, and III as defined by 
 R06 are also indicated in each diagram. The label 'ALL' mark the 
 regions where models of all evolutionary stages can be present. The 
 big star represents normal unreddened photospheres. The majority of 
 PMS stars in Cha~II fall in regions corresponding to Stage II sources. 
 \label{cccmd}}
\end{figure}

About 60\% of the sample reported in Table~\ref{tab:yso} have sufficient 
data to perform an analysis in CC and CM diagrams similar to the one 
presented in previous c2d papers \citep[e.g.][]{Har07}. Figure~\ref{cccmd} 
shows CC and CM diagrams in the Spitzer and 2MASS bands. The diagrams 
are compared with those derived from the models by R06, adjusted 
in order to fit the distance and mass distribution of Cha~II and matching 
the sensitivity cutoff for the 2MASS \citep{Cut03} and Spitzer bands 
\citep{Har06, Jor06}. Though the number of objects in Cha~II for which 
we can perform this analysis is small, the range of colors and magnitudes 
are those expected for YSOs in different evolutionary stages. For instance, 
it is evident that only a handful of Stage~I sources exist in Cha~II. 
Only two YSOs have colors similar to Stage~I sources; these are 
Iso-ChaII~28 and IRAS~12500$-$7658, which were classified as 
Class~I IR sources in previous works \citep[][and references therein]{You05}. 
The comparison with the models indicates that the majority of the previously 
known PMS objects have colors consistent with those belonging to Stage~II, 
while most of the candidate PMS objects have IR colors similar to 
those of Stage~III; the exception is the new Spitzer YSO candidate 
SSTc2d~J130521.7$-$773810, which has colors consistent with Stage~II 
objects. 

Differently to what is found in  H07 for Serpens, we do not have sources 
with very red colors in the [8.0] vs. [3.8]-[8.0] diagram. The smaller 
percentage of very red objects might in principle be due to a more evolved 
evolutionary status of the YSO population; however, these evolved objects 
are not found in big numbers in Cha~II. Thus, the lack of very red 
objects is probably a consequence of the low number statistics. 
On the other hand, while the number of Stage~I sources is quite small
relative to the number of Stage~II sources, the disk fraction is quite 
high (see \S~\ref{disk_frac}). This may mean that either {\it i)} star
formation in Chamaeleon has occurred in a burst a few million years 
ago or {\it ii)} that the disk evolution time-scale may be different
as in other regions like Serpens, Perseus and Ophiuchus. The discussion 
on triggered star formation in Chamaeleon presented in \S~\ref{spa_distr_class} 
seems to be consistent with the former hypothesis. 

The IR colors of the very low-mass objects ($M < 0.2 M_{\odot}$) in Cha~II 
are very similar to those of Class~II sources. We use here the average 
mass for each individual object as determined by S07b. We note that the 
colors of these very low-mass objects are also consistent with those of 
Stage~II. This shows, as previous studies do \citep[cf.][]{All06a}, that 
disks are also common in very low-mass objects, down to the sub-stellar 
regime. 

The analysis of the SEDs reported in \S~\ref{sec_seds} implies that the 
conclusions drawn from the CC and CM diagrams discussed above can be 
extrapolated to the whole sample of 62 objects. Therefore we conclude 
that most objects in Cha~II have IR colors similar to those of Class~II 
sources, with only a few corresponding to Class~I and Class~III sources 
and that most PMS objects in this cloud are in the evolutionary Stage~II. 
A few Class~III objects may be missed, however, by the multi-color criteria 
by  H07 in the IRAC~3.6-8\,$\mu$m/MIPS~24\,$\mu$m overlapping area. More 
discussion on the source IR classes is provided in the next sections.

\subsection{Spectral Energy Distributions}
\label{sec_seds}

In order to study in more detail the circumstellar matter around
the certified and candidate PMS objects, we have determined their 
SEDs from the optical to mm wavelengths depending on available data. 
These SEDs, corrected for interstellar extinction, can be used to 
model the structure of the PMS circumstellar material. To this aim, 
the flux at each wavelength must be corrected for reddening.
The extinction has been determined for each individual PMS object 
and candidate using the spectral types provided in S07b and the 
extinction law by \citet{Wei01} for R$_V$=5.5 (hereafter WD5.5, see 
also \S~\ref{SFE}). For the details we refer the reader to S07b.
The uncertainties on $A_V$, derived from the uncertainties in 
temperature, range from 0.1 to less than 1~mag. 
Our extinction determinations are in general agreement with the 
values reported by the large-scale extinction maps of the Cha~II 
cloud by \citet{Cam99}, Kainulainen (private communication) and the 
c2d extinction map (discussed in \S~\ref{SFE}) within less than 2~mag
for objects with $A_V$ lower than about 4~mag, while for those with 
higher values the maps tend to provide higher $A_V$ values. This 
might mean that the certified and candidate PMS objects are not 
behind the cloud or may be a consequence of the large beams used 
to make the extinction maps. 

A different approach was followed for the objects lacking spectroscopy.
In these cases, the $A_V$ and temperature values were derived following 
the prescription by S07a, but using the WD5.5 extinction law. 
These methods yield consistent temperatures with those determined from
spectroscopy, with residuals less than 200~K provided the extinction 
is not high, which is the case for most of the candidate PMS objects 
in Cha~II. 
Moreover, since most of these candidates are expected to be Class~III 
sources (see \S~\ref{ccdiags} and \S~\ref{alpha}), those methods to estimate 
effective temperature can be applied confidently. 

In the case of the embedded Class~I sources IRAS~12500$-$7658 and 
Iso-ChaII~28, for which we do not know the temperature, we cannot 
apply the above procedures. Since most of the emission of these sources 
is in the IR, one way to determine their properties is by fitting an 
disk+envelope accretion model to their SED. 
From such fit it is possible to provide an estimate of the temperature 
and other physical parameters of the central object. This will be
done in \S~\ref{disk_pars}.   

\begin{figure}[!h]
\epsscale{1.0}
\plotone{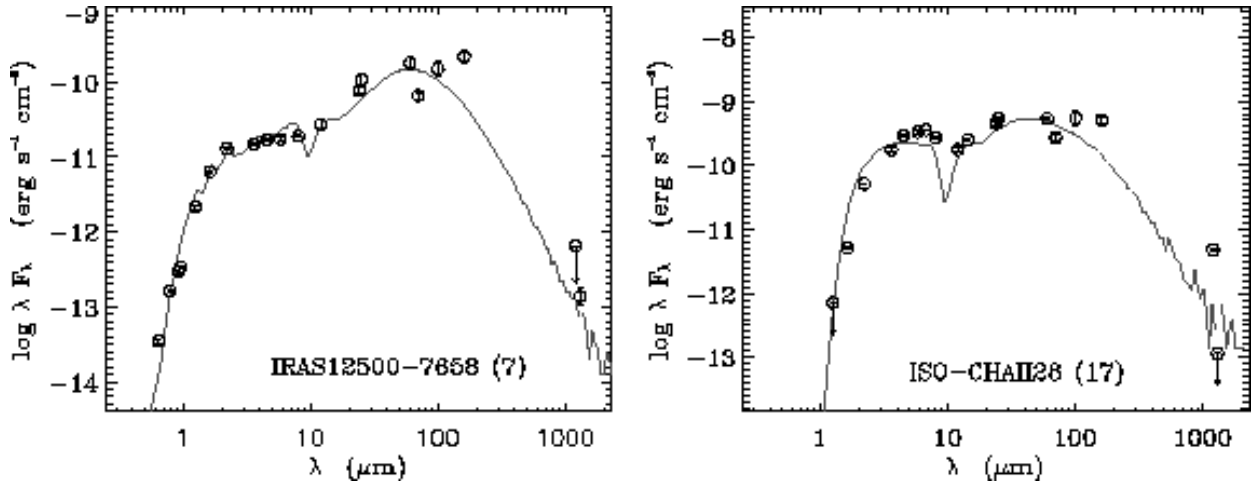}
\caption{Spectral energy distributions of the two Class~I YSOs in Cha~II.
       Complementary ancillary data in mm wavelengths for IRAS~12500$-$7658 
       are from \citet{Hen93}. The best fit of the SEDs derived from the
       R06 models as explained in \S~\ref{disk_pars} is 
       over-plotted as a solid line on each plot.
       \label{seds_classI} }
\end{figure}

\begin{figure} 
\epsscale{1.0}
\plotone{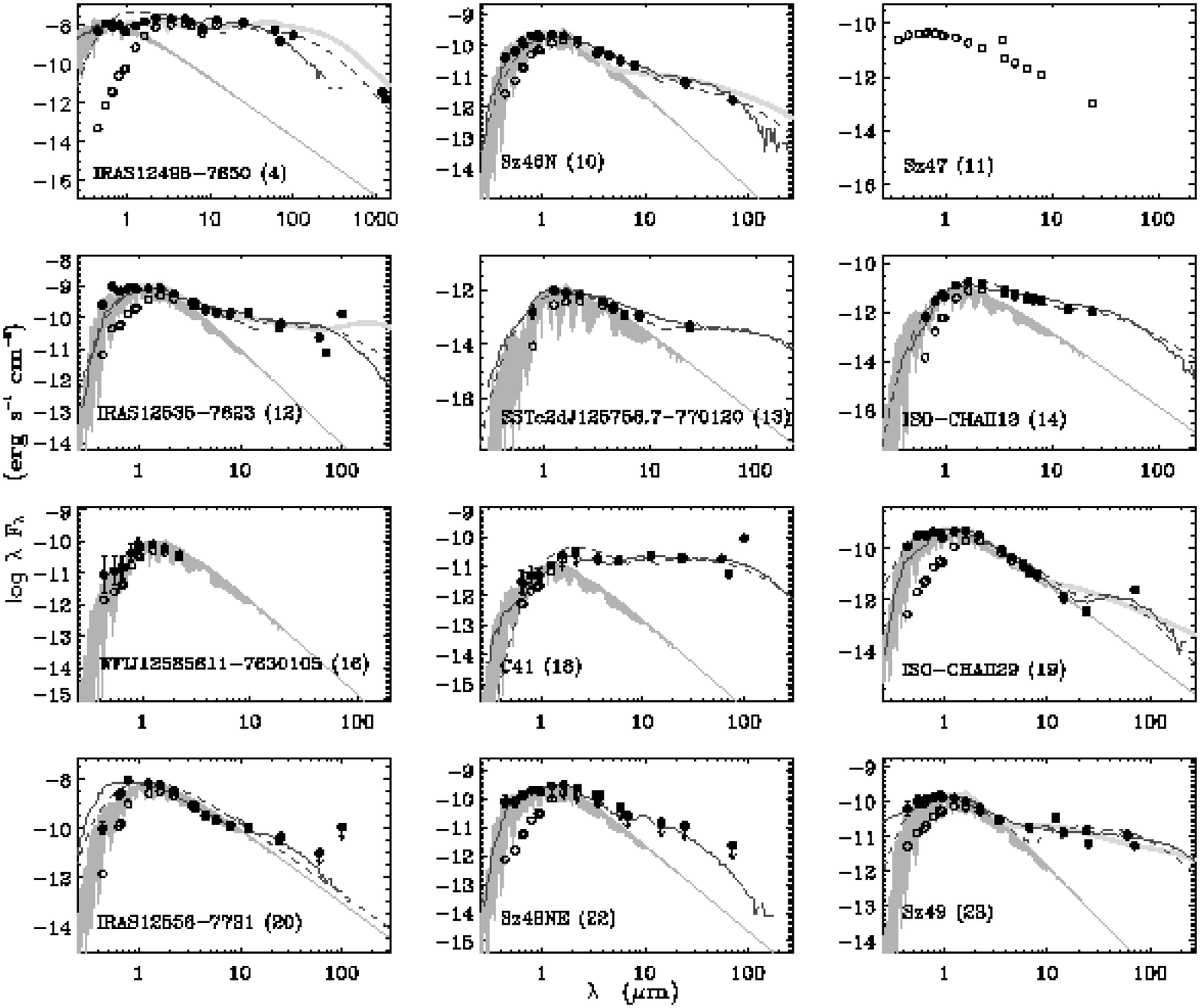}
\caption{Spectral energy distributions of the Class~II and Class~III 
       PMS objects in Cha~II. The dereddened fluxes are represented 
       with solid dots, while the observed fluxes are represented 
       with open circles. The best-fitNextGen spectrum 
       \citep[by][for objects with T$_{\rm eff } >$ 4000~K]{Hau99}
       or StarDusty spectrum 
       \citep[by][for objects with T$_{\rm eff } <$ 4000~K]{All00} 
       is over-plotted on each SED, representing the stellar flux. 
       The best fits of the SEDs derived from both the R06 and D01 
       models are over-plotted as a solid and dashed red lines, 
       respectively, while those derived from the \citet{Dal05} 
       models are represented with the thick pink lines. Note that 
       a reasonable fit to the stellar and dust emission of the 
       veiled object Sz~47 could not be obtained, while for 
       WFI~J12585611$-$7630105 the dust emission could not be modeled 
       because of insufficient IR photometry. 
       The modeling was not attempted for those sources showing 
       pure-photospheric Class~III SEDs (cf. \S~\ref{disk_pars}). 
       Each panel is labeled with the object's name and its entry 
       number from Table~\ref{tab:yso} in parenthesis.
       \label{seds}}
\end{figure}

The observed SEDs of the two Class~I objects are presented in 
Figure~\ref{seds_classI}, while in Figure~\ref{seds} those for the Class~II 
and Class~III objects are shown. The SEDs for the candidate PMS objects 
are presented in Figure~\ref{seds_cand}. Although the SEDs for a few of 
these sources have already been presented in previous works that use 
c2d data \citep[][and S07a]{You05, Cie05, All06a, Alc06}, here we include 
all of them for completeness. 
Moreover, we have incorporated in the SEDs additional data in the optical 
from the NOMAD dataset \citep{Zac05}, in the IR from the ISO observations 
by \citet{Per03} and in mm wavelength from the observations by \citet{Hen93}. 
For objects saturated in the R-band, we also used the the data corresponding 
to the continuum H$\alpha$ filter reported in S07a.

\begin{figure}[!h]
\epsscale{1.0}
\plotone{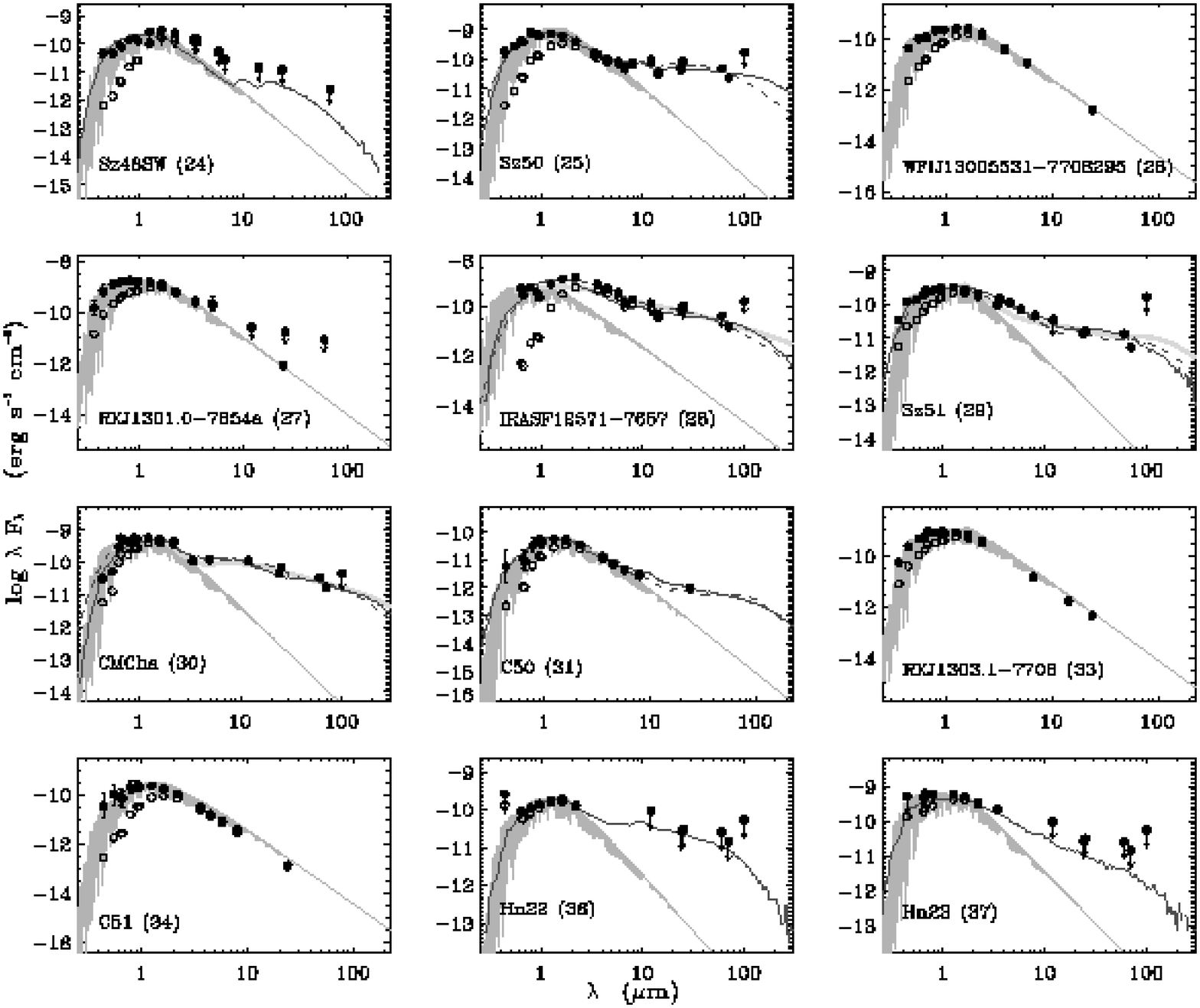}
Fig.~\ref{seds} - Continued 
\end{figure}


\begin{figure}[!h]
\epsscale{1.0}
\plotone{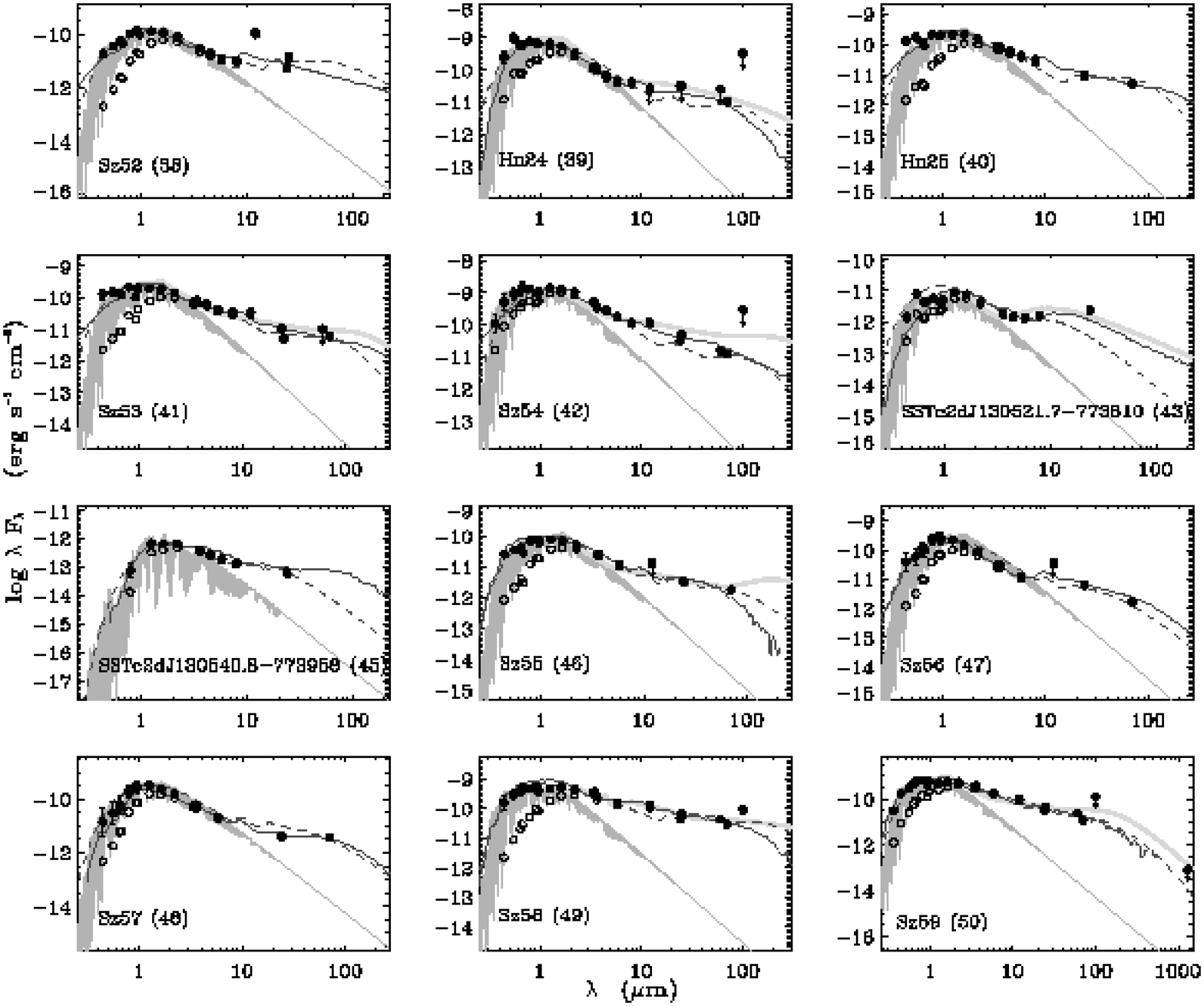}
Fig.~\ref{seds} - Continued
\end{figure}


\begin{figure}[!h]
\epsscale{1.0}
\plotone{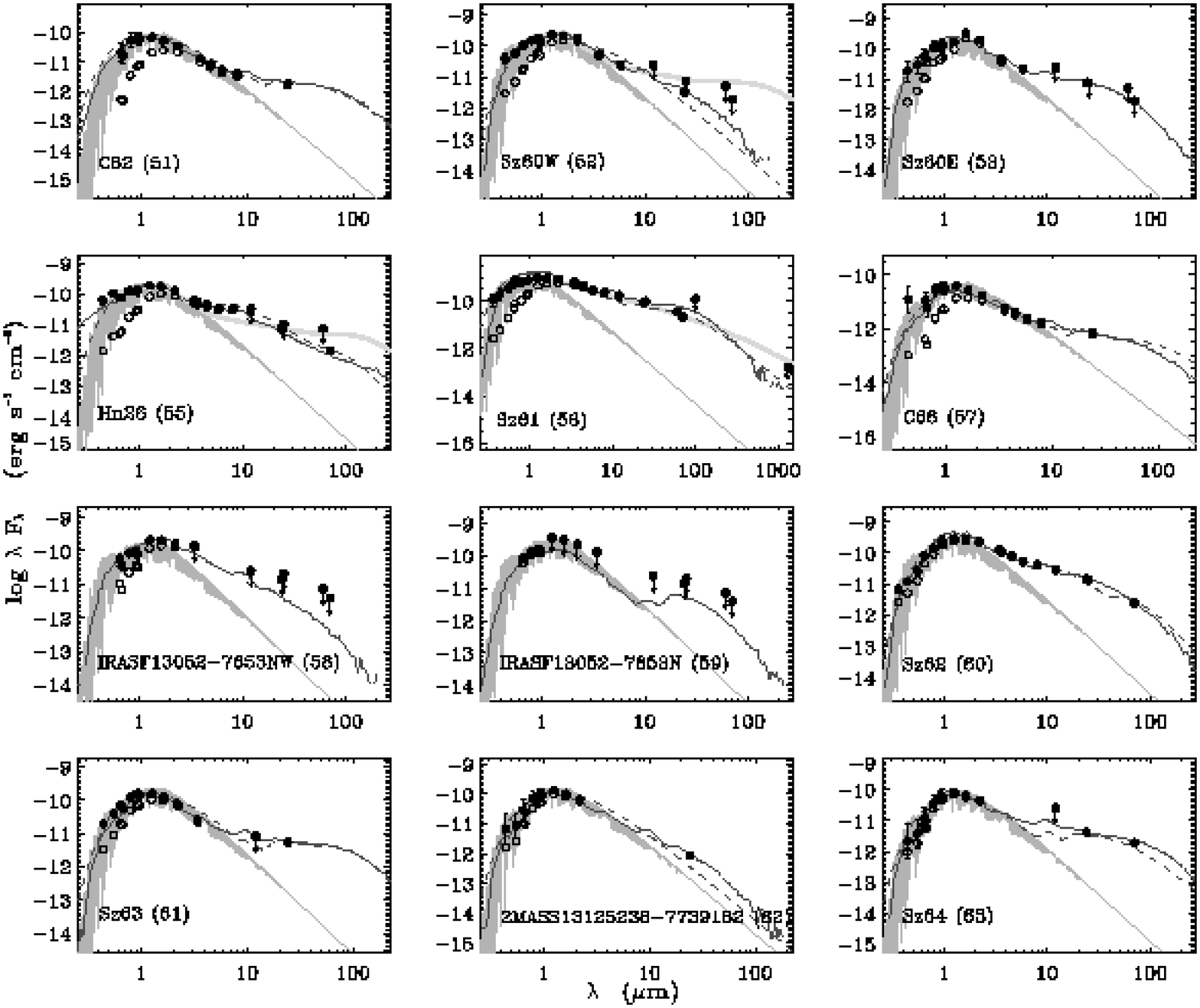}
Fig.~\ref{seds} - Continued
\end{figure}

\begin{figure}[!h]
\epsscale{1.0}
\plotone{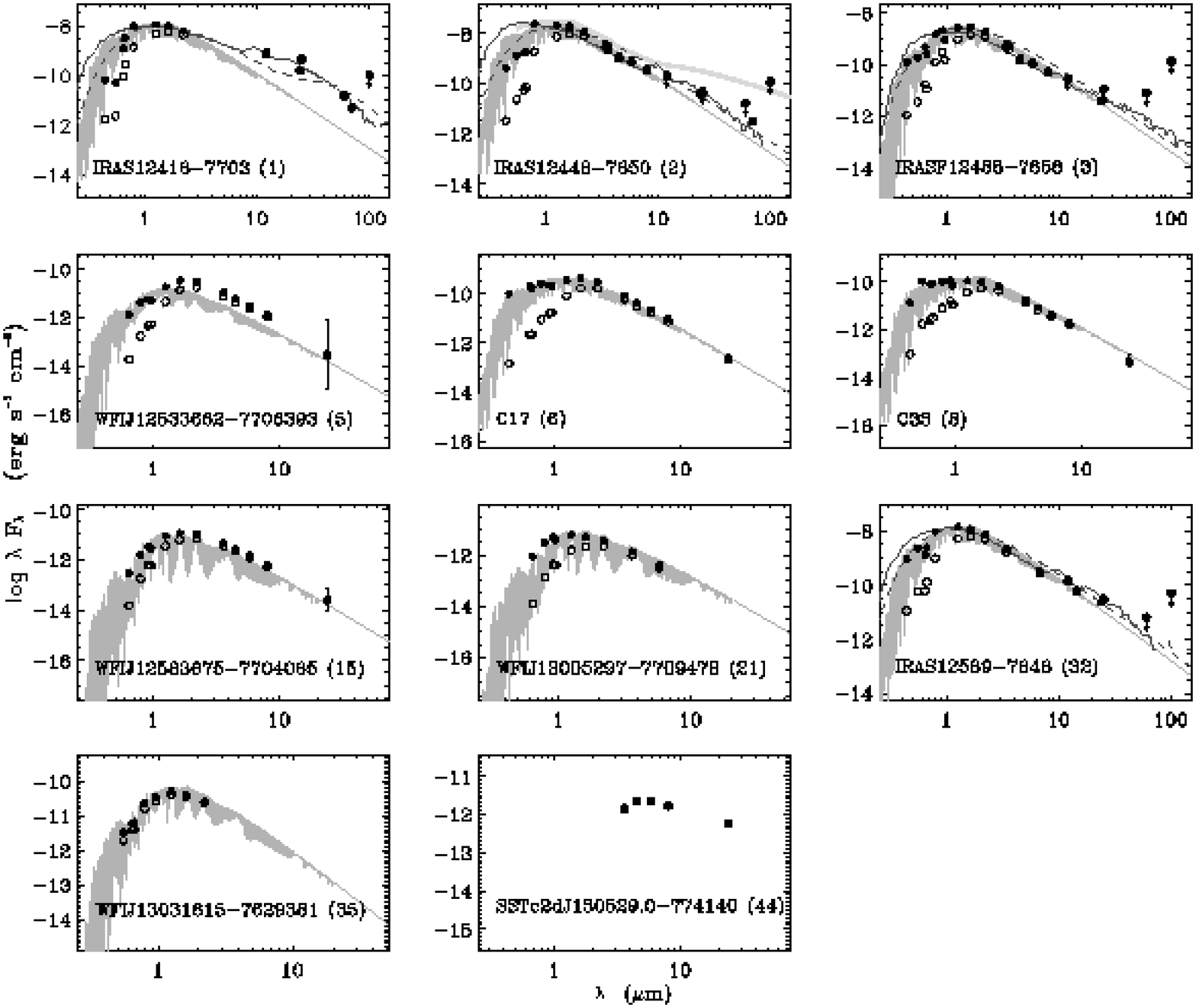}
\caption{Spectral energy distributions of the candidate PMS objects in 
       Cha~II. The symbols are as in Figure~\ref{seds}. The best-fit 
       NextGen spectrum \citep[by][for objects with 
       T$_{\rm eff } >$ 4000~K]{Hau99} or StarDusty spectrum 
       \citep[by][for objects with T$_{\rm eff } <$ 4000~K]{All00} 
       is over-plotted on each SED. The best fits of the SEDs derived 
       from both the R06 and D01 models are over-plotted as a solid 
       and dashed red lines, respectively, while those derived from the 
       \citet{Dal05} models are represented with the thick pink lines. 
       Fits were not possible for SSTc2d~J130529.0$-$774140 and 
       WFI~J13031615$-$7629381 because of insufficient IR photometry. 
       Likewise, the non detection of SSTc2d~J1305290$-$774140 in  
       optical bands prevents any estimate of its stellar 
       parameters. The modeling was not attempted for those 
       sources showing pure-photospheric Class~III SEDs 
       (cf. \S~\ref{disk_pars}).
       \label{seds_cand}}
\end{figure}

Except for a few flat SEDs, most objects in Cha~II possess typical 
SEDs where the flux smoothly drops with wavelength, depending on the 
Lada Class. This may be interpreted in terms of disks that evolve 
approximately homogeneously from flared to flat \citep{Lad06}. However, 
there may be objects in which the SED, after dropping normally to a 
certain wavelength, rises at longer wavelengths to a flux level 
comparable to that of flat-spectrum or Class~II SEDs. 
An example of such type of objects in our sample is Iso-ChaII~29,
with its SED rising beyond 24$\mu$m (see Figure~\ref{seds}). 
Such SED behavior is explained in terms of an inner disk hole \citep{Lad06}. 
These type of sources, better known as transition objects, are interesting 
because they represent a class of objects where the inner part of the disk 
may have already been cleared \citep[][and references therein]{Cal05}.
The number of these type of objects may depend on the detection limit 
at $\lambda \ge$ 24$\mu$m \citep{Lad06}. In Cha~II the number statistics 
are rather poor to make an estimate of their frequency.

\subsection{The $\alpha$ index and Lada Classes}
\label{alpha}

The spectral index $\alpha$ can be used to further investigate the different
object classes \citep{Lad06} in Cha~II. Traditionally, this index is determined 
as the slope of the SED at wavelengths longer than 2$\mu$m. However, the 
index may vary significantly depending on the range of wavelengths and method 
used to derived it. R06 discuss several ways to determine the index and the 
consequences of using the various combinations of fluxes at different 
wavelengths. 
We have followed their prescription in order to determine the index 
$\alpha_{[K \& {\rm MIPS~24\,\mu m}]}$ for the Cha~II certified and candidate 
PMS objects. We stress that one of the c2d products is the determination 
and inclusion of the spectral index $\alpha$ in the catalogs. Within the c2d, 
the index is calculated as described in the final c2d delivery document 
\citep{Eva07}, i.e. the slope of the least-squares fit line to all the flux 
measurements between the K and MIPS~24\,$\mu$m  bands, while 
$\alpha_{[K \& {\rm MIPS~24\,\mu m}]}$ is calculated as the slope of the line 
joining the flux measurements at K and MIPS~24\,$\mu$m. 
We have compared the c2d $\alpha$ index with $\alpha_{[K \& {\rm MIPS~24\,\mu m}]}$ 
and find a very good agreement, within an RMS$<$0.16, for all the sources  
in Cha~II detected with a S/N $>$ 3 in the bands between $K$ and MIPS~24\,$\mu$m. 
Thus, since the R06 models are used in \S~\ref{disk_pars} for the 
SED fitting, for the sake of homogeneity and for discussions regarding 
the different Lada classes in Cha~II, we use the reference index, 
$\alpha_{[K \& {\rm MIPS~24\,\mu m}]}$, as defined by R06, but adopting 
the Lada class separation as extended by \citet{Gre94}, i.e. 
$\alpha_{[K \& {\rm MIPS~24\,\mu m}]} \geq 0.3$ for Class~I, 
$-0.3 \leq \alpha_{[K \& {\rm MIPS~24\,\mu m}]} < 0.3$ for Flat sources,
$-1.6 \leq \alpha_{[K \& {\rm MIPS~24\,\mu m}]} < -0.3$ for Class~II sources 
and $\alpha_{[K \& {\rm MIPS~24\,\mu m}]} < -1.6$ for Class~III sources. 
The $\alpha_{[K \& {\rm MIPS~24\,\mu m}]}$ slope is reported in the third 
column of Table~\ref{disk_results_lum}.

In Table~\ref{ir_classes}, the statistics of the Lada classes for the 
YSOs and PMS objects and candidates in Cha~II is given. For our purpose, 
IRAS~13036$-$7644 (or BHR~86) is included in this table as a Class~I 
source \citep[see][]{Leh05}. 
It is interesting to note that including all the PMS objects nearly 
doubles the number of the Class~II sources over the YSOs. The reason 
is that the IRAC~3.6-8\,$\mu$m overlap coverage missed about half of 
the Class~II objects in Cha~II. Assigning a Lada class to those PMS 
objects off the IRAC~3.6-8\,$\mu$m overlap area is not a problem because 
only the $K$ and MIPS~24\,$\mu$m  fluxes are needed to calculate the 
$\alpha_{[K \& {\rm MIPS~24\,\mu m}]}$ index\footnote{We have, in 
any case, verified the Class~II classification of these objects using 
IR photometry from the literature and, when available, also the 
IRAC~3.6\,$\mu$m/IRAC~5.8\,$\mu$m or IRAC~4.5\,$\mu$m/IRAC~8\,$\mu$m 
data. We remind the reader that, due to the 7~arc-min East-West shift 
between the IRAC~3.6\,$\mu$m/IRAC~5.8\,$\mu$m and IRAC~4.5\,$\mu$m/IRAC~8\,$\mu$m 
observations \citep[see end of \S~\ref{sample} and right panel of Figure~1 by][]{Por07}, 
some objects located off the IRAC~3.6-8\,$\mu$m overlap area, but on the 
eastern border of the IRAC~3.6\,$\mu$m/IRAC~5.8\,$\mu$m observations, 
possess only IRAC~3.6\,$\mu$m/IRAC~5.8\,$\mu$m data while others, on 
the western border of the IRAC~4.5\,$\mu$m/IRAC~8\,$\mu$m observations, 
possess only IRAC~4.5\,$\mu$m/IRAC~8\,$\mu$m data.}. Thus we can classify 
those too. Only for 2 out of the 62 PMS objects and candidates it was 
not possible to assign a Lada Class because of insufficient photometry
(see Tables~\ref{tab:flux_opt}, \ref{tab:flux_J_IR3}, \ref{tab:flux_iso6.7_iras60}, 
and \ref{tab:flux_70_1300}).  
One, WFI~J12585611$-$7630105, is a certified PMS object, whereas the 
other, WFI~J13031615$-$7629381, is a candidate PMS object (S07a). Based 
on their position in the $J-H$ vs. $H-K$ diagram we can classify
both objects as Class~III. These two objects are, in any case, 
indicated with questions marks in Table~\ref{ir_classes}.
Regardless of the inclusion or not of the candidates PMS objects, the 
dominant objects in Cha~II are those of Class~II, with only a few 
Class~I sources. Another 
interesting quantity is the ratio of the number of Class~I and flat sources 
to the number of Class~II YSOs and Class~III sources; regardless of the 
inclusion of the candidate PMS objects, this ratio is on the order of 0.1, 
i.e. a factor of about two less than in Serpens (H07). Note that, in 
addition, some Class~III sources in the IRAC~3.6-8\,$\mu$m/MIPS~24\,$\mu$m 
overlapping area might be missed by the multi-color selection criteria.

In summary, in Cha~II the Class~II sources seem to represent $\sim$60\% 
of the population, the Class~III sources $\sim$30\%, with only a 
minority (less than 10\%) being Class~I or flat-spectrum sources.
These conclusions are in very good agreement with those drawn from the 
color-color and color-magnitude diagrams reported in \S~\ref{ccdiags}. 
We also conclude that in Cha~II the Lada Classes are quite consistent 
with the IR Stages as defined by R06. Therefore,  
$\alpha_{[K \& {\rm MIPS~24\,\mu m}]} = -1.6$  is a good definition 
for the borderline between optically thick and thin disks in Cha~II. 
This limit will be used in \S~\ref{disk_frac} to estimate the 
fraction of optically thick disks.

\subsection{Spatial distribution and clustering of the PMS objects}
\label{spa_distr_class}

\begin{figure}[!h]
\epsscale{1.0}
\plotone{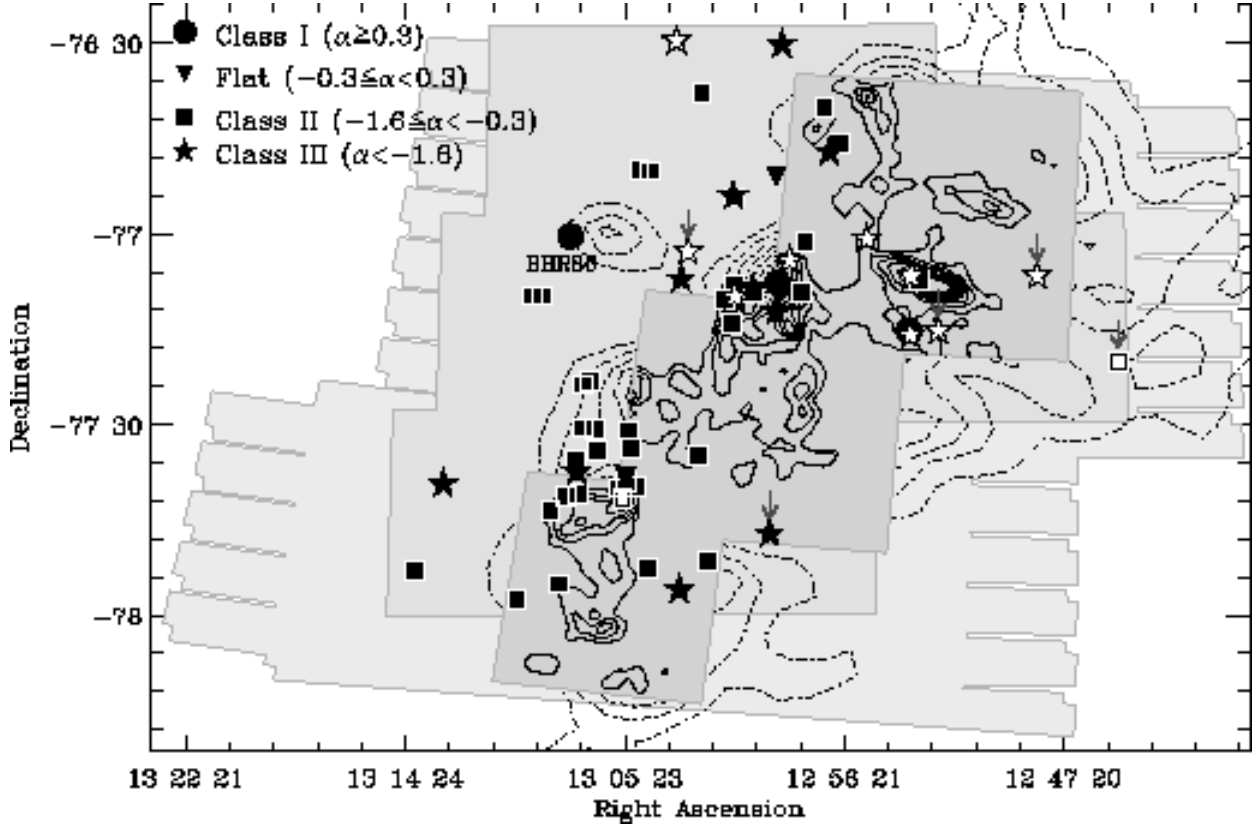}
\caption{Spatial distribution of the certified (filled symbols) and 
       candidate (open symbols) PMS objects in Cha~II as function 
       of Lada Class, over-plotted on the contours from the c2d 
       extinction map (continuous lines, see \S~\ref{SFE}). 
       The contour levels of extinction are from 2~mag to 20~mag,
       in steps of 2~mag. The down arrows mark the five optically
       bright IRAS sources discussed in the text (cf. \S~\ref{disk_pars}). 
       The shaded areas, from light to dark-gray, display the regions 
       observed with MIPS, WFI and the IRAC~3.6-8\,$\mu$m overlap, 
       respectively. The dashed lines outside the IRAC area are the 
       contour levels of extinction from \citet{Cam99}, from 1~mag 
       to 6~mag in step of 0.35~mag. The higher resolution of the 
       c2d extinction map with respect to the \citet{Cam99} map can 
       be appreciated.
	\label{spa_distr}}
\end{figure}

The spatial distribution of the different classes of objects in Cha~II 
is shown in Figure~\ref{spa_distr} over-plotted on the c2d extinction map, 
derived as explained in \S~\ref{SFE}. Note that the latter covers only the 
overlap area observed in all IRAC bands. However, from other maps 
\citep{Cam99} it is evident that there is only one additional extinction 
peak off the area covered with IRAC, which is close to IRAS~13036$-$7644 
or BHR~86, the easternmost Class~I source in Cha~II (see Figure~\ref{spa_distr}). 
Note that the Class~I sources coincide or are located close to the 
sites of highest extinction. 

The arrows in Figure~\ref{spa_distr} indicate the five optically bright 
sources IRAS~12416$-$7703, IRAS~12448$-$7650, IRASF~12488$-$7658, 
IRAS~12556$-$7731 and IRAS~12589$-$7646. These sources are scattered over 
a large area, in regions of quite low extinction. The former four are 
mainly located to the west of the dust-lanes of the main cloud body, 
while the latter one lies to the east, between the main cloud and BHR~86. 
As demonstrated in S07b, these five sources are over-luminous 
in the HR diagram when adopting the Cha~II distance to compute 
their luminosity and hence, might be unrelated to Cha~II. However, 
strong Lithium absorption at 6708\AA~ in the optical spectrum of 
IRAS~12556$-$7731 has been detected (S07b); the presence of 
such line is a necessary condition to assess the youth of low-mass 
stars and is used as a diagnostic for the sub-stellar nature of 
objects below the Hydrogen burning limit. Further discussion on 
the five IRAS sources will be presented in S07b.

The vast majority of the certified and candidate PMS objects are located 
to the Eastern edge of the cloud, with several of them being off the cloud 
boundaries as marked by the extinction maps. This may lead to the idea 
that some agent triggering star formation from the East of the cloud 
might have occurred.  
For the Chamaeleon cloud complex there is no evidence of supernova events 
or the presence of OB associations, but a scenario in which a high-velocity 
cloud with a velocity of about 100 km/s collides with the galactic plane 
at an inclination angle of about 60$^{\circ}$ has been proposed by \citet{Lep94}. 
These authors claim that such event could be the responsible for the large 
scale filamentary structure of the Chamaeleon clouds, and hence also of 
Cha~II. It is suspected that the young stars in the cloud complex are moving 
towards lower galactic longitudes, approximately in direction $l=300^{\circ}$ 
to $l=280^{\circ}$, with a few km/s, almost parallel to the galactic plane. 
The spatial distribution of the stars in Chamaeleon shown in Figure~10 
by \citet{Duc05} somewhat traces the direction of the velocity vector of 
the hypothetic high-velocity cloud. Proper motion studies 
\citep[e.g.][and references therein]{Duc05} seem to confirm this, although 
the errors are quite large and the results may be blurred by the solar 
reflex motion \citep{Tei00}. 

\begin{figure}[!h]
\epsscale{0.75}
\plotone{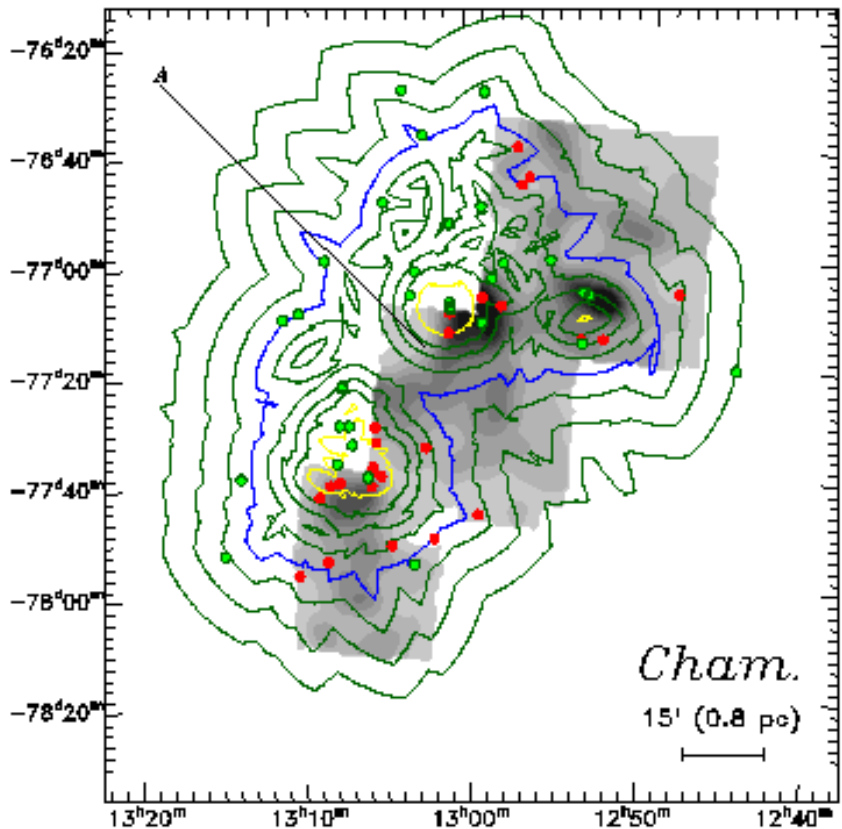}
\caption{Volume density plot of Cha~II as determined from the nearest-neighbor 
       algorithm described in \citet{Jor07}. The derived density contours are 
       compared with the c2d extinction map (in gray scale). The blue contour
       corresponds to 1.0~$M_{\odot}$\,pc$^{-3}$ density level, while the 
       green ones correspond to levels of 0.125, 0.25, 0.5, 2, 4 and 8 
       times this level. The yellow contour corresponds to the 
       25~$M_{\odot}$\,pc$^{-3}$ density level. The two tight groups
       in Cha~II are marked. Red dots represent the YSOs, while the green 
       dots represent the rest of the PMS objects.
	\label{surf_dens_fig_chaII}}
\end{figure}

To compare the sample of clouds mapped by c2d it is important to examine 
the distribution of YSOs in a uniform way. \citet[][hereafter LL03]{Lad03}
suggested that a cluster should be a group of some 35 members with a total
mass density larger than 1.0~$M_{\odot}$\,pc$^{-3}$. To test this criterion 
we calculated the volume density of YSOs for each of the c2d clouds using 
a nearest-neighbor algorithm similar to the one applied by \cite{Gut05}
\citep[see][for details]{Jor07}, assuming that the distribution of sources 
is locally spherical. Generally the LL03 criterion is found to be
loose to define sub-structures and pick out known clusters in the larger 
scale clouds. We thus adopt the tighter level of 25 times the LL03 criterion 
(i.e. a mass density of 25~$M_{\odot}$\,pc$^{-3}$), which normally provides 
the already established cluster and group boundaries. 
We refer to ``clusters'' as regions with more than 35 YSOs within 
a given volume density contour and ``groups'' as regions with less. 
The lowest number of YSOs that is considered to constitute a separate 
entity is 5. ``Clusters'' and ``groups'' can be either ``tight'' or ``loose'' 
depending on whether their volume densities, $\rho$, are higher than 
25~$M_\odot {\rm pc}^{-3}$ or 1~$M_\odot {\rm pc}^{-3}$ (corresponding 
to 50 and 2 YSOs~pc$^{-3}$, respectively assuming an average YSO mass of
0.5~$M_\odot$). We note that although these criteria are useful as a way of
making direct comparison between regions within clouds and across different
clouds such as those in the c2d sample, this method is empirical and should
not be taken, e.g., as evidence in discussions on whether the star formation
process is hierachical or not. For Cha~II we applied the algorithm to the
whole sample of 62 certified and candidate PMS objects. The overall results,
reported in Table~\ref{surf_dens}, are shown in Figure~\ref{surf_dens_fig_chaII}. 
For the sake of comparison with the other c2d clouds, we use the boundaries
defined by the whole sample to investigate also the clustering of the 
26 YSO selected from the c2d criteria and also report their statistics 
in Table~\ref{surf_dens}.

The clustering analysis gives two tight groups in Cha~II at the level of 
25 times the LL03 criterion; neither have enough members to qualify them 
as separate clusters. 
These groups, that we call A and B in Table~\ref{surf_dens}, can be identified 
by eye in Figure~\ref{spa_distr}. Group A, at around RA=13:01 and DEC=$-$77:06, 
lies close to the position of the Class~I source Iso-ChaII~28, while group B 
at around RA=13:07 and DEC=$-$77:35, is further to the South-East of group A. 
The two groups apparently coincide with regions of high extinction and 
several Class~II sources are distributed around them, while the Class~III 
sources seem to be more scattered all over the region. On the other hand, 
Cha~II as a whole can be defined as a loose cluster at the level of one time 
the LL03 criterion. Most of the sources (48 out of 62; 75\%) in ChaII are 
associated with this cluster which naturally also includes both the "A" 
and "B" group from above (1/3 of the cluster members). The overall results are 
also given in Table~\ref{surf_dens}. At this level, the remaining 14 sources can 
be considered as part of the extended population in Cha~II. Finally, by applying
the 1$\times$LL03 criterion to the restricted YSO sample but using the group
boundaries identified from the larger sample of sources, we recover only 
group B, but with 6 members.

Interestingly, the Class~III sources are more abundant in the loose cluster 
than in the other sub-structures. This may lead to the idea that those objects 
may have formed more closely together and then dispersed during the evolution 
of the star forming region. Assuming the typical velocity dispersion of 
1~km~s$^{-1}$ of young clusters, 2-3~Myr (i.e. the typical age of Cha~II members, 
see S07b) old stars would move about 2\,pc from their birth sites. 
However, the more scattered Class~III sources are located at less than 1.5\,pc 
from the tight groups. Therefore, it is more likely that those objects were 
formed closer to their present position rather than in one of the tight groups. 
 

\subsection{Luminosities}
\label{lum}

One way to determine the degree of accretion activity of a PMS population 
is by means of the frequency of the disk/envelope luminosity of its members. 
To do this we calculated the luminosity of each PMS object as follows: 
{\em i)} the ''total" luminosity was first computed by direct integration 
of its dereddened SED, assuming isotropic radiation and the distance 
$d=178$~pc \citep{Whi97}. We have extrapolated the SED beyond the available 
terminal wavelength using the method by \citet{Coh73};
{\em ii)} for each PMS object the stellar luminosity was also computed 
by integrating the model spectrum \citep{All95, All00} corresponding to 
the PMS object temperature or, in the case of the candidate PMS objects, 
to the model spectrum which best fits the dereddened SED as explained 
in S07a. 

\begin{figure}[!h]
\epsscale{0.7}
\plotone{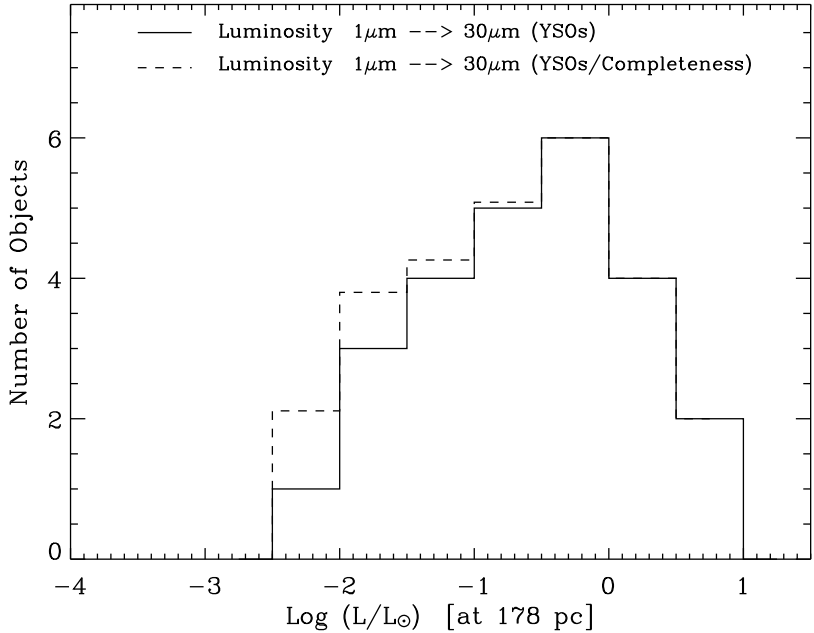}
\caption{Luminosity distribution for the YSOs in the overlapping 
area surveyed with IRAC and MIPS in Cha~II (solid histogram). 
The plotted luminosities were determined as in H07, i.e. by integration
of the SEDs from 1 to 30,$\mu$m. The corrected luminosity distribution, 
determined by applying completeness factors at each luminosity bin as 
in  H07, is over-plotted (dashed histogram). Note that IRAS~12496$-$7650 
is not considered in these histograms, becuase it is saturated in 
IRAC~3.6\,$\mu$m and IRAC~4.5\,$\mu$m (see also \S~\ref{sel_ysos}).
\label{completeness}}
\end{figure}

Using the total luminosity calculated as explained in {\em i)} above, 
we have investigated the expected number of low-luminosity objects 
and completeness in the IRAC~3.6-8\,$\mu$m/MIPS\-~24\,$\mu$m overlap area 
in Cha~II. To this aim, we explored similar ideas to those exposed in  
H07 regarding luminosity histograms and completeness limits in the 
1-30~$\mu$m range. 
Since the full SWIRE catalog is much deeper than the c2d one, it can 
be considered as a reference catalog with 100\% completeness. Both the 
full SWIRE catalog and a "trimmed" version of it, which matches the 
sensitivity of the c2d observations in Cha~II, are thus used for 
completeness determinations.
In Figure~\ref{completeness} two luminosity histograms are shown: one for 
the YSOs, while the other one is the luminosity histogram corrected for 
completeness according to the guidelines described in H07. 
Both histograms are produced assuming the distance of 178~pc for Cha~II. 
The luminosity histograms show that only a very small number ($\sim$2) 
of additional low-luminosity ($log L/L_{\odot} < - 1.7$) YSOs are still 
expected to be found. Therefore, we conclude that the sample of YSO 
candidates in Cha~II, as provided by the observations in the 
IRAC~3.6-8\,$\mu$m/MIPS~24\,$\mu$m overlap area, is fairly complete. 
We recall, in addition, that the two very low-luminosity objects 
discovered by \citet{All06a} were not selected from the multi-color 
criteria by H07 (cf. Table~\ref{tab:yso}) because they are too faint 
in MIPS~24\,$\mu$m\footnote{MIPS~24\,$\mu$m  fluxes for these two objects 
are provided by the band-filling process, but errors are large to be 
accepted by the H07 criteria.}. 
Hence, they are not among the objects producing the solid histogram 
in Figure~\ref{completeness}. These two objects are, in any case, 
included in our census making our sample of YSOs complete down 
to $log L/L_{\odot} \approx - 2.5$. Though slighly higher, this limit 
is consistent with the one derived from the limiting magnitudes for 
a complete sample in Cha~II, based on the histograms of source 
number counts, reported in the basic data papers by \citet{Por07} 
and \citet{You05} for the IRAC and MIPS observations, respectively. 

\begin{figure}[!h]
\epsscale{0.8}
\plotone{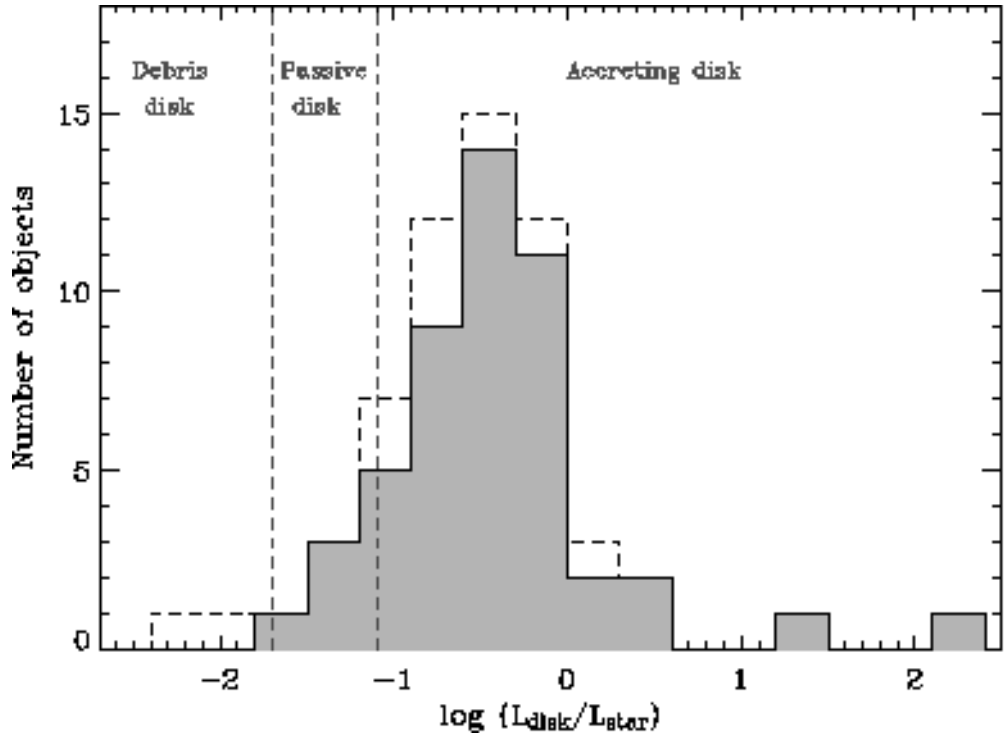}
\caption{Distribution of the L$_{disk}$/L$_{star}$ ratio of the confirmed
       PMS objects (solid histogram) and when adding the candidate PMS 
       objects (dashed histogram). The dashed vertical lines mark the 
       typical boundaries of active, passive and debris disks as defined 
       by \citet{Ken87}. Most objects in Cha~II are in the accretion 
       phase. \label{Ldisk}}
\end{figure}

In order to study the excess luminosity, we compare the total luminosity 
with the one emitted only by the central object: the difference between 
the two luminosities derived as in {\em i)} and {\em ii)} above, can be 
attributed to the circumstellar material. Figure~\ref{Ldisk} shows the 
distribution of the residual luminosity, i.e. of the L$_{disk}$/L$_{star}$ 
ratio, for the Cha~II certified and candidate PMS objects. From this 
figure it is evident that the Cha~II population is dominated by objects 
with active accretion, in agreement with the conclusions of \S~\ref{alpha}.
It is also evident that the average PMS object in Cha~II is the one for 
which the disk luminosity is about one third of the stellar luminosity. 
In the case of the Class~I sources we have derived the
L$_{disk+envelope}$/L$_{star}$ ratio after the SED model fits performed 
in \S~\ref{disk_pars}. The models provide the approximate temperature of 
the central object from which it is possible to derive a luminosity. 

\subsection{Disk modeling}
\label{disk_pars}
 
Understanding the evolution of accretion disks can provide strong 
constraints on theories of planet formation and evolution 
\citep[][and references therein]{Lyn74, Ada87, Ken87, Bac93, Lag00, 
Zuc01, Dul07} and measuring the lowest mass at which young objects 
harbor circumstellar disks is crucial for determining whether 
planets can form around low-mass BDs \citep{Luh05a, Luh05b, Lad06}.

In order to further investigate the circumstellar material of the 
Cha~II PMS objects, we have modeled the SEDs using the 
\citet[][hereafter D01]{Dul01}, \citet{Dal05} and R06 SED models. 
While the D01 models consider reprocessing 
flared disks, \citet{Dal05} and R06 simulate the SEDs of PMS stars 
with different combinations of accretion disk/envelope parameters. 
The sampling of the parameter space is quite different among the 
three prescriptions. For instance, in the \citet{Dal05} models the 
central object temperature is in steps of 500~K, with a minimum 
star temperature of 4000~K and a minimum stellar mass of 0.7 M$_{\odot}$; 
in addition, such models are provided only for two disk inclination 
angles (30$^\circ$ and 60$^\circ$).
The sampling is much better in the D01 and R06 model grids, but the 
minimum stellar mass for which the R06 models are valid is 0.1~M$_{\odot}$; 
thus, for objects far below this, the modeling results must be treated 
with caution. 
Flux upper limits are taken into account in the SED fitting tool by 
\citet[][hereafter R07]{Rob07}, while in the D01 and \citet{Dal05} 
prescriptions no upper limits are considered; however, when using the 
D01 and \citet{Dal05} models, we checked the consistency between the best 
fit model and possible flux upper limits. 

Taking all this into account, the SED fits were performed using 
the following criteria:

{\it i)} the models are done for $star+disk/envelope$ systems; thus, 
SED fits of pure-photospheric Class~III SEDs  cannot be attempted;

{\it ii)} SEDs with only flux upper limits in their IR part were 
fitted using only the R06 models. The D01 and \citet{Dal05} models 
were fitted only if a reasonable number of IR absolute fluxes are 
available; 

{\it iii) } for the fits, the temperature of the central object 
was fixed using the values derived in S07b\footnote{Fixing 
the star temperature is not foreseen in the R07 
web-interface. Thus, an IDL procedure was used which reads 
the grid of output models and selects the one that, matching 
input stellar parameters, best fits the SED.}. 
The \citet{Dal05} model fits were attempted only for objects with 
T$_{\rm eff} >$3500~K, selecting the model object temperature 
closest to the one provided in S07b;

{\it iv)} the fits of the \citet{Dal05} models were performed
by normalizing the modeled SEDs to the derreddened I-band flux,
where the possibility of excess is minimised;

{\it v)} the fits of the Class~I SEDs were done using only 
the R06 models;

{\it vi)}  for objects not detected in the optical, like the Class~I 
source Iso-ChaII~28, we have set all R07 input parameters 
free. 

In Figures~\ref{seds_classI}, \ref{seds} and \ref{seds_cand} the resulting 
best fits are over-plotted as red continuous and dashed lines
for the R06 and  D01 models respectively, while those 
derived from the \citet{Dal05} models are represented with thick pink 
lines. The general results are reported in Tables~\ref{disk_results_lum} 
and \ref{disk_results_rob_dal}. 
We caution that the results on the mass-accretion rate are not reliable 
for Class~II sources lacking $U$ or $B$ photometry. Therefore, 
this parameter is provided only for such objects possessing fluxes in 
at least one of these two bands. For some of these objects the resulting  
mass-accretion rate is zero.
We also caution that the output disk parameters refer to the best 
fit model SED to each source, and that there may be a number of model 
SEDs providing a good fit. In this sense, the parameter values quoted 
in Tables~\ref{disk_results_lum} and \ref{disk_results_rob_dal} should 
be treated with caution.

A reasonable fit with the R06 models could be performed basically 
for all the sources with IR excess in the sample.
Exceptions are the strongly veiled object Sz~47 and the YSO candidate
SSTc2d~J130529.0$-$774140, the latter with insufficient photometry 
for the analysis.
Noteworthy are the results for the Class~I source IRAS~12500$-$7658: 
the SED is consistent with an embedded, very low-mass object with  
T$_{\rm eff}$ = 2900~K, mass accretion rate $\sim$2$\times$10$^{-6}$~M$_{\odot}$/yr, 
and circumstellar dust parameters typical for very low-mass forming 
objects. If the low temperature of this object will be spectroscopically 
confirmed, it will be an interesting case of a Class~I source, 
very close or below the sub-stellar limit. 
It is worth noting that the object is detected in the $R$ and $I$ 
bands and that using the techniques described in S07a 
an effective temperature T$_{\rm eff}=$3000$\pm$200~K is independently 
estimated. However, the S07a methods may underestimate T$_{\rm eff}$
if the extinction is high. The integrated luminosity, $L_{bol}=0.92 L_{\odot}$, 
is about one order of magnitude above the luminosity limit defining 
the so called {\it VeLLOs} \citep[very-low luminosity objects, see e.g.][]{Dun06}. 
Thus, another possibility is that what we are seeing in the optical 
is a very low-mass companion, with the main luminosity source being 
invisible. 

\begin{figure}[!h]
\epsscale{1.0}
\plotone{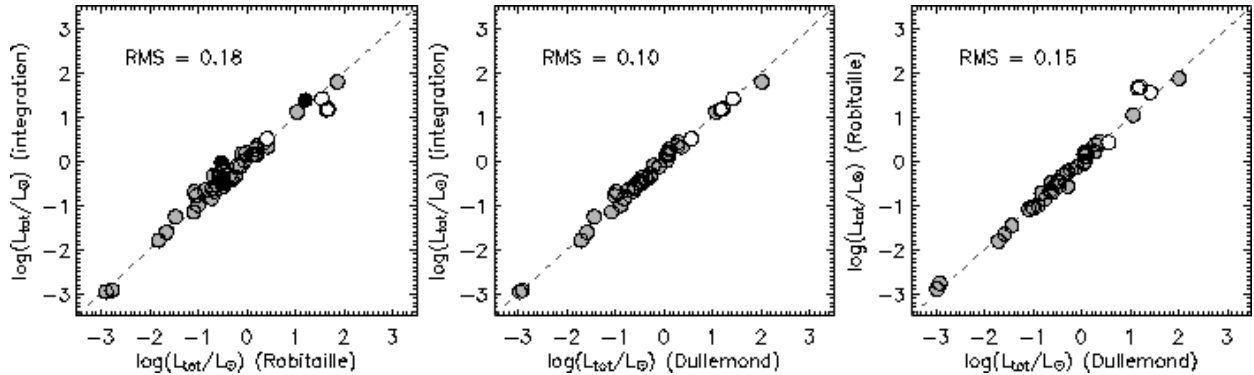}
\caption{Comparison of the total luminosity derived with different methods
       as indicated in the axes labels. Symbols are as in Figure~\ref{cccmd}.
       \label{comp_Ltot}}
\end{figure}
   
This section aims for a quantitative description of the circumstellar 
dust-envelope properties found for the Cha~II PMS population, which may 
provide some constraints for the disk evolution and planet formation models. 
However, since the disk/envelope parameters are strongly model-dependent, 
we will just compare the results among the different models in order to 
provide statistical ranges of these parameters and to quantify the deviations 
among different methods. Again, we caution that the output disk 
parameters are only the best fit values and that a more careful analysis
would require a number of model SEDs providing a good fit. This is, 
however, beyond the scope of this paper and is deferred to a future work.

An important output parameter from the models is the total object 
luminosity. We find a very good agreement between the total luminosities 
as derived in \S~\ref{lum} and the total output luminosity from the
D01 and R06 models: the RMS difference relative to the former is within 
0.1~dex in $log(L/L_{\odot})$, while relative to the latter is on the order 
of 0.18~dex (cf. Figure~\ref{comp_Ltot}); for some objects the total 
luminosity from the R06 fits is slightly lower (cf. Figure~\ref{comp_Ltot}), 
which is mainly a consequence of the fact that the R07 tool considers 
upper limits for the SED fits.  The worse parameter sampling of the
\citet{Dal05} models leads to some difficultities for a reasonable 
fit in some cases. The RMS differences in luminosity relative to 
these models can be up to 0.5~dex. 
Differences in the total luminosity among models may also be due to the 
lack of data at short wavelengths in optical bands; for instance, the R06 
fits tend to slightly overestimate the fluxes at wavelengths bluer than 
about 0.45\,$\mu$m relative to the D01 fits, when optical data are 
lacking. This happens in the SEDs of IRAS~12496$-$7650, Sz~49 and Hn~26 
(cf. Figure~\ref{seds}). Similarily, for Iso-ChaII~29 the R06 best-fit 
model flux is slightly underestimated relative to the SED data points 
at $\lambda<$0.8\,$\mu$m, while the \citet{Dal05} models are unable to 
reproduce its essentially photospheric points at around 10\,$\mu$m.

\begin{figure}[!h]
\epsscale{0.75}
\plotone{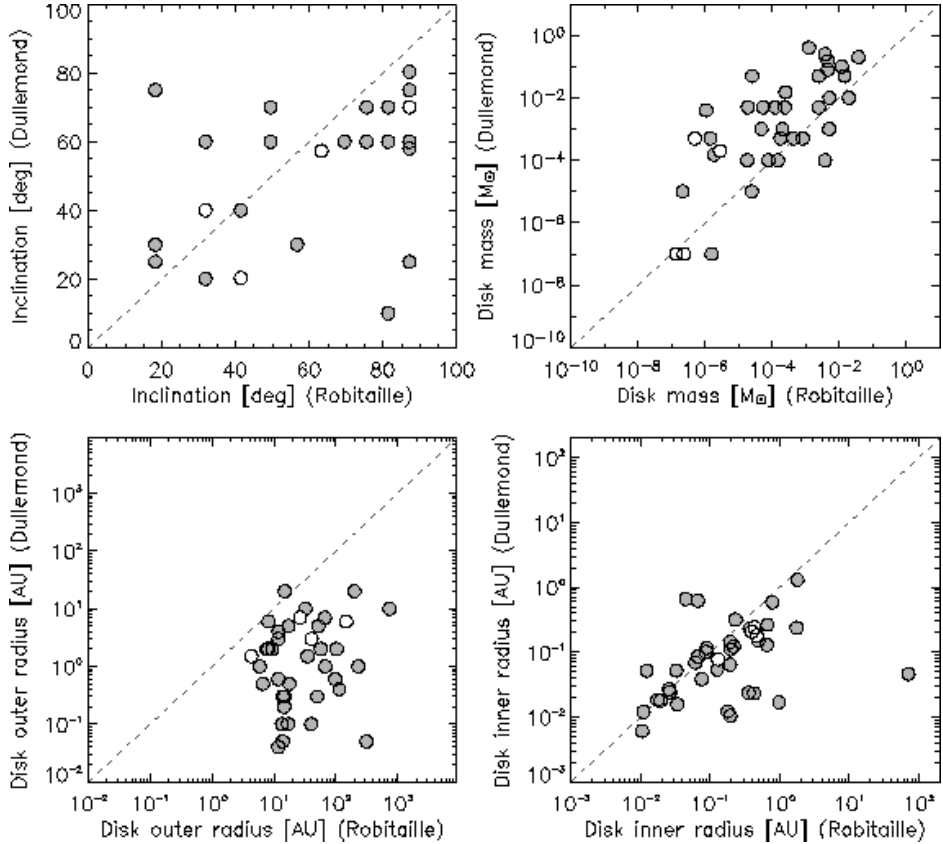}
\caption{Comparison of some of the output parameters derived from the 
       best fits to the SEDs using the D01 and R06 
       models as indicated in the axes labels. Symbols are as in 
       Figure~\ref{cccmd}.      
	\label{comp_model_pars}}
\end{figure}

\begin{figure}[!h]
\epsscale{1.0}
\plotone{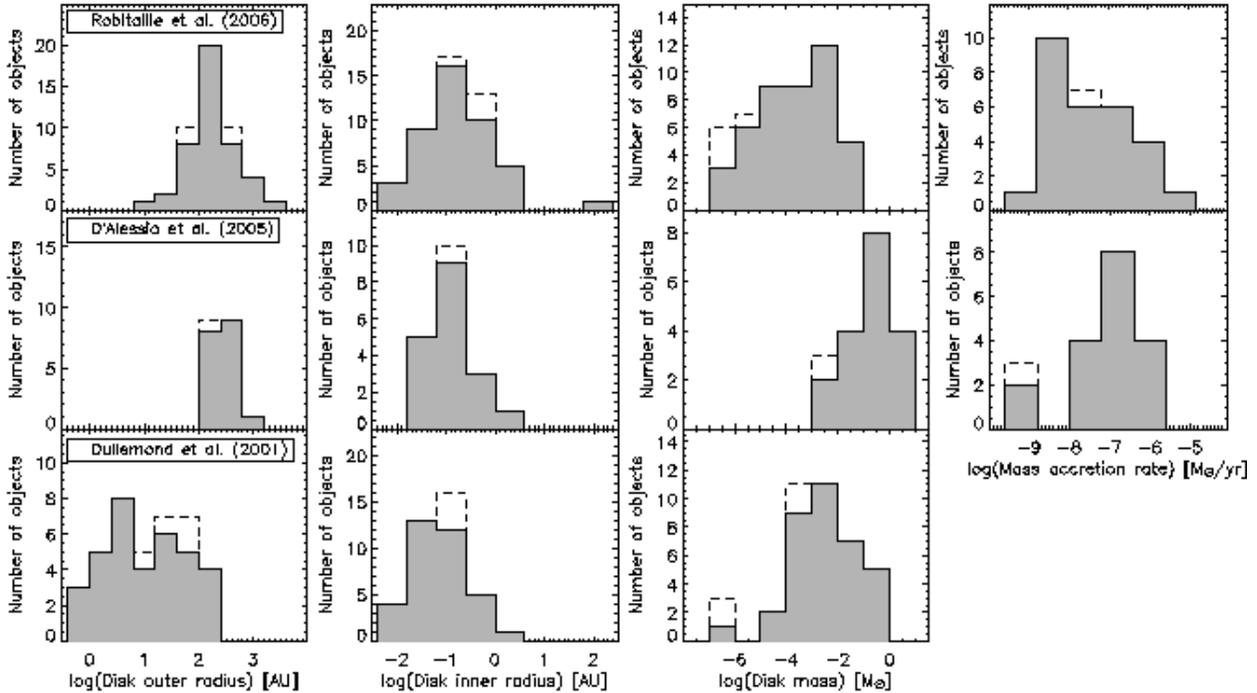}
\caption{Distributions of relevant disk parameters resulting from the
       models as labeled in the leftmost panels. The solid histograms
       refer to the confirmed PMS objects, while the dashed ones
       represent the distributions when adding the candidate PMS objects.
	\label{hist_model_pars}}
\end{figure}

A comparison of some of the output parameters resulting from the 
best fits of the D01 and R06 models is shown in the scatter plots 
in Figure~\ref{comp_model_pars} and a synthesis of the distribution 
of their values derived from each model is shown in the histograms 
in Figure~\ref{hist_model_pars}. There seems to be a fairly good 
correlation of the disk mass between the D01 and R06 models; however, 
the former models tend to overestimate the disk mass by about one 
order of magnitude relative to the latter. Geometrical parameters 
like the outer disk radius seem to be quite different among the
different models; 
on the other hand, there seems to be a fair agreement of the inner 
disk radius for most objects, while for some the disk accretion models 
tend to overestimate this parameter relative to the reprocessing flared 
disk models; the difference is highest for Iso-ChaII~29, which is 
represented by the rightmost point in the lower right panel of 
Figure~\ref{comp_model_pars}. From a visual inspection of the fits,
the most reliable disk parameters for this object seem to be those 
derived from the R06 models.

The discrepancies among the disk parameters derived from the different 
prescriptions might be in principle attributed to differences in the 
sampling of the parameter space covered by the models. This may be the 
case for the \citet{Dal05} models, but not for the R06 models that cover 
a wide range of disk parameters with a good sampling; also, the D01 tool 
calculates analytically the disk parameters based on the provided inputs. 
Thus, the differences on the disk parameters from the D01 and R06 models
are most likely related to the fact that the two sets of models use 
different opacities and disk geometries, and that D01 considers flared 
reprocessing disks, while \citet{Dal05} and R06 reproduce the SEDs of 
accretion disks. Another possible source for the discrepancy among disk 
outer radii is linked to the lack of fluxes at millimeter wavelengths 
that allow stronger constrains on this parameter.
The IRAC and MIPS wavelengths preferentially probe the inner 
parts of the disks; a good wavelength coverage should provide good 
measurements of the inner disk structures, differently with the 
outer disk radius, for which the mid-IR data are not sensitive.
For the three models, the typical value for the disk inner radius 
is on the order of 0.1~AU (cf. Figure~\ref{hist_model_pars}) 
with a similar range of this parameter among the different models, while 
the distribution of outer radius peaks at around 100~AU for the accretion 
models, but results smaller for the reprocessing models, again showing 
that this parameter is not well constrained because of the lack of millimeter 
data. Similarily, Figure~\ref{comp_model_pars} confirms that the
inclination angle is mostly undertermined from the SED alone, a well 
known result from previous works \citep[e.g.][and references therein]{Dal99, Chi99, Rob07}. 
We recall that differences may also be due to the fact that 
in our analysis the output disk parameters are only those resulting
from the best fit SED to each source.

The disk mass for the sample ranges from about 1 to 10$^{-4}$~$M_{\odot}$
with the D01 models and from 10$^{-2}$ to 10$^{-6}$~$M_{\odot}$ with the R06 
models. The disk mass distribution from both models roughly peaks at around 
10$^{-3}$~$M_{\odot}$. The \citet{Dal05} models provide higher disk masses 
than the D01 and the R06 models. 
  
Although the sample of objects for which we can provide the mass-accretion 
rate is small, it is worthwhile to compare the value of this parameter as 
derived from the two accretion models. For some objects the accretion 
models agree within less than about one order of magnitude, but the 
statistics are small for any solid conclusion. The range of mass acretion 
rate determined from both accretion models is similar and goes from about 
10$^{-9}$ to about 10$^{-6}$~$M_{\odot}/yr$.

\subsection{Comparison with previous SED fits}

A SED of IRAS~12496$-$7650 (or DK~Cha) has been reported by \citet{Hen93}.
These authors obtained 1.3 mm fluxes  and performed the fit to the
SED of the source by assuming both spherical and disk geometries. 
They found that the former model best fits the SED, except for the 
millimeter flux, while the disk model provides too much stellar flux 
in the near IR due to the assumption of a thin disk. Their disk-model
results for the disk radius ($R_d \approx$ 65~AU) and disk inner 
radius ($R_{in} \approx$ 0.13~AU) are smaller than the values derived 
from both the R06 ($R_d \approx$ 150~AU; $R_{in} \approx$ 1.8~AU) 
and D01 ($R_d \approx$ 200~AU; $R_{in} \approx$ 1.3~AU).
The results of the disk mass are different among the different 
fits ($M_d(Henning) \approx$ 0.04~$M_{\odot}$; 
$M_d(Robitaille) \approx$ 0.0013~$M_{\odot}$ ; 
$M_d(Dullemond) \approx$ 0.4~$M_{\odot}$). 
Using the 1.3\,mm flux and formula 3 in \citet{Hen93} a circumstellar
mass of 0.07~$M_{\odot}$ is derived.

A first fit of the SED of the Class~I source Iso-ChaII~28 was performed 
by \citet{Per03} using the DUSTY code by \citet{Ive99}, which assumes an
embedded source with a spherically symmetric dust envelope. The outer 
envelope radius they report is $\approx$ 1600~AU, while the value derived 
from the R06 fit is $\approx$ 2500~AU. The difference is most likely due 
to the assumption by \citet{Per03} of a spherically symmetric dust envelope, 
whereas the R06 model assumes a source with a disk and an envelope.
Moreover, the envelope outer radius is not well constrained in the \citet{Per03} 
fit because they include only an upper limit for the (sub)millimeter 
flux. On the other hand, the envelope inner radius they fix ($\approx$ 0.16~AU) 
is much lowen than the one derived from the R06 fit ($\approx$ 1~AU). 
Note also that the temperature of 4000~K assumed by \citet{Per03} for the 
central object is lower than the value derived from the R06 fit 
($T \approx$ 4500~K), but such parameter is not constrained 
enough to say whether the difference of 500~K is really significant 
for such an embedded source. In conclusion, the differences on the 
resulting parameters come from the fact that the models are fundamentally 
different. This example only illustrates that fits to the SED of objects 
that are transitionary between envelope and disk stages are very 
degenerate, in particular when (sub)millimeter data are missing.

In addition, the SEDs of 10 classical T~Tauri stars in Cha~II were 
modelled by \citet{Cie05} using the D01 flared-disk models. Their results on 
the disk inner radius are in very good agreement with those derived by us
using the same models, except for the star Sz~62 for which \citet{Cie05} 
derived $R_{in} \approx$ 0.08~AU and we obtain a value a factor of about 
8 larger. We stress that the 70~$\mu$m fluxes, which further constrain 
the SED fit, were not included in the SEDs by \citet{Cie05}. When dropping 
the 70~$\mu$m flux from the SED of Sz~62 we obtain a closer value 
($R_{in} \approx$ 0.26~AU) to the one reported by \citet{Cie05}, but 
still larger by a factor of about three. 

More recently, \citet{All06a} modelled the SEDs of six Cha~II objects 
using the D01 models. The agreement between their results
and ours using the same models is also quite good, except for 
SSTc2d~J130540.8$-$773958 and C66, for which we get a slightly larger 
disk inner radius. We note also that the 70~$\mu$m fluxes were not 
included in the SEDs by \citet{All06a}, which might explain the 
small differences.

Finally, the SED of the sub-stellar object Iso-ChaII~13 was modelled
by \citet{Alc06} also using the D01 flared-disk models. The results are 
in agreement with those reported here for the D01 models.

\subsection{Disk evolution issues}

\begin{figure}[!h]
\epsscale{0.8}
\plotone{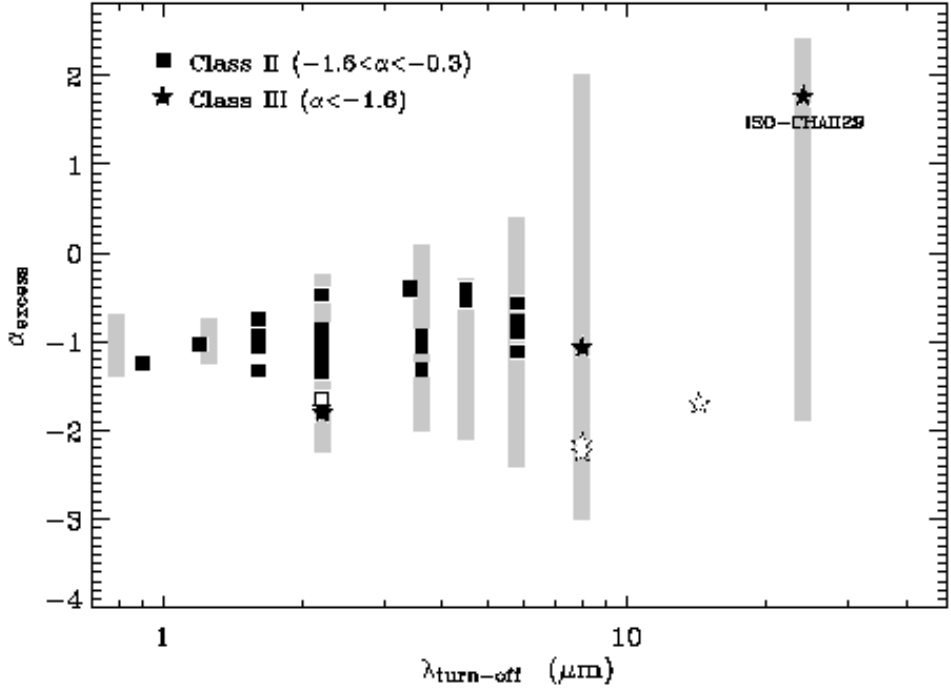}
\caption{Distribution of excess slopes, $\alpha _{excess}$, with respect 
       to the wavelength at which the IR excess begins, $\lambda _{turn-off}$, 
       for the sample of confirmed (filled symbols) and candidate (open sympols) 
       PMS objects in Cha~II. The symbols are as in Figure~\ref{spa_distr}. 
       The intervals of $\alpha _{excess}$ for the classical and weak T~Tauri 
       stars were taken from Figure~22 by H07 and are represented with shaded 
       bars.
       \label{alpha_excess}}
\end{figure}

An interesting diagnostic to study disk evolution in young low-mass objects 
is the diagram of the excess slope, $\alpha_{excess}$, versus the wavelength 
at which the IR excess begins, $\lambda_{turn-off}$ 
\citep[see][and references therein]{Cie07}. 
The $\alpha_{excess}$ index is the SED slope determined for wavelengths 
equal to or longer than $\lambda_{turn-off}$.  H07 compared the
position of the Serpens YSOs in such a diagram with that of several
classical and weak T~Tauri stars and concluded that the Serpens objects 
follow the trend previously observed by \citet{Cie07}, i.e., a wider range 
of excess slopes for later evolutionary phases. We have performed a 
similar analysis which is shown in Figure~\ref{alpha_excess}. For the 
sake of clarity, the range of $\alpha_{excess}$  at each wavelength 
for the classical and weak T~Tauri stars from \citet{Cie07} are represented 
here with shaded strips. The analysis is done only for the certified and 
candidate PMS objects with IR excess.
Despite the low number statistics, the resulting $\alpha_{excess}$ 
and $\lambda_{turn-off}$ for the Cha~II objects follow a similar 
trend as observed in Serpens by H07. The $\alpha_{excess}$ 
and $\lambda_{turn-off}$ values are reported in Table~\ref{disk_results_lum}.

The sources with a large $\alpha_{excess}$ and $\lambda_{turn-off}$, 
i.e. objects on the upper right of the diagram, are interpreted in 
terms of disks with large inner holes. One example in Cha~II
is Iso-ChaII~29. Its SED rises beyond 24$\mu$m (see Figure~\ref{seds}).
This object deserves further investigation with IR spectroscopy in 
order to assess possible signs of dust grain growth and settling to 
the disk mid-plane.
This led us to investigate the relationship between the reference 
$\alpha$ index (i.e. $\alpha_{[K \& {\rm MIPS~24\,\mu m}]}$) and the 
disk inner radius. 
To this aim and considering the poor statistics, we divide the PMS 
objects and candidates in two samples according to $\lambda_{turn-off}$ 
in the following way: from the $\alpha_{excess}$ versus $\lambda_{turn-off}$ 
diagram we note that the dispersion of $\alpha_{excess}$ starts 
increasing at $\lambda_{turn-off}\sim$4-5$\mu$m; under the assumption 
of thermal emission from a thin disk with black body grains, this 
corresponds to a distance of about 0.2~AU from the central object. 
We thus use this limit to define two samples to investigate 
the dispersion of their $\alpha_{[K \& {\rm MIPS~24\,\mu m}]}$ index.
We find  that objects with disk inner radius larger than 0.2~AU 
show a dispersion of $\alpha_{[K \& {\rm MIPS~24\,\mu m}]}$ a factor 
2.5 larger than those with disk inner radius less than 0.2~AU.
This may be interpreted in terms of larger inner disk radii for 
later evolutionary disk phases.

\subsection{Disk fraction}
\label{disk_frac}

As concluded in \S~\ref{alpha}, Lada Classes in Cha~II are quite consistent 
with their Stages as defined by R06 and using 
$\alpha_{[K \& {\rm MIPS~24\,\mu m}]} = -1.6$  is a good separation between 
thick and thin disks. Thus, sources with optically thick disks should have 
$\alpha_{[K \& {\rm MIPS~24\,\mu m}]} >$~-1.6, while for anemic disks 
-2.5~$< \alpha_{[K \& {\rm MIPS~24\,\mu m}]} <$~-1.6. 
Objects with $\alpha_{[K \& {\rm MIPS~24\,\mu m}]} <$~-2.5 should be 
characterized by  normal late-type photospheres. From this classification 
the total fraction of objects with disks, either anemic or thick, would 
be on the order of 80\%, while those with thick disks would represent 
about 70\% of the sample. 

\begin{figure}[!h]
\epsscale{0.7}
\plotone{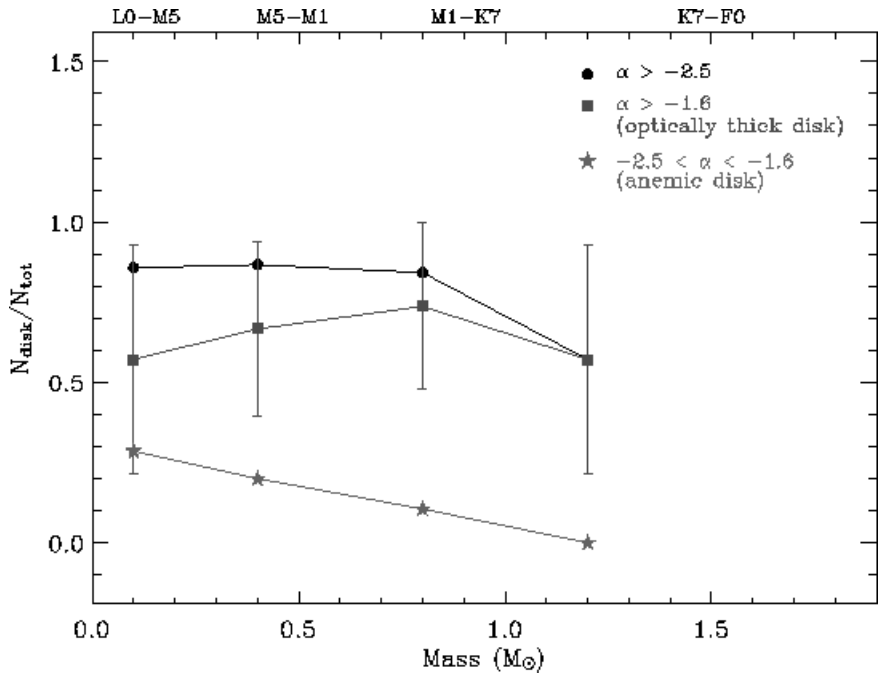}
\caption{Disk fraction versus mass of the central object. The symbols represent 
       the different disk types, namely anemic and thick, as indicated in the 
       labels. 
       The error bars, plotted only for the thick disks, represent the
       propagated Poissonian errors. The approximate spectral types 
       corresponding to the four mass bins adopted are indicated in the 
       top. 
        \label{disk_f_mass}}
\end{figure}

The disk fraction in Cha~II is considerably higher than the values derived 
in other regions \citep[i.e. in IC~348 by][]{Lad06} and would mean that if 
all the stars in Cha~II were born with circumstellar disks, only a minor 
fraction of about 20-30\% have lost their primordial disks in about 4~Myr, 
the average age for the Cha~II members (S07b). 
One may wonder whether our census might have missed a significant number 
of Class~III objects, but this is unlikely. Although the c2d criteria 
select only IR-excess objects, the other surveys are rather complete, 
both in space and flux. For instance, covering 1.75 square degrees, 
the survey by S07a picks up basically all low-mass objects (M$<$1~M$_{\odot}$)
without significant IR excess and is 95\% complete down to a mass 
M=0.03~M$_{\odot}$, for a visual extinction A$_V\approx$2~mag. 
In addition, our spatially unbiased spectroscopic survey with FLAMES
(down to $R=$18\,mag, S07b) would have discovered at least some 
more Class~III objects, if they existed. Furthermore, if many more 
Class~III sources had been missed in optical and IR surveys, they would 
have been detected in X-rays, as they were in other regions, but only two  
\citep[RXJ1301.0-7654a and RXJ1303.1-7706,][]{Alc95, Alc00} were discovered. 
Although such X-ray surveys are not extremely deep they would have 
detected many more Class~III sources, if they were present. A further 
confirmation of this result comes from the negative detections by a 
deep XMM pointing in the North-West of the cloud (B. Stelzer, private 
communication). Thus, unless all the Class~III sources are located 
behind the cloud, which is very unlikely, the result on the high disk 
fraction in Cha~II seems to be real.

The disk fraction as function of the central object mass in Cha~II is 
shown in Figure~\ref{disk_f_mass}. The average stellar mass, as derived 
by S07b for each individual object, was used for this analysis. 
In order to try to minimize the errors we have divided the sample 
in four mass bins\footnote{For this exercise we used the masses derived 
from the \citet{Bar98} \& \citet{Cha00} tracks for stellar masses less 
than 1.4~$M_{\odot}$, while for higher mass objects we use the masses 
derived from the \citet{Pal99} tracks, as the highest mass in the 
\citet{Bar98} \& \citet{Cha00} tracks is 1.4~$M_{\odot}$. These values 
are reported in S07b.}. 
Although the frequency of thick disks is apparently highest close 
to one solar-mass, the low number statistics and hence, the large 
errors prevent us from drawing any conclusion. However, this would 
be consistent with the results by \citet{Lad06} that thick disks 
in IC~348 are more frequent for solar-mass stars.
On the other hand, there seems to be a decline of anemic disks 
with stellar mass, but again the numbers are quite small.

It is also interesting to investigate the disk frequency in the
sub-stellar regime, but since only 3 out of the 7 sub-stellar 
objects (i.e. 43\%) are spectroscopically confirmed as members 
of Cha~II (see S07b) the figures are not statistically
significant; all the 3 confirmed sub-stellar objects possess 
optically thick disks according to the above definition, while 
the 4 sub-stellar candidates posses anemic or no disks. 
Therefore, the disk fraction below the H-burning limit in Cha~II 
strongly depends on the nature of these 4 candidates. 
If the latter will be all confirmed as members, the sub-stellar 
disk fraction would be on the order of 40\%, i.e. similar to other 
regions; otherwise it would be exceptionally high. We remind the 
reader that recent Spitzer studies in IC~348 by \citet{Luh05c}  
may indicate an increase of the disk fraction below the H-burning 
limit relative to that in stars \citep{Lad06}.

\section{Overall results on star formation in Cha~II}
\label{overallsatarformation}

The physical parameters of the certified and candidate PMS objects are 
necessary for the study of the rate and efficiency of star formation 
of the cloud. We use here those determined by S07b.

\subsection{The Star Formation Rate}
\label{SFR}

One of the c2d goals is to compare the star formation rate (SF~rate) 
among the clouds of the Legacy survey. The results by  H07  showed that 
the Serpens clouds turned some 60~$M_{\odot}$ into YSOs every Myr.  
In order to homogeneously compare how much of the Cha~II cloud-mass
is being transformed into YSOs, we have to use the total mass of 
the objects selected from the multi-color criteria by  H07. 
The overall results of star formation in Cha~II are summarised
in Table~\ref{starform}. As mentioned in the previous sections, the 
number of YSOs in the 1.038~deg$^2$ IRAC~3.6-8\,$\mu$m/MIPS~24\,$\mu$m 
overlap area in Cha~II is 26 (see \S~\ref{sample}, and Table~\ref{tab:yso}). 
Assuming a distance of 178~pc, the resulting IRAC~3.6-8\,$\mu$m/MIPS~24\,$\mu$m 
overlap surveyed area in Cha~II is about 10.02~pc$^2$. This means a 
density of about 2-3 YSOs per pc$^2$, i.e. a factor of about 4-6 less 
than in Serpens. 
The average mass of the Cha~II members is 0.52~$M_{\odot}$ \citep{Spe07b}, 
which means a total YSOs mass of $\sim$14~$M_{\odot}$. The typical age 
of these objects, determined from the \citet{DAn98} and \citet{Pal99} 
theoretical isochrones, is on the order of 2~Myr (S07b). 
This would mean that the Cha~II cloud is turning some 7~$M_{\odot}$
into YSOs every Myr, i.e. much less than the estimate in Serpens.
A slightly higher star formation rate, but still less than in Serpens, 
is derived for the whole sample of certified and candidate PMS objects 
in the 1.75~deg$^2$ surveyed area, i.e. in $\sim$17~pc$^2$: assuming the 
same average individual mass we estimate a total mass of certified 
and candidate PMS objects on the order of 32~$M_{\odot}$. Using the same 
theoretical isochrones, the resulting average age of the whole sample is 
on the order of 3~Myr (S07b). Therefore, the star formation rate would 
be on the order of 10~$M_{\odot}$ per Myr. The star formation rate in 
the other c2d clouds seem to be also higher than in Cha~II 
(Evans et al. 2007, in preparation). It is worth noting that the 
star formation rate per unit area is quite similar for both PMS and 
YSO samples after correction for the larger area and older age of 
the PMS sample.

Using the results on clustering analysis in \S~\ref{spa_distr_class},
we can also investigate the SF rate for the sub-structures in Cha~II.
The results are summaried also in Table~\ref{starform}. The SF rate
per unit area and volume for the tight groups A and B, estimated 
by assuming an age of 2~Myr, is higher, but yet much lower than the 
values for the sub-groups in Serpens and the other c2d clouds. 
On the other hand, the SF rate of the Cha~II loose cluster resembles 
the global SF rate of the cloud determined above for the whole sample 
of PMS objects and candidates, as one should expect.

We therefore conclude that Cha~II has the lowest SF rate in comparison 
to the other four clouds investigated in the Legacy survey and hence, 
it is currently the least active star-forming region among the c2d 
clouds.

\subsection{Cloud mass and the Star Formation Efficiency}
\label{SFE}

To derive the star formation efficiency, we need an estimate of the 
cloud mass. \citet{Bou98} combined CO observations with IRAS images, 
H~I observations and extinction data to study the distribution of mass 
in the Chamaeleon cloud complex. The Cha~II cloud has also been mapped 
in molecules such as $^{12}$CO \citep{Bou98, Miz01}, $^{13}$CO \citep{Miz98, Hay01} 
and C$^{18}$O \citep{Miz99, Hay01}. Estimates of the Cha~II cloud mass 
based on these observations are reported in Table~\ref{cloud_mass}. 

Another way to estimate the cloud mass is by means of extinction 
maps. Various combinations of maps in molecules, far-IR continuum, 
H~I, extinction, and polarization toward field stars have been used 
to study the properties of the interstellar medium and magnetic field 
in the Chamaeleon clouds \citep{Cov97b}. The $R_V$ ratio is found to be 
essentially normal (i.e. $\sim 3.1$) in the outer parts of the clouds, 
but several stars near the opaque core of Cha~I show an unusually high 
value of $R_V$ which ranges between 5 and 6, and there is some evidence 
that the Cha~II cloud might show similar features indicating that the 
two clouds are probably part of the same structure 
\citep{Gra75, Ryd80, Vrb84, Ste89, Whi87, Whi94, Whi97, Luh04}. 
The extinction has been previously mapped in Cha~II with a resolution 
of a few arcminutes by using IR star counts \citep{Het88, Cam99}. The 
extinction has also been mapped using the c2d data. The c2d extinction 
maps were created at different spatial resolutions that range from 60~arcsec 
to 300~arcsec. The details on how such maps were derived for each one of 
the five clouds of the c2d project can be found in the final data delivery 
document \citep{Eva07} and in H07. From the c2d map, the maximum extinction 
in the Cha~II cloud is $A_V\approx$ 20~mag and occurs close to 
IRAS~12496$-$7650 or DK~Cha (see Figure~\ref{spa_distr}). The cloud mass
as determined from the c2d extinction map is reported in 
Table~\ref{cloud_mass}. In addition, following the prescription by \citet{Kai05}, 
we also calculated the cloud mass using the extinction maps by \citet{Cam99} 
and Kainulainen (private communication). There is a remarkable agreement 
of the mass estimates from the extinction maps and the 100$\mu$m emission.

Based on these mass estimates, we derived the global Star Formation Efficiency 
(SFE) in Cha~II as $SFE=\frac{M_{stars}}{M_{cloud}+M_{stars}}$ where $M_{cloud}$ 
is the cloud mass, and $M_{star}$ is the total mass in PMS objects
or YSOs,  which we estimate to be on the order of 32~$M_{\odot}$ for the former 
and about 14~$M_{\odot}$ for the latter (see \S~\ref{SFR} above). 
Note that this is the SFE {\it so far} as further star formation is possible. 
There is also the caveat that some of the gas that was used for the 
formation of some of the more evolved PMS objects has already been 
dispersed.
The results on the efficiency of star formation are provided in Table~\ref{cloud_mass}. 
Consistently with the only previous estimate in literature \citep[$\sim$1\%,][]{Miz99}, 
we find that the SFE in Cha~II ranges between 1 and 4\% and that it is similar 
to the estimates for other T associations like Taurus \citep[1-2\%,][]{Miz95} and 
Lupus \citep[0.4-3.8\%,][]{Tac96}, but surprisingly lower than the 
value derived for Cha~I \citep[13\%,][]{Miz99}. 
 
As suggested by \citet{Miz99}, the different star-formation activities 
in the Chamaeleon clouds may reflect a different history of star formation,
as a consequence of Cha~I being older than than Cha~II. This hypothesis 
is not quite supported by our results, since the the average age of the 
Cha~I population is similar to that of Cha~II (see results by S07b). 
The SFE estimates by \citet{Miz99} assume an average stellar mass of 
0.7 $M_{\odot}$ and 0.5 $M_{\odot}$ for Cha~I and Cha~II, respectively, 
i.e. the average values determined by \citet{Alc97}. While for Cha~II this 
assumption is consistent with the results by S07b, i.e. $\sim$0.52~$M_{\odot}$, 
the average value assumed for Cha~I is significantly higher than measured 
by S07b, i.e. $\sim$0.45~$M_{\odot}$, based on the more recent census of 
the Cha~I population by \citet{Luh04}. Our estimate of the SFE for Cha~I 
is on the order of 7.5\%, i.e. lower than the value derived by \citet{Miz99}, 
but yet significantly higher than our estimate for Cha~II.
 
On the other hand, as mentioned in \S~\ref{spa_distr_class}, \citet{Lep94} 
proposed arguments in favor of a common explanation for the displacement 
of the prominent nearby star-formation regions with respect to the 
Galactic plane in terms of the impact of high-velocity clouds 
(v$\sim$100 km/s). This scenario would imply a common triggering agent 
for the star formation in Cha~I and Cha~II and hence, a similar star 
formation activity, which is not supported by our SFE estimates.

At the level of the sub-structures, the SFE estimates are higher. 
Using the c2d extinction map, we calculated the mass enclosed within each
structure as defined by the clustering analysis described in 
\S~\ref{spa_distr_class}. By using the the number of YSOs corresponding to 
each structure (see Table~\ref{surf_dens}), we find that the efficiencies 
for the two tight A and B groups are as high as 20--25\%, i.e. comparable 
to the efficiencies of dense clusters \citep[e.g.,][]{Lad03}, but the SFE
is only a few percent when averaged over the entire Cha~II cluster and 
distributed group.

\section{Cha~II outflows}
\label{outflows}
In this section our data set is used to investigate the HH~52-54 outflow 
and its exciting source, as well as to search for other outflow candidates 
in Cha~II.

\subsection{HH~52, 53 \& 54}
\begin{figure}[!h]
\epsscale{0.85}
\plotone{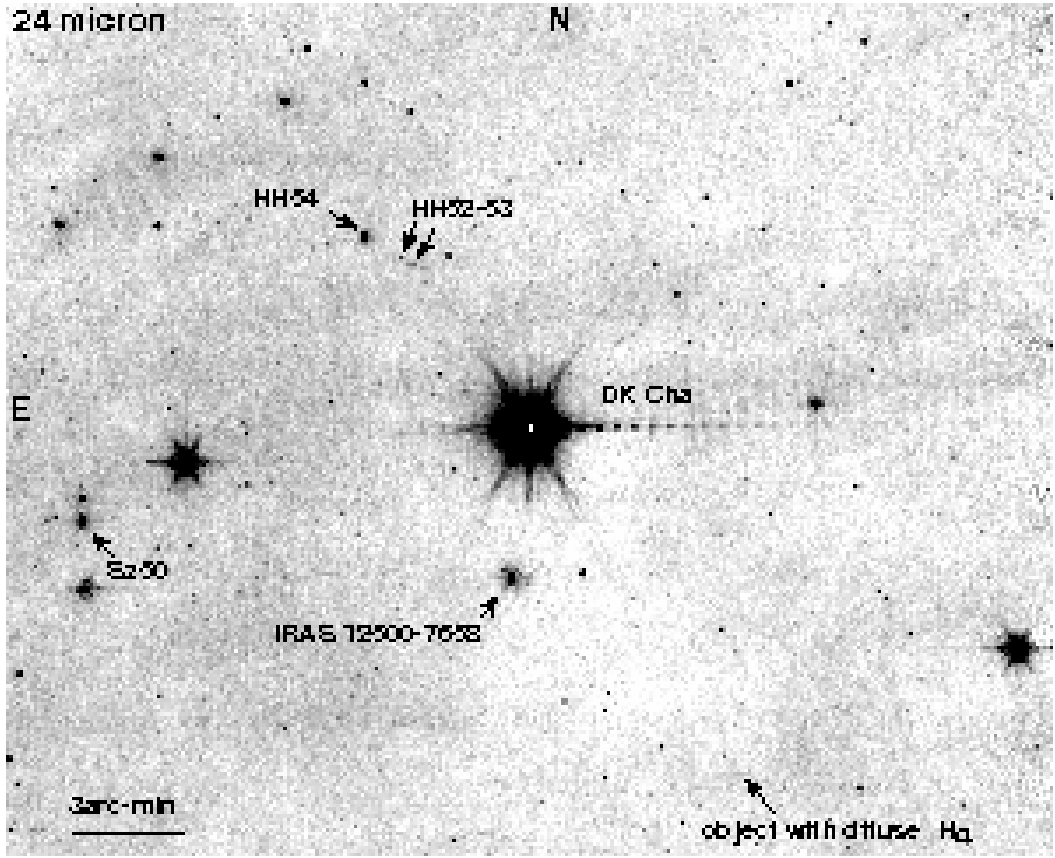}
\caption{The image shows part of the 24~$\mu$m c2d mosaic in Cha~II
       in gray-scale levels of intensity. Some of the sources 
       discussed in the text are indicated. The object with diffused
       H$\alpha$ emission discussed in \S~\ref{new_outflows} is also
       marked. 
       \label{dkcha_24mu}
	}
\end{figure}

The  known outflow in Cha~II is located at the northern edge of the cloud 
\citep[see Spitzer images in][and Figure~\ref{dkcha_24mu}]{Por07} and 
consists of several structures identified as the Herbig-Haro (HH) 
objects HH~52, HH~53 \& HH~54 \citep{San87}. There has been controversy 
about the source driving the outflow: \citet{San87} proposed several possible 
sources that were later rejected in a spectroscopic study by \citet{Gra88}. 
However, the latter study revealed blue-shifted gas with a radial velocity 
that becomes increasingly bluer along the stream from the North-East to 
the South-West. Based on these observations, \citet{Gra88} proposed that
IRAS~12496$-$7650, i.e. the embedded Herbig Ae star DK~Cha located at about
14 arc-min from HH~54, might be the driving source of the outflow. Based 
on CO observations, \citet{Kne92} concluded that IRAS~12496$-$7650 is, however,
unrelated to HH~52-54, suggesting IRAS~12515$-$7641\footnote{This source, 
identified as IRAS~C12515$-$7641 in SIMBAD and as IRAS~F12514$-$7641 in 
\citet{Pru92}, is also associated with the object C30 by \citet{Vuo01}.} 
and IRAS~12522$-$7640 as the driving sources of HH~52-53 and HH~54, 
respectively.

\begin{figure}[!h]
\epsscale{0.77}
\plotone{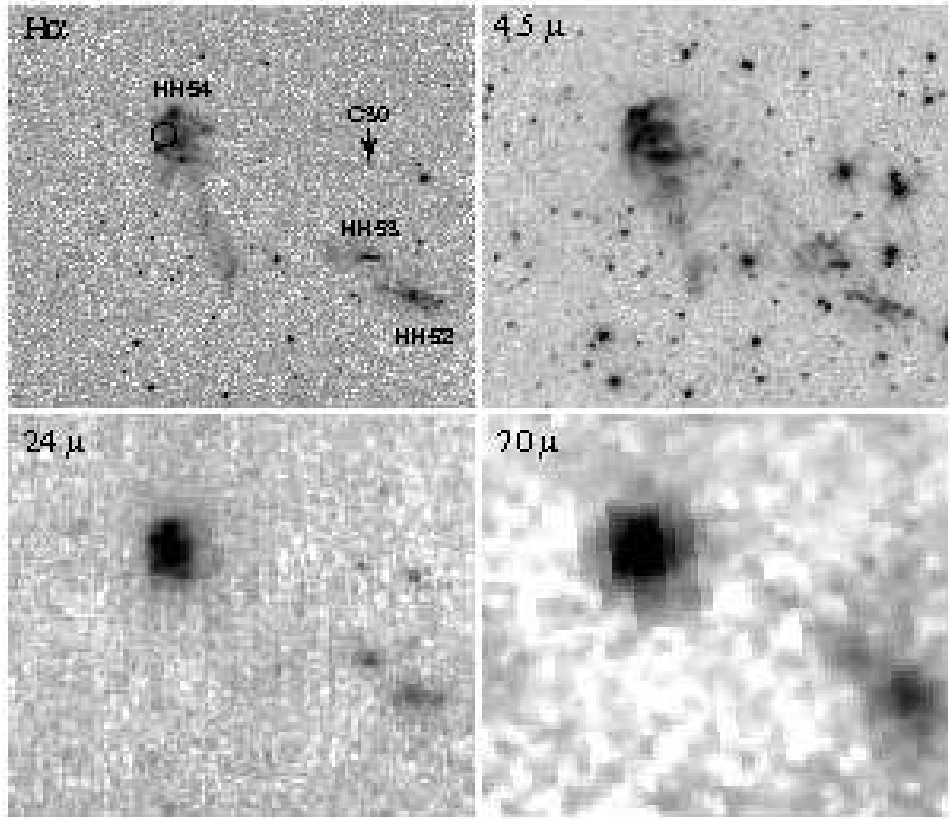}
\caption{Images of the HH~52-54 outflow at H$\alpha$ (upper left),
       4.5~$\mu$m (upper right), 24~$\mu$m (lower left) and 
       70~$\mu$m (lower right). The HH objects and IRAS~12515$-$7641
       (or C30) are indicated in the WFI H$\alpha$ image. The circle
       in this image marks the position of IRAS~12522$-$7640.
       The images cover an area of approximately 4.5$\times$4
       arc-min square and North is up and the East to the left.
        \label{hh54_irac_mips}}
\end{figure}

We have investigated in all our images, from optical to 70$\mu$m, both 
IRAS~12515$-$7641 and IRAS~12522$-$7640. The latter coincides with the head
of the HH~54 bowshock (see Figure~\ref{hh54_irac_mips}), but in none of 
the images appears as a point-like object. Note also that, as pointed out 
by \citet{Hug89}, the source may have been mistakenly classified as a point 
source because of the line emission from the HH nebulosity itself. 
Moreover, its IRAS colors are incompatible with those of T~Tauri stars 
or embedded objects \citep{Pru92}. 
Therefore, it is most likely produced by the IR emission of HH~54 itself. 
On the other hand, IRAS~12515$-$7641, or C~30 \citep{Vuo01}, appears as a 
point-like object in all our images (cf. Figure~\ref{hh54_irac_mips}). 
This source was spectroscopically observed by us with FLAMES. Its optical 
spectrum is  that of a late K-type, most likely a background giant. Thus, 
contrary to the conclusions drawn from the CO study by \citet{Kne92}, none 
of these two IRAS sources seems to be the PMS star driving the outflow. 
Moreover, a proper motion analysis of the infrared knots in HH~54 \citep{Car06} 
led to the conclusion that its driving source should be located further 
to the South-West. 

If~ IRAS~12496$-$7650 is not the driving source of the outflow, as 
concluded by \citet{Kne92}, which source is then exciting HH~52-54?
Since our census for PMS objects in Cha~II is rather complete, we 
inspected all our images to search for other possible candidates 
driving the outflow; we searched from the North-East to the South-West 
direction of the stream of HH~52-54, as suggested by the proper motion 
study by \citet{Car06}. 

Figure~\ref{dkcha_24mu} shows part of the c2d mosaic at 24~$\mu$m with 
some interesting objects indicated. Besides IRAS~12496$-$7650, other 
relatively nearby candidates are C33 and WFI~\-J12533662$-$7706393, located 
at about 2.3 arc-min and 6.5 arc-min respectively; these are, however, 
Class~III candidate PMS objects without significant IR excess or strong winds. 
Thus, they are not expected to drive any outflow. 
Apart from IRAS~12496$-$7650 (DK~Cha), the "nearest" object with strong IR 
emission in that direction is IRAS~12500$-$7658, one of the Class~I sources 
in Cha~II; its angular distance from HH~54 is about 20 arc-min, i.e. 
$\sim$1~pc at the distance of Cha~II; its mass accretion rate, determined 
from the SED fit, is on the order of 2$\times$10$^{-6}$M$_\odot$~yr$^{-1}$, 
i.e. typical for low-mass PMS stars driving outflows. 
We remind the reader that IRAS~12500$-$7658 might be a very low-mass 
object (see \S~\ref{disk_pars} and results by S07b) and that 
the velocity of a jet is expected to correlate with the depth of the 
potential well; thus, in principle, one would expect that such a 
parsec-scale outflow is driven by a more massive star.
However, no correlation exists between the mass of the driving 
source and the size of the outflows \citep[see][for a review]{Rei01}. 
We have thus searched for other possible HH-features further to the 
South-West of IRAS~12500$-$7658, but none was found. Nevertheless, 
an object with diffuse H$\alpha$ emission (see \S~\ref{new_outflows}) 
located some 20~arc-min to the South-West of DK~Cha 
(see Figure~\ref{dkcha_24mu}) was discovered, which would still favor 
this star as the driving source of the outflow. This is further 
discussed in \S~\ref{new_outflows} below.

Noteworthy, HH~54 is very bright at 24~$\mu$m. This can be seen in 
Figure~\ref{dkcha_24mu} and in its zoomed version in Figure~\ref{hh54_irac_mips},
where a comparison of the H$\alpha$, 4.5~$\mu$m, 24~$\mu$m and 70~$\mu$m 
images is shown. Most of the 24~$\mu$m emission comes from the [S I] 
and [Fe II] lines, as shown in the Spitzer spectrum of the object by 
\citet{Neu06}. As can be seen from Figure~\ref{hh54_irac_mips}, the 
object is also well detected at 70~$\mu$m. We determined that HH~54 
has a peak surface brightness of about 24~MJy/sr above the 70~$\mu$m 
background. The total 70~$\mu$m flux density is 0.68~Jy in an 
arcmin-square aperture, which is comparable to the 60~$\mu$m flux 
density of IRAS~12522$-$7640 \citep[cf.][]{Lis96}. If this flux density 
all came from the [O~I] line, the line flux would be on the order of 
$1.1 \times 10^{-11}$~erg~s$^{-1}$~cm$^{-2}$, i.e. close, 
but slightly lower than the value derived by \citet{Nis96} based on  
long-wavelength spectrometer (LWS) ISO data. Under the assumption 
of dissociative J-shocks, the [O I] luminosity is a very good 
diagnostic for mass-loss rate \citep{Hol85, Cec97}. Assuming the
distance of 178~pc and according to \citet{Hol85}, the mass-loss rate 
determined from the above [O I] line flux would be on the order of 
$1.1\times 10^{-6} M_{\odot}/yr$, which is a factor of about 
2 less than derived by \citet{Nis96}, but the difference is mainly 
due to the larger distance (250~pc) these authors assume  
for Cha~II.
Based on measurements of the [O I] line flux, \citet{Nis96} derived 
values of the mass-loss rate for HH~52 and HH~53, which are similar 
between each other, but lower than the one derived for HH~54. 
This suggests, as these authors conclude, that the wind-source 
driving HH~54 should be other than the one driving HH~52 and HH~53. 
The mass-loss rate they measure for IRAS~12496$-$7650 (DK~Cha), also 
from a LWS spectrum, is very similar to the one they derived for 
HH~52 and HH~53. Therefore, they cannot exclude that the driving 
source for at least the latter two is precisely IRAS~12496$-$7650. 
The mass-loss rate of the order of $10^{-6} M_{\odot}/yr$ 
would place the three HH objects among the most powerful known and, 
according to the measurements by \citet{Har95}, a driving source 
with an accretion rate on the order of a few $10^{-5} M_{\odot}/yr$ 
would be necessary to produce such a mass loss. The accretion rate 
of the Class~I source IRAS~12500$-$7658 would be probably high enough.
Thus, a scenario in which IRAS~12500$-$7658 is driving HH~54, while 
IRAS~12496$-$7650 drives HH~52 and HH~53, might be feasible.
However, given the uncertainties involved, one cannot completely 
rule out that IRAS~12496$-$7650, i.e. DK~Cha, is the driving source 
of the three HH objects. The HH~52-54 outflow is not associated 
with any of the tight groups A or B identified in \S~\ref{spa_distr_class},
but both DK~Cha and the outflow can be associated with the loose
cluster.

\subsection{New outflows} 
\label{new_outflows}

\begin{figure}[!h]
\epsscale{0.9}
 \plotone{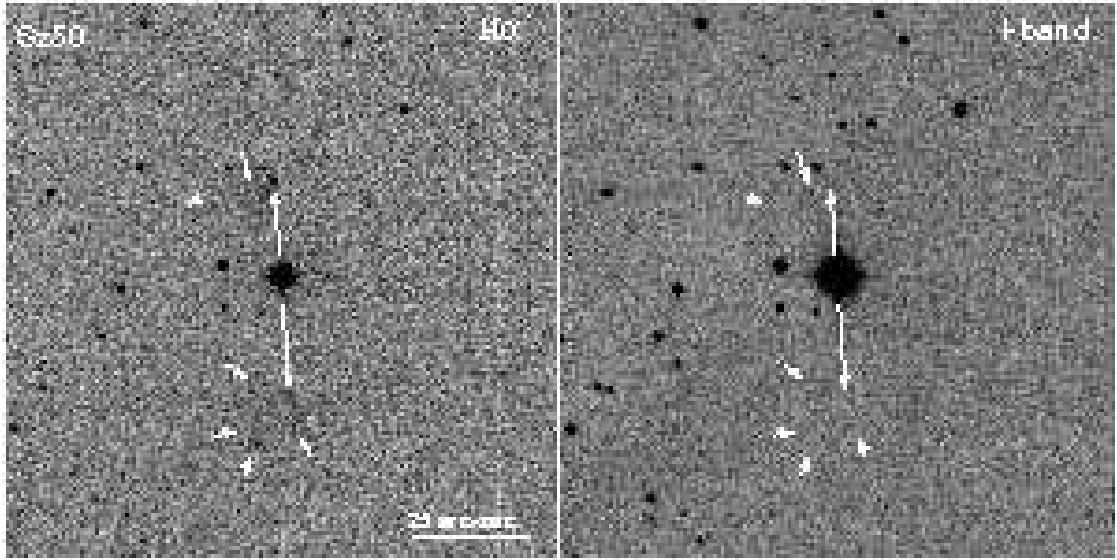}
\caption{Images of the region around the classical T Tauri star Sz~50
       in H$\alpha$ (left) and in the I-band (right). Sz~50 is in the 
       center of the images. The arrows indicate the knots of HH~939 
       and the diffuse emission in H$\alpha$. North is up and East to 
       the left. Both images cover an area of about 2$\times$2 arc-min 
       square.
        \label{Sz50_outflow}}
\end{figure}

\begin{figure}[!h]
\epsscale{0.8}
\plotone{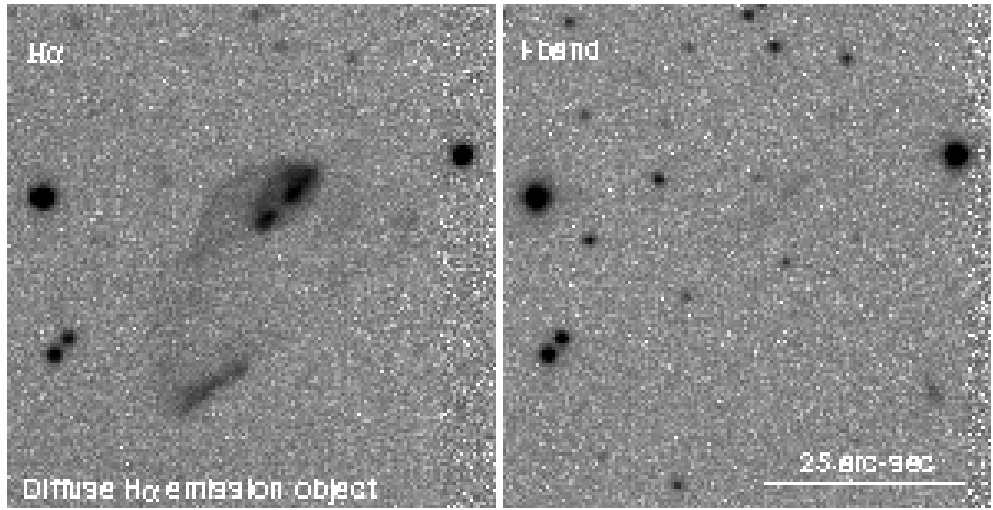}
\caption{Images of the diffused H$\alpha$ emission object in H$\alpha$ (left)
       and in the I-band (right). North is up and East to the left. Both 
       images cover an area of about 1$\times$1 arc-min square.
        \label{anon_outflow}}
\end{figure}

By inspection of our images, and in particular those in H$\alpha$, 
it was possible to identify other possible candidates to outflows 
in Cha~II. These candidates are shown in Figures~\ref{Sz50_outflow} 
and ~\ref{anon_outflow} where the WFI H$\alpha$ narrow-band images 
are compared with those in the I-band. In both cases, the diffuse 
H$\alpha$ emission is not detected in the I-band, which demonstrates 
that the structures shown cannot be of extragalactic nature.

Two compact H$\alpha$ emission knots are found on opposite sides of 
the classical T Tauri star Sz~50 (cf. Figure~\ref{Sz50_outflow}). 
Both are roughly circular with FWHM 20\% larger than nearby point 
sources, implying deconvolved source sizes of 0.5-1 arcsec.  
These two knots presumably trace a bipolar Herbig-Haro flow from 
Sz~50, which is consistent with the detection of [O I] emission
at 6300~\AA~ in the stellar spectrum by \citet{Hug92}. No counterparts 
to these knots are detected in the Spitzer IRAC images, which are 
sensitive to rotational H$_2$ emission; this suggests that the shocked 
material is primarily atomic. The flow axis follows PA 4 degrees, 
with one knot 21.2~arcsec North of the star and the other 26.7~arcsec 
to the South. 
Diffuse H$\alpha$ emission is seen to the East of both knots, 
suggesting that deeper images might detect a greater flow extent. 
The bipolar flow spans a projected distance of $\sim$ 8500 AU.  
For a typical outflow velocity of 200 km/sec, the observed flow 
features would have an age of $\sim$ 100 years. 
This object has been assigned the number HH~939, following the 
criteria for the HH numbers designation \citep{Rei94}. The positions 
of the compact H$\alpha$ emission knots of HH~939 are reported in 
Table~\ref{HH939}.  The HH~939 outflow and its driving source fall
right in the center of the tight group A identified in 
\S~\ref{spa_distr_class}.

Another object with diffuse H$\alpha$ emission has been found at some 
20 arc-min to the South-West of IRAS~12496$-$7650 (DK~Cha). Images in 
H$\alpha$ and in the I band are shown in Figure~\ref{anon_outflow}. 
The object, also marked in Figure~\ref{dkcha_24mu}, is barely detected at 
24~$\mu$m; unfortunately, we cannot verify possible H$_2$ emission because 
the object falls outside the area covered with IRAC. While searching in 
all our images, we found no obvious driving source for this object. Note, 
however, that IRAS~12496$-$7650 (DK~Cha) is located about half the way 
between HH~54 and this H$\alpha$ object (see Figure~\ref{dkcha_24mu}). 
Such geometry would suggest that it could be the counterpart of HH~52-54, 
supporting the hypothesis that the embedded Herbig star is driving both 
HH features.
However, assuming that the new one is indeed an HH object, the object's 
morphology (see Figure~\ref{anon_outflow}) is not what one would expect 
for a flow coming from the North-East, casting some doubts that it may 
be part of the famous outflow. We extensively inspected all the images 
to search for other possible HH-like objects to the South-West of 
IRAS~12496$-$7650 (DK~Cha), but no other interesting features were found. 
More data in IR bands and in CO are necessary to investigate in more 
depth the nature of the diffuse H$\alpha$ emitting object and eventually, 
confirm whether it is part of the HH~52-54 outflow.

\section{Summary}
\label{sum}

Based on Spitzer c2d IRAC and MIPS observations, complemented 
with optical and near-IR data, we have performed a census of 
the PMS population in Cha~II. The properties of the PMS objects 
and candidates, as well as of their circumstellar matter have 
been studied. The population consists of 51 certified 
and 11 candidate PMS objects  in about 1.75 square degrees. The census is 
complete down to $M \sim0.03 M_{\odot}$. 

We have investigated both, the YSOs with infrared excess selected 
according to the c2d criteria, as well as the other PMS cloud 
members and candidate PMS objects from the previous surveys. The c2d multi-color
criteria select 26 YSOs in the IRAC~3.6-8\,$\mu$m/MIPS~24\,$\mu$m overlapping area of 
1.038~square-degree in Cha~II, where A$_{\rm V}>$2; the new c2d selection 
criteria work very well in cleaning the samples of YSO candidates 
from background extra-galactic contaminants. Most PMS objects and
candidates off the IRAC~3.6-8\,$\mu$m/MIPS~24\,$\mu$m overlapping area 
show IR excess; thus, they would have been selected with such criteria 
if IRAC~3.6-8\,$\mu$m/MIPS~24\,$\mu$m  data had been available. Our spectroscopic 
follow-ups, published elsewhere, confirmed the PMS nature of most of the 
selected candidates, supporting the reliability of the new c2d selection 
criteria and of previous selections.

We studied the volume density of the 62 PMS objects and candidates
with the following results: two tight groups, with volume densities 
higher than 25~M$_{\odot}$~pc$^{-3}$, can be identified but neither have 
enough members to qualify them as separate clusters. The spatial distribution 
of these groups is well correlated with the regions of high extinction
as derived from the c2d extinction map. On the other hand, at 
1~M$_{\odot}$~pc$^{-3}$, corresponding to the level proposed by \citet{Lad03} 
to define clusters, Cha~II as a whole can be defined as a loose cluster. 
Most (75\%) of the sources in Cha~II are associated with this cluster, 
which also includes both tight sub-groups.

Seven of the PMS objects and candidates are visual multiples with separations 
0.8-6.0 arcsec and are thus  blended in the IRAC or MIPS images. This gives a 
fraction of the order of 13$\pm$3\% in the separation range of 140-1060 AU, 
which is consistent with the observations in PMS clusters.
We did not find significant variability, on a timescale of 6 hours, 
in any of the Cha~II certified or candidate PMS objects, in line with other 
c2d investigations.

The analysis of the colors and SEDs shows that the Cha~II population is 
dominated by objects with active accretion, with only a minority being 
systems with passive disks. The average PMS object in Cha~II is the one 
for which the disk luminosity is about one third of the stellar luminosity. 
Most objects in Cha~II possess typical SEDs corresponding 
to disks that evolve more or less homogeneously from flared to flat. 
The disk fraction in Cha~II of 70-80\% is exceptionally  high in comparison 
with other star formation regions. If all the stars in Cha~II were born with 
circumstellar disks, only a minor fraction of about 20-30\% have lost their 
disks in about 2-3~Myr, i.e. the typical age of Cha~II members. 
Such high disk fraction and the very small number of Class~I sources relative 
to Class~II sources may indicate that star formation in Chamaeleon has occured 
rapidly a few million years ago. 

Many of the PMS stars close to the H-burning limit and all the legitimate 
sub-stellar objects possess optically thick disks, confirming that disks 
are common down to the very low-mass regime, in agreement with  previous 
findings in other regions of star formation. 
A trend for the optically thick disks to be more frequent in solar-mass 
stars is observed, but it is blurred by the poor statistics in Cha~II.
No evidence is found that more luminous stars, being more likely 
multiples, possess less massive disks with big holes as a consequence 
of rapid disk dissipation due to multiplicity.

The cloud mass determined from the extinction maps is on the order 
of 1000 $M_{\odot}$ and the SFE of 1-4\% is similar to our estimates 
for other T associations like Taurus and Lupus, but significantly 
lower than for Cha~I ($\sim$7\%). This suggests that different 
star-formation activities in the Chamaeleon clouds may reflect a 
different history of star formation. The Cha~II cloud turned about 
7$M_{\odot}$ into stars every Myr, i.e. less than the star formation 
rate in other star forming regions. At the level of the tight groups,
the star formation rate and efficiency are much higher, but yet
lower than in the other c2d clouds. 
 
Finally, an extensive investigation on the controversial driving 
source of the HH~52-54 outflow has been performed. Our dataset seem to 
still favor DK~Cha as the driving source, but large-scale sensitive
observations in CO are necessary to firmly establish this issue. 
The HH~52-54 outflow is very well detected at 70~$\mu$m with a 
flux density comparable to that of IRAS~12522$-$7640 at 60~$\mu$m. 
If this flux density all came from the [O I] line, the line flux 
would be on the order of 
$1.1 \times 10^{-11} erg \cdot s^{-1} \cdot cm^{-2} $.
A new Herbig-Haro outflow, HH~939, has been discovered based on
our WFI H$\alpha$ images. The driving source of the HH knots is 
the classical T~Tauri star Sz~50.

\acknowledgments
We thank the anonymous referee for his careful reading and useful 
comments/suggestions.
Support for this work, part of the Spitzer Legacy Science Program, was 
provided by NASA through contract 1224608 issued by the Jet Propulsion 
Laboratory, California Institute of Technology, under NASA contract 1407. 
This work was partially financed by the Istituto Nazionale 
di Astrofisica (INAF) through PRIN-INAF-2005.
L.S. acknowledges financial support from PRIN-INAF-2005 (Stellar clusters: 
a benchmark for star formation and stellar evolution) and 
B.M. thanks the Fundaci\'on Ram\'on Areces for financial support. 
We thank  T. Robitaille for discussions and help on the use of his 
accretion models and J. Kainulainen for providing an unpublished 
extinction map of Cha~II and for discussions on extinction issues.
We are also grateful to Ewine van Dishoeck for several discussions, 
comments and suggestions on an earlier version of the paper and to 
Paul Harvey for the many discussions on the YSO selection and for 
his comments to the paper. We also thank Bo Reipurth for discussions 
on HH objects and B. Stelzer for communications on unpublished results 
of XMM observations in Cha~II. We also thank Tyler Bourke for his 
comments. A special thank to pussycat Matula for her assiduous and warm 
assistantship during the preparation of this manuscript. 
We acknowledges the c2d collaborators for the many discussions and 
suggestions during the telecons.
This paper is based on observations carried out at the 
European Southern Observatory, La Silla and Paranal (Chile), under 
proposals numbers 67.C-0225, 68.C-0311, 076.C-0385 and 078.C-0293.
This publication makes use of data products from the Two Micron All Sky 
Survey, which is a joint project of the University of Massachusetts and 
the Infrared Processing and Analysis Center/California Institute of 
Technology, funded by NASA and the National Science Foundation. We also 
acknowledge extensive use of the SIMBAD data base.

{}

\begin{deluxetable}{lllllrlll}
\tabletypesize{\scriptsize}

\rotate

\tablecaption{\label{tab:yso} Pre-Main Sequence objects and candidates in Cha~II }

\tablewidth{0pt}

\tablehead{
\colhead{No.} & \colhead{c2d Name} & \colhead{Object ids.} & \colhead{R.A. (2000)} & \colhead{DEC. (2000)}  &
\colhead{Sel. Flag$^\dag$}   &  \colhead{Lada} & \colhead{Posi.} & \colhead{Refs.} \\

\colhead{} & \colhead{} & \colhead{} & \colhead{} & \colhead{}  &
\colhead{}   & \colhead{Class} & \colhead{Flag$^\ddag$} & \colhead{} 
}

\startdata

 1  &  SSTc2d~J124506.3$-$772013  &  IRAS~12416-7703 / AX~Cha	     &  12:45:06.43  &  $-$77:20:13.52  &   *	CND  &   II  &  off  &  a    \\
 2  &  SSTc2d~J124825.7$-$770637  &  IRAS~12448-7650		     &  12:48:25.70  &  $-$77:06:36.72  &  YSO  CND  &  III  &  in   &  b    \\
 3  &  SSTc2d~J125230.7$-$771513  &  IRAS~F12488-7658 / C~13	     &  12:52:30.49  &  $-$77:15:12.92  &  YSO  CND  &  III  &  in   &  b, c \\
 4  &  SSTc2d~J125317.2$-$770711  &  IRAS~12496-7650 / DK~Cha	     &  12:53:17.17  &  $-$77:07:10.63  &  YSO  PMS  &   II  &  in   &  b    \\
 5  &  SSTc2d~J125336.7$-$770639  &  WFI~J12533662-7706393	     &  12:53:36.62  &  $-$77:06:39.31  &       CND  &  III  &  in   &  d    \\
 6  &  SSTc2d~J125339.0$-$771553  &  C~17			     &  12:53:38.84  &  $-$77:15:53.21  &       CND  &  III  &  in   &  c    \\
 7  &  SSTc2d~J125342.9$-$771512  &  IRAS~12500-7658		     &  12:53:42.79  &  $-$77:15:11.59  &  YSO  PMS  &    I  &  in   &  b    \\
 8  &  SSTc2d~J125525.8$-$770047  &  C~33			     &  12:55:25.72  &  $-$77:00:46.62  &       CND  &  III  &  in   &  c    \\
\enddata

\tablecomments{Most coordinates are from optical images, where the multiple objects are better 
           resolved. Units of right ascension are hours, minuts and seconds, and units of 
	   declination are degrees, arcminutes and arcseconds. The c2d names are as in the
	   c2d final delivery document by \citet{Eva07}. We note that, because of  
	   reprocessing during the source extraction for new catalog releases, 
	   some of these names may be slightly different than those reported 
	   in \citet{You05} for the same sources.}

\tablenotetext{\dag :}{PMS = Pre-Main Sequence Object; CND = Candidate Pre-Main Sequence Object; 
		   YSO  = object with IRAC/MIPS~24\,$\mu$m  colors that passed the  H07  criteria.
		   * = would have met the c2d selection criteria if IRAC observations 
		   had been available. 
		   IRAS~12496-7650 is definitely a YSO, but not selected by the  H07  
		   criteria because it is saturated in IRAC~3.6\,$\mu$m and IRAC~4.5\,$\mu$m.}

\tablenotetext{\ddag  :}{in= object inside the IRAC~3.6-8\,$\mu$m/MIPS~24\,$\mu$m overlapping area;
                     off= object off the IRAC~3.6-8\,$\mu$m/MIPS~24\,$\mu$m overlapping area. }

\tablerefs{ 
(a) \citealt{Gau92};
(b) \citealt{Pru92};
(c) \citealt{Vuo01};
(d) \citealt{Spe07a};
(e) \citealt{Kne92};
(f) \citealt{Sch91};
(g) \citealt{Alc00};
(h) \citealt{All06a};
(i) \citealt{Per03};
(j) \citealt{Alc95};
(k) \citealt{Har93};
(l) \citealt{Bou95};
(m) \citealt{You05};
(n) \citealt{Lar98};
(o) this work
}
		   
\end{deluxetable}



\begin{deluxetable}{llrrrrrrrr}
\tabletypesize{\tiny}

\rotate

\tablecaption{\label{tab:flux_opt} Available fluxes, in milli Jy, from 0.36$\mu$m to 0.96$\mu$m }

\tablewidth{0pt}

\tablehead{
\colhead{No.} & \colhead{Object id.} &   \colhead{$U$} & \colhead{$B$} & \colhead{$V$} & \colhead{$R$} & 
\colhead{$Ha12$} & \colhead{$I$} & \colhead{$m914$}     & \colhead{$z$}  \\

              &                     & \colhead{0.36$\mu$m} & \colhead{0.44$\mu$m} & \colhead{0.55$\mu$m} & \colhead{0.64$\mu$m} & 
\colhead{0.665$\mu$m} & \colhead{0.79$\mu$m} & \colhead{$0.915\mu$m}     & \colhead{0.96$\mu$m}   
}

\startdata

 1  &   IRAS~12416-7703 	 &   --  	   &  0.24$\pm$0.06   &  0.43$\pm$0.10   &   19.08$\pm$4.57  &  61.29$\pm$7.34   &  363.22$\pm$6.69   &     --  	 &      --	       \\ 
 2  &   IRAS~12448-7650 	 &   --  	   &  0.48$\pm$0.12   &  4.22$\pm$1.01   &   11.82$\pm$2.83  &  16.12$\pm$1.93   &  510.70$\pm$18.81  &     --  	 &      --	       \\ 
 3  &   IRAS~F12488-7658	 &   --  	   &  0.16$\pm$0.04   &  0.65$\pm$0.16   &    3.57$\pm$0.03  &   2.44$\pm$0.29   &   32.52$\pm$0.60   &  90.11$\pm$6.64  &    49.38$\pm$2.27   \\ 
 4  &   IRAS~12496-7650 	 &   --  	   &  0.010$\pm$0.002 &  0.140$\pm$0.004 &    0.83$\pm$0.01  &   0.78$\pm$0.10   &    6.73$\pm$0.06   &  16.55$\pm$1.22  &    19.30$\pm$0.36   \\ 
 5  &   WFI~J12533662-7706393	 &   --  	   &   --	      &   --		 &   0.004$\pm$0.001 &    --		 &    0.050$\pm$0.002 &   0.14$\pm$0.02  &     0.16$\pm$0.01   \\ 
 6  &   C~17			 &   --  	   &  0.02$\pm$0.01   &   --		 &   0.450$\pm$0.004 &   0.48$\pm$0.06   &    2.29$\pm$0.02   &   4.64$\pm$0.34  &     5.27$\pm$0.05   \\ 
 7  &   IRAS~12500-7658 	 &   --  	   &   --	      &   --		 &   0.010$\pm$0.001 &   0.01$\pm$0.01   &    0.040$\pm$0.002 &   0.09$\pm$0.01  &     0.11$\pm$0.01   \\ 
 8  &   C~33			 &   --  	   & 0.010$\pm$0.003  &  0.32$\pm$0.08   &   0.56 $\pm$0.01  &   0.72$\pm$0.09   &    2.15$\pm$0.02   &   4.64$\pm$0.34  &     3.54$\pm$0.03   \\
\enddata

\tablecomments{$U$ fluxes from photometry by \citet{Hug92}. }
\tablecomments{$B$ \& $V$ fluxes from photometry extracted from the NOMAD catalog \citep{Zac05}. }
\tablecomments{$R$, $Ha12$, $I$, $m914$, \& $z$ fluxes from photometry by S07a. }
\tablecomments{For SSTc2d~J125758.7-770120 and  SSTc2d~J130540.8-773958 the $I$ fluxes from photometry by \citet{All06a}.}
	
\end{deluxetable}



\begin{deluxetable}{llrrrrrrr}
\tabletypesize{\tiny}

\rotate

\tablecaption{\label{tab:flux_J_IR3} Available fluxes, in milli Jy, from 1.25$\mu$m to 5.0$\mu$m}

\tablewidth{0pt}

\tablehead{
\colhead{No.} & \colhead{Object id.} &  \colhead{$J$} & \colhead{$H$} &  \colhead{$K_S$} &
\colhead{$L$} & \colhead{IRAC}  & \colhead{IRAC}  & \colhead{$M$}  \\

              &                     &  \colhead{1.25$\mu$m} & \colhead{1.65$\mu$m} &  \colhead{2.2$\mu$m} &
\colhead{3.4$\mu$m} & \colhead{3.6$\mu$m}  & \colhead{4.5$\mu$m}  & \colhead{5.0$\mu$m} 

}

\startdata

 1 &  IRAS~12416-7703		& 2035.96$\pm$53.03  &  3012.31$\pm$88.89  &  3277.39$\pm$100.10 &   	--	       &      --             &    --		 &      --               \\ 
 2 &  IRAS~12448-7650		& 3136.21$\pm$28.89  &  5351.05$\pm$98.57  &  5858.32$\pm$53.96  &  3456.39$\pm$31.83  &  2560.00$\pm$177.00 & 1510.00$\pm$92.10 &      --               \\ 
 3 &  IRAS~F12488-7658  	&  385.71$\pm$10.86  &   770.03$\pm$25.45  &   876.40$\pm$31.38  &   542.78$\pm$5.00   &   456.00$\pm$24.30  &  213.00$\pm$11.60 &      --               \\ 
 4 &  IRAS~12496-7650		&  304.88$\pm$8.15   &  1569.12$\pm$74.59  &  5828.77$\pm$169.32 & 12320.32$\pm$226.95 &      --             &     --		 & 21455.67$\pm$592.84   \\ 
 5 &  WFI~J12533662-7706393	&    1.91$\pm$0.08   &     7.63$\pm$0.24   & 	12.96$\pm$0.38   &   	--	       &     9.29$\pm$0.46   &    6.51$\pm$0.33  &      --               \\ 
 6 &  C~17			&   32.84$\pm$0.88   &    87.46$\pm$2.46   &   117.20$\pm$3.32   &   	--	       &    69.70$\pm$3.43   &   44.10$\pm$2.11  &      --               \\ 
 7 &  IRAS~12500-7658		&    0.89$\pm$0.07   &     3.45$\pm$0.15   & 	 9.53$\pm$0.29   &   	--	       &    17.90$\pm$0.87   &   25.40$\pm$1.21  &      --               \\ 
 8 &  C~33			&   14.94$\pm$0.44   &    27.99$\pm$0.77   & 	30.70$\pm$0.83   &   	--	       &    16.00$\pm$0.79   &   10.00$\pm$0.48  &      --               \\
\enddata

\tablecomments{$J$, $H$, \& $K_S$ fluxes from 2MASS photometry \citep{Cut03}.}
\tablecomments{$L$ \& $M$ fluxes from photometry by \citet{Hug92}.}
\tablecomments{IRAC~3.6\,$\mu$m \& IRAC~4.5\,$\mu$m fluxes from c2d photometry \citep{Eva07}.}

\end{deluxetable}


\begin{deluxetable}{llrrrrrrr}
\tabletypesize{\tiny}

\rotate

\tablecaption{\label{tab:flux_iso6.7_iras60} Available fluxes, in milli Jy, from 5.8$\mu$m to 25$\mu$m}

\tablewidth{0pt}

\tablehead{
\colhead{No.} & \colhead{Object id.} & \colhead{IRAC} &  \colhead{ISO} & \colhead{IRAC} & \colhead{IRAS} & \colhead{ISO} & 
\colhead{MIPS} &  \colhead{IRAS}    \\

              &                      &  \colhead{5.8$\mu$m} & \colhead{6.7$\mu$m} & \colhead{8.0$\mu$m} & \colhead{12$\mu$m} & \colhead{14.3$\mu$m} & 
\colhead{24$\mu$m} &  \colhead{25$\mu$m}  
}

\startdata

 1 &  IRAS~12416-7703		&	--	      &    --		 &	 --		 &  2950.00$\pm$250.00  &     --	    & 1240.00$\pm$165.00 &    3640.00$\pm$180.00   \\
 2 &  IRAS~12448-7650		&  1380.00$\pm$67.60  &    --		 &    868.00$\pm$42.00   &   $<$760.00  	&     --	    &  301.00$\pm$27.90  &     $<$380.00	   \\
 3 &  IRAS~F12488-7658  	&   211.00$\pm$10.40  &    --		 &    128.00$\pm$6.10	 &   $<$100.00  	&     --	    &	30.80$\pm$2.85   &     $<$90.00 	   \\
 4 &  IRAS~12496-7650		& 18700.00$\pm$3620.00&    --		 &  11000.00$\pm$3460.00 & 44170.00$\pm$2210.00 &     --	    &	  --		 & 102800.00$\pm$5140.00   \\
 5 &  WFI~J12533662-7706393	&     4.78$\pm$0.25   &    --		 &	2.89$\pm$0.16	 &	--		&     --	    &	 0.22$\pm$0.83   &	  --		   \\
 6 &  C~17			&    33.30$\pm$1.58   &    --		 &     19.40$\pm$0.92	 &	--		&     --	    &	 1.63$\pm$0.24   &	  --		   \\
 7 &  IRAS~12500-7658		&    33.50$\pm$1.59   &    --		 &     51.20$\pm$2.42	 &   110.00$\pm$20.00	&     --	    &  612.00$\pm$56.60  &     900.00$\pm$180.00   \\
 8 &  C~33			&     7.13$\pm$0.34   &    --		 &	4.44$\pm$0.21	 &	--		&     --	    &	 0.34$\pm$0.22   &	  --		   \\
\enddata

\tablecomments{IRAC~5.8\,$\mu$m , IRAC~8\,$\mu$m \& MIPS~24\,$\mu$m  fluxes from c2d photometry  \citep{Eva07}.}
\tablecomments{ISO6.7 \& ISO14.3 fluxes from \citet{Per03}.}
\tablecomments{IRAS12 \& IRAS25 fluxes from \citet{Bei88}.}
\tablecomments{The MIPS~24\,$\mu$m flux for Sz~63 is taken from \citet{You05}.}

\end{deluxetable}



\begin{deluxetable}{llrrrrrl}
\tabletypesize{\tiny}

\rotate

\tablecaption{\label{tab:flux_70_1300} Available fluxes, in milli Jy, from 60$\mu$m to 1300$\mu$m}

\tablewidth{0pt}

\tablehead{
\colhead{No.} & \colhead{Object id.} & \colhead{IRAS}      & \colhead{MIPS}     & \colhead{IRAS}      & \colhead{MIPS}      &  \colhead{1200$\mu$m} & \colhead{1300$\mu$m}  \\
\colhead{ }   &  \colhead{ }         &  \colhead{60$\mu$m} & \colhead{70$\mu$m} & \colhead{100$\mu$m} & \colhead{160$\mu$m} &  \colhead{ }          & \colhead{ }  
}

\startdata

 1 &  IRAS~12416-7703	      &    300.00$\pm$ 15.00	 &  116.00$\pm$12.20   &     $<$3440.00 	  &	 -		 &    - 	      &    -		   \\ 
 2 &  IRAS~12448-7650	      &      $<$330.00  	 &   75.00$\pm$8.18    &     $<$3980.00 	  &	 -		 &    - 	      &    -		   \\ 
 3 &  IRAS~F12488-7658        &      $<$160.00  	 &     -	       &     $<$4500.00 	  &	 -		 &    - 	      &    -		   \\ 
 4 &  IRAS~12496-7650	      & 105650.00$\pm$10660.00   & 36500.00$\pm$3420.00 & 100140.00$\pm$10010.00   &	 -		 & 1470.00$\pm$120.00 & 680.00$\pm$22.00   \\ 
 5 &  WFI~J12533662-7706393   & 	--		 &     -	       &	  -		  &	 -		 &    - 	      &    -		   \\ 
 6 &  C~17		      & 	--		 &     -	       &	  -		  &	 -		 &    - 	      &    -		   \\ 
 7 &  IRAS~12500-7658	      &     3630.00$\pm$730.00   & 1540.00$\pm$ 150.00 &   5060.00$\pm$1100.00    & 11800.00$\pm$2400.00 &  $<$260.00	      &  59.70$\pm$15.00   \\ 
 8 &  C~33		      & 	--		 &     -	       &	  -		  &	 -		 &    - 	      &    -		   \\
\enddata

\tablecomments{IRAS60 \& IRAS100 fluxes from \citet{Bei88}.}
\tablecomments{MIPS~70\,$\mu$m fluxes from c2d photometry  \citep{Eva07}.}
\tablecomments{MIPS~160\,$\mu$m \& 1200$\mu$m fluxes from \citet{You05}.}
\tablecomments{1300$\mu$m fluxes from \citet{Hen93}.}

\end{deluxetable}


\begin{table}  
\begin{flushleft}
\center{\caption{Summary of Lada classes in Cha~II. 
		\label{ir_classes}}}

\begin{tabular}{c|cccc|cccc}
\tableline  
\tableline
Lada  & certified & YSO   &  \multicolumn{2}{c|}{\underline{Total YSOs}} & certified & PMS   &  \multicolumn{2}{c}{\underline{Total PMS}} \\
Class & YSOs      & cand. &	           &                         & PMS       & cand. &     &         \\
\tableline
I          &  2 & 0 & 2  & (8\% )  &  3     &  0   & 3  & (5\% )  \\
Flat       &  1 & 0 & 1  & (4\% )  &  2     &  0   & 2  & (3\% )  \\
II         & 18 & 1 & 19 & (73\% ) &  36    &  2   & 38 & (61\% ) \\
III        &  2 & 2 & 4  & (15\% ) &  9+1?  & 8+1? & 19 & (31\% ) \\
\tableline
total      & 23 & 3 & 26 &	   & 51     & 11 & 62 & 	\\
\tableline

\end{tabular}
\end{flushleft}
\end{table}

\begin{table}  
\begin{center}
\caption{Summary of groups vs. Lada classes in Cha~II. The number of c2d identified YSOs 
is given in parentheses. \label{surf_dens} } 

\begin{tabular}{lccccc}

\tableline  
\tableline  
Region  &   Tot. ($N_{\rm YSO}$)  & Class~I & Flat  &  Class~II  &  Class~III  \\                      
        &                         &         &       &           &              \\                      
\tableline  
  Cha~II loose cluster ($\rho >$1$\times$LL03)    &  48 (22) &  2 (2) &  2 (1) & 31 (17) & 13 (4) \\   
  $\ldots$A tight group ($\rho >$25$\times$LL03)  &  6 (2)   &  0 (0) &  0 (0) &  4  (2) &  2 (0) \\   
  $\ldots$B tight group ($\rho >$25$\times$LL03)  &  10 (6)  &  0 (0) &  1 (1) &  8  (5) &  1 (0) \\    
  Distributed ($\rho <$1$\times$LL03)             &  14 (4)  &  1 (0) &  0 (0) &  7  (2) &  6 (2) \\   
\tableline

\end{tabular}
\end{center}
\end{table}

\begin{deluxetable}{ll|ccccl|ccc}
\tabletypesize{\scriptsize}

\rotate

\tablecaption{\label{disk_results_lum} SED parameters and disk modeling results from D01 models.}

\tablewidth{0pt}

\tablehead{

\colhead{No.} &  \colhead{Object id.}  &  \colhead{$\alpha$} &
\colhead{$log(L_d/L_*)$} & \colhead{$log(L_d)$}    &  
\colhead{$\lambda_{turn-off}$} &  \colhead{$\alpha_{excess}$} & 
\colhead{$R_{disk}^{Dull}$}  & \colhead{$R_{in}^{Dull}$}  & \colhead{$log(M_{disk}^{Dull})$}  \\

\colhead{}    &   \colhead{}   &  \colhead{$_{[K\&MIPS~24\,\mu m}]$} &
\colhead{}    & \colhead{($L_{\odot}$)} &  
\colhead{($\mu$m)}    &     \colhead{}            &  
\colhead{(AU)}     & \colhead{(AU)}        & \colhead{($M_{\odot}$)}   
  }

\startdata
 1 & 	      IRAS12416-7703 &   -1.51  &  -0.90 &  0.21 &   2.2   &   -1.64  &     15.0    &	 0.24  &   -3.7   \\ 
 2 & 	      IRAS12448-7650 &   -2.38  &  -2.12 & -0.72 &   8.0   &   -2.15  &     70.0    &	 0.21  &   -3.3   \\ 
 3 & 	     IRASF12488-7658 &   -2.54  &  -0.75 & -0.32 &   8.0   &   -2.24  &     60.0    &	 0.08  &   -7.0   \\ 
 4 & 	      IRAS12496-7650 &   -0.72  &   0.36 &  1.63 &   0.9   &	0.05  &    200.0    &	 1.30  &   -0.4   \\ 
 5 & 	WFIJ12533662-7706393 &   -2.88  &   0.26 & -1.69 &   -     &	 -    &     -	    &	  -    &      -   \\ 
 6 & 			 C17 &   -2.98  &  -1.15 & -1.49 &   -     &	 -    &     -	    &	  -    &      -   \\ 
 7 & 	      IRAS12500-7658 &    0.62  &   2.17 & -0.04 &   -     &	 -    &     -	    &	  -    &      -   \\ 
 8 & 			 C33 &   -3.03  &  -0.77 & -1.55 &   -     &	 -    &     -	    &	  -    &      -   \\
\enddata

\tablenotetext{\dag :}{object with insufficient photometry to derive the $\alpha_{[K\&MIPS~24\,\mu m}]$ index.}
\tablenotetext{\ast :}{$\alpha_{excess}$ index relies on upper limits. }

\end{deluxetable}

\begin{deluxetable}{ll|cccc|cccc}
\tabletypesize{\scriptsize}

\rotate

\tablecaption{\label{disk_results_rob_dal} Disk modeling results from R07 and \citet{Dal05} models.}

\tablewidth{0pt}

\tablehead{
\colhead{No.} &  \colhead{Object id.} & \colhead{$R_{disk}^{Rob}$}   & \colhead{$R_{in}^{Rob}$}   & \colhead{$log(M_{disk}^{Rob})$}   & \colhead{$M_{accr}^{Rob}$$^\dag$}   & \colhead{$R_{disk}^{D'Ale}$} & \colhead{$R_{in}^{D'Ale}$} & \colhead{$log(M_{disk}^{D'Ale})$} & \colhead{$M_{accr}^{D'Ale}$$^\dag$} \\
\colhead{}    &  \colhead{}           & \colhead{(AU)}               & \colhead{(AU)}             & \colhead{($M_{\odot}$)}           &   \colhead{($M_{\odot} / yr$)}      & \colhead{(AU)}               & \colhead{(AU)}             & \colhead{($M_{\odot}$)}           &   \colhead{($M_{\odot} / yr$)} 
}
\startdata
 1 & 	      IRAS12416-7703 &     42	&   0.43   &   -5.56  &   0.0e+00  &	 -  &	   -	 &	 -   &     -    \\
 2 & 	      IRAS12448-7650 &    263	&   0.40   &   -6.31  &   1.0e-08  &   100  &	 0.11	 &    -3.00  &  1.e-09  \\
 3 & 	     IRASF12488-7658 &   1805	&   0.13   &   -6.63  &   3.6e-08  &	 -  &	   -	 &	 -   &     -    \\
 4 & 	      IRAS12496-7650 &    149	&   1.79   &   -2.90  &   1.8e-07  &   800  &	 1.01	 &     0.93  &  1.e-06  \\
 5 & 	WFIJ12533662-7706393 &    -	&   -	   &	  -   &      -     &	 -  &	   -	 &	 -   &     -    \\
 6 & 			 C17 &    -	&   -	   &	  -   &      -     &	 -  &	   -	 &	 -   &     -    \\
 7 & 	      IRAS12500-7658 &     50	&   0.65   &   -2.19  &   2.0e-06  &	 -  &	   -	 &	 -   &     -    \\
 8 & 			 C33 &    -	&   -	   &	  -   &      -     &	 -  &	   -	 &	 -   &     -    \\
\enddata
\tablenotetext{\dag :}{The mass-accretion rate is given only for sources with $U$ and/or $B$ photometry (see \S~\ref{disk_pars}).}
\end{deluxetable}

\begin{center}
\begin{table}  
\tabletypesize{\scriptsize}
\caption{Numbers, Densities and Star Formation Rates.  \label{starform}}
\begin{tabular}{lcccccccc}
\tableline
\tableline 
 Type    & $\Omega$     &   Area  &  N  &  N/$\Omega$  &  N/Area     &  N/Vol$^\dagger$   & SF~rate		      &  SF~rate/Area   \\
         & (deg$^{2}$)  & (pc$^2$) &    & (deg$^{-2}$) & (pc$^{-2}$)  & (pc$^{-3}$) & (M$_{\odot}$ Myr$^{-1}$) & (M$_{\odot}$ Myr$^{-1}$ pc$^{-2}$) \\
\tableline 
 YSOs    &      1.038  & 10.02    & 26  &	25	&    2.6      &    1.1   &	   7.0        &   0.68    \\
 PMS     &      1.750  & 17.0     & 62  &	35	&    3.7      &    1.2   &	   10.7       &   0.63    \\
\tableline

Cha~II loose cluster  &  0.944 &  9.11   & 48  &       51     &    5.3      &    2.3   &         12.5       &   0.90    \\
$\ldots$A tight group &  0.022 &  0.211  &  6  &       275    &    28       &    82    &          1.6       &   9.5     \\
$\ldots$B tight group &  0.035 &  0.333  & 10  &       289    &    30       &    69    &          2.6       &   9.0     \\
  
\tableline
\end{tabular}
\tablenotetext{\dag}{Volume estimated by assuming that the distribution of sources 
is locally spherical.}
\end{table}
\end{center}

\begin{table}  
\begin{center}
\caption{Estimates of the cloud mass and SFE in Cha~II. \label{cloud_mass}}
\begin{tabular}{lllcc}
\tableline 
\tableline
Tracer     &     Mass      &  Ref.  &   \multicolumn{2}{c}{\underline{SFE (\%) } } \\
           & ($M_{\odot}$) &        &         YSOs  &        PMS     \\
\tableline
100$\mu$m  & 900$^a$       & \citet{Bou98}     & 1.5 & 3.4   \\
$^{12}$CO  & 1250$^b$      & \citet{Bou98}     & 1.1 & 2.5  \\
$^{12}$CO  & 1860$^b$      & \citet{Miz01}     & 0.7 & 1.7  \\
$^{13}$CO  & 960$^c$       & \citet{Miz99}     & 1.4 & 3.2  \\
Extinction & 670$^d$       & this work       & 2.0 & 4.6 \\
Extinction & 830$^e$       & \citet{Cam99}     & 1.7 & 3.7  \\
Extinction & 780$^e$       & Kainulainen (private communication)  & 1.8 & 3.9 \\

\tableline

\end{tabular}

\tablenotetext{^a}{Total mass of $H_2$ derived from the 100$\mu$m luminosity}
\tablenotetext{^b}{Total mass of $H_2$ derived from the $^{12}$CO luminosity}
\tablenotetext{^c}{Total mass of $H_2$ derived from the $^{13}$CO luminosity}
\tablenotetext{^d}{Cloud mass for A$_{\rm V}>$ 2 mag}
\tablenotetext{^e}{Cloud mass calculated using prescription by \citet{Kai05} for A$_{\rm V}>$ 2 mag}

\end{center}
\end{table}

\begin{table}  
\begin{flushleft}
\caption{Estimates of the SFE for the sub-groups in Cha~II. The values for the 
c2d identified YSOs are given in parentheses.\label{SFE_substructures}}
\begin{tabular}{l|cc|c|cc|cc}
\tableline 
\tableline
Region          & \multicolumn{2}{c|}{\underline{No.}} & M$_{\rm sub-cloud}$    &  \multicolumn{2}{c|}{M$_{\rm objects}$} & \multicolumn{2}{c}{SFE (\%)} \\
                &       &                         &  ($M_{\odot}$)           &         &  ($M_{\odot}$)         &       &	     \\
\tableline
Cha~II loose cluster  & 48 & (22)         & 400.7 	        &  25.0 & (11.4)     &  5.8 & (2.8)      \\
$\ldots$A tight group &  6 & (2)          & 11.9  	        &  3.1	& (1.0)      & 20.7 & (7.8)      \\
$\ldots$B tight group & 10 & (6)          & 14.7  	        &  5.2	& (3.1)      & 26.1 & (17.4)     \\
distributed           & 14 & (4)          & 275.2 	        &  7.3	& (2.1)      &  2.6 & (0.8)      \\
\tableline
\end{tabular}

\tablecomments{M$_{\rm sub-cloud}$ is the mass within volume density contour from c2d dust extinction maps}
\tablecomments{M$_{objects}$ is the total mass in objects estimated as $0.52 M_\odot \cdot N_{\rm objetcs} $}

\end{flushleft}
\end{table}

\begin{table} 
\begin{center}
\caption{Components of the HH~939 outflow. \label{HH939}}
\begin{tabular}{lll}
\tableline 
\tableline
HH         &     R.A. (2000)  &  DEC. (2000)   \\
           &                  &                \\
\tableline

HH~939~N   &    13:00:55.92   & -77:10:01.0     \\
HH~939~S   &    13:00:54.92   & -77:10:49.0     \\

\tableline

\end{tabular}
\end{center}
\end{table}


\begin{thebibliography}{}

\bibitem[\protect\citeauthoryear{Adams et al.}{1987}]{Ada87} Adams, F.C., Lada, C.J. \& Shu, F.H., ApJ 312, 788
\bibitem[\protect\citeauthoryear{Alcal\'{a} et al.}{1995}]{Alc95} Alcal\'{a}, J.M., Krautter, J., Schmitt, J., et al. 1995, A\&A Supp. Ser. 114, 109
\bibitem[\protect\citeauthoryear{Alcal\'{a} et al.}{1997}]{Alc97} Alcal\'{a}, J.M., Krautter, J., Covino, E., et al. 1997, A\&A 319, 18
\bibitem[\protect\citeauthoryear{Alcal\'{a} et al.}{2000}]{Alc00} Alcal\'{a}, J.M., Covino, E., Sterzik, M.F., et al. 2000, A\&A 355, 629
\bibitem[\protect\citeauthoryear{Alcal\'{a} et al.}{2006}]{Alc06} Alcal\'{a}, J.M., Spezzi, L., Frasca, A., \& Covino, E., 2006, A\&A 453, L1		 
\bibitem[\protect\citeauthoryear{Allard \& Hauschildt}{1995}]{All95} Allard, F., \& Hauschildt, P.H., 1995, ApJ 445, 433
\bibitem[\protect\citeauthoryear{Allard et al.}{2000}]{All00} Allard F., Hauschildt, P.H., \& Schwenke, D., 2000, ApJ 540, 1005
\bibitem[\protect\citeauthoryear{Allen et al.}{2004}]{All04} Allen, L.E., Calvet, N., D'Alessio, P., et al. 2004, ApJS 154, 363
\bibitem[\protect\citeauthoryear{Allers et al.}{2006}]{All06a} Allers, K.N., Kessler-Silacci, J.E., Cieza, L.A., \& Jaffe, D.T., 2006, ApJ 644, 364
\bibitem[\protect\citeauthoryear{Allers}{2006}]{All06b} Allers, K.N., 2006, Ph. D. Dissertation, The University of Texas at Austin   
\bibitem[\protect\citeauthoryear{Allers et al.}{2007}]{All07} Allers, K.N., Jaffe, D.T., Luhman, K. L., et al. 2007, ApJ 657, 511   
\bibitem[\protect\citeauthoryear{Backman \& Paresce}{1993}]{Bac93} Backman, D.E., \& Paresce, F., 1993, proceedings of the PPIII conference, pag. 1253 
\bibitem[\protect\citeauthoryear{Baraffe et al.}{1998}]{Bar98} Baraffe, I., Chabrier, G., Allard, F., \& Hauschildt, P.H., 1998, A\&A 337, 403 
\bibitem[\protect\citeauthoryear{Barrado y Navascu\'{e}s et al.}{2004}]{Bar04a} Barrado Y Navascu\'{e}s, D., \& Jayawardhana, R., 2004, AJ 615, 840
\bibitem[\protect\citeauthoryear{Beichmann et al.}{1988}]{Bei88} Beichmann, C.A., Helou, G., \& Walker, D.W., 1988, IRAS Point Source Catalog (NASA RP-1190; Washington: GPO)
\bibitem[\protect\citeauthoryear{Bertin \& Arnouts}{1996}]{Ber96}  Bertin, E., \& Arnouts, S., 1996, A\&AS 117, 393
\bibitem[\protect\citeauthoryear{Boulanger et al.}{1998}]{Bou98} Boulanger, F., Bronfman, L., Dame, T.M., \& Thaddeus, P., 1998, A\&A 332, 273
\bibitem[\protect\citeauthoryear{Brandner \& Zinnecker}{1999}]{Bra97} Brandner, W., \& Zinnecker, H., 1997, A\&A 321, 220
\bibitem[\protect\citeauthoryear{Bourke et al.}{1995}]{Bou95}  Bourke, T.L., Hyland, A.R., \& Robinson, G., 1995, MNRAS 276, 1052   
\bibitem[\protect\citeauthoryear{Calvet et al.}{2005}]{Cal05} Calvet, N., D'Alessio, P., Watson, D.M., et al. 2005, ApJ 630, 185 
\bibitem[\protect\citeauthoryear{Cambr\`esy}{1999}]{Cam99} Cambr\'esy, L., 1999, A\&A 345, 965
\bibitem[\protect\citeauthoryear{Caratti o Garatti et al.}{2006}]{Car06} Caratti o Garatti, A., Giannini, T., Nisini, B., \& Lorenzetti, D., 2006, A\&A 449, 1077
\bibitem[\protect\citeauthoryear{Ceccarelli et al.}{1997}]{Cec97} Ceccarelli, C., Haas, M.R., Hollenbach, D.J., \& Rudolph, A.L., 1997, ApJ 476, 771
\bibitem[\protect\citeauthoryear{Cieza et al.}{2005}]{Cie05} Cieza, L., Kessler-Silacci, J.E., Jaffe, D.T., et al. 2005, ApJ 635, 422
\bibitem[\protect\citeauthoryear{Cieza et al.}{2007}]{Cie07} Cieza, L. et al. 2007, ApJ, in press
\bibitem[\protect\citeauthoryear{Chabrier et al.}{2000}]{Cha00} Chabrier, G., Baraffe, I., Allard, F., \& Hauschildt, P., 2000, ApJ 542, 464   
\bibitem[\protect\citeauthoryear{Chiang \& Goldreich}{1999}]{Chi99} Chiang, E.I., \& Goldreich, P., 1999, ApJ 519, 279
\bibitem[\protect\citeauthoryear{Cohen}{1973}]{Coh73} Cohen, M., 1973, Mon. Not. R. Astron. Soc. Vol. 164, p. 417
\bibitem[\protect\citeauthoryear{Covino et al.}{1997a}]{Cov97a} Covino, E., Alcal\'{a}, J.M., Allain, S., et al. 1997, A\&A 328, 187
\bibitem[\protect\citeauthoryear{Covino et al.}{1997b}]{Cov97b} Covino, E., Palazzi, E., Penprase, B., et al. 1997, A\&AS 122, 95 
\bibitem[\protect\citeauthoryear{Cutri et al.}{2003}]{Cut03} Cutri, R.M., Skrutskie, M.F., Van Dyk, S., et al. 2003, The IRSA 2MASS Point Source Catalog, NASA/IPAC Infrared Science Archive, http://irsa.ipac.caltech.edu/applications/Gator
\bibitem[\protect\citeauthoryear{D'Antona \& Mazzitelli}{1998}]{DAn98} D'Antona, F., \& Mazzitelli, I., 1997, Mem. S.A.It. 68, 807
\bibitem[\protect\citeauthoryear{D'Alessio et al.}{2005}]{Dal05} D'Alessio, P., Mer\'{i}n, B., Calvet, N., at al. 2005, Rev. Mex. de Aston. y Astrof. 41, 61
\bibitem[\protect\citeauthoryear{D'Alessio et al.}{1999}]{Dal99} D'Alessio, P., Calvet, N., Hartmann, L., et al. 1999, ApJ 527, 893
\bibitem[\protect\citeauthoryear{Ducourant et al.}{2005}]{Duc05} Ducourant, C., Teixeira, R., P\'eri\'e, J.P.,, et al. 2005, A\&A 438, 769 
\bibitem[\protect\citeauthoryear{Dullemond et al.}{2001}]{Dul01} Dullemond, C.P., Dominik, C., \& Natta, A., 2001, ApJ 560, 957 (D01)
\bibitem[\protect\citeauthoryear{Dullemond et al.}{2007}]{Dul07} Dullemond, C.P., Hollenbach, D., Kamp, I., D'Alessio, P., 2007, in Protostars and Planets V, B. Reipurth, D. Jewitt, and K. Keil (eds.), University of Arizona Press, p. 555
\bibitem[\protect\citeauthoryear{Dunham et al.}{2006}]{Dun06} Dunham, M.M., Evans II, N.J., Bourke, L.T., et al. 2006, ApJ 651, 945
\bibitem[\protect\citeauthoryear{Evans et al.}{2003}]{Eva03} Evans II, N.J., and the c2d Team, PASP 115, 965 
\bibitem[\protect\citeauthoryear{Evans et al.}{2005}]{Eva05} Evans II, N.J., Harvey, P., \& Dunham, M.M., 2005, Third Delivery of Data from the c2d Legacy Project: IRAC and MIPS (Pasadena, SSC)   
\bibitem[\protect\citeauthoryear{Evans et al.}{2007}]{Eva07} Evans II, N.J., Harvey, P., \& Dunham, M.M., 2007, Final Delivery of Data from the c2d Legacy Project: IRAC and MIPS (Pasadena, SSC)   
\bibitem[\protect\citeauthoryear{Gauvin \& Strom}{1992}]{Gau92} Gauvin, L.S., \& Strom, K.M., 1992, AJ 385, 217
\bibitem[\protect\citeauthoryear{Graham \& Hartigan}{1988}]{Gra88} Graham, J.A., \& Hartigan, P., 1988, AJ 95, 1197
\bibitem[\protect\citeauthoryear{Grasdalen et al.}{1975}]{Gra75} Grasdalen, G., Joyce, R., Knacke, R.F., et al. 1975, AJ 80, 117
\bibitem[\protect\citeauthoryear{Greene et al.}{1994}]{Gre94} Greene, T.P., Wilking, B.A., Andr\'e, P., et al. 1994, ApJ 434, 614
\bibitem[\protect\citeauthoryear{Gregorio-Hetem et al.}{1988}]{Het88} Gregorio Hetem, J.C., Sanzovo, G.C., \& Lepine, J.R.D., 1988, A\&AS 76, 374
\bibitem[\protect\citeauthoryear{Gutermuth et al.}{2005}]{Gut05} Gutermuth, R.A., Megeath, S.T., Pipher, J.L., et al. 2005, ApJ 632, 397 
\bibitem[\protect\citeauthoryear{Hartigan}{1993}]{Har93} Hartigan, P., 1993, AJ 105, 1511
\bibitem[\protect\citeauthoryear{Hartigan et al.}{1995}]{Har95} Hartigan, P., Edwards, S., \& Ghandour, L., 1995, ApJ 452, 736
\bibitem[\protect\citeauthoryear{Hartmann et al.}{2005}]{Hart05} Hartmann, L., Megeath, S.T., Allen, L.E., et al. 2005, ApJ 629, 881
\bibitem[\protect\citeauthoryear{Harvey et al.}{2006}]{Har06} Harvey, P., Chapman, N., Shih-Ping, L., et al. 2006, ApJ 644, 307
\bibitem[\protect\citeauthoryear{Harvey et al.}{2007}]{Har07} Harvey, P., Mer\'in, B., Huard, T.L., et al. 2007, ApJ, 663, 1149 (H07)
\bibitem[\protect\citeauthoryear{Hauschildt et al.}{1999}]{Hau99} Hauschildt, P.H., Allard, F., \& Baron, E., 1999, ApJ 512, 377
\bibitem[\protect\citeauthoryear{Hayakawa et al.}{2001}]{Hay01} Hayakawa, T., Cambr\`esy, L., Onishi, T., et al. 2001, PASJ 53, 1109
\bibitem[\protect\citeauthoryear{Henning et al.}{1993}]{Hen93} Henning, T., Pfau, W., Zinnecker, H., \& Prusti, T., 1993, A\&A 276, 129
\bibitem[\protect\citeauthoryear{Hollenbach}{1985}]{Hol85} Hollenbach, D.J., Icarus 61, 36
\bibitem[\protect\citeauthoryear{Huard et al.}{2007}]{Hua07} Huard, T., et al. 2007, in preparation    
\bibitem[\protect\citeauthoryear{Hughes et al.}{1989}]{Hug89} Hughes, J.H., Emerson, J.P., Zinnecker, H., \& Whitelock, P.A., 1989, M.N.R.A.S. 236, 117
\bibitem[\protect\citeauthoryear{Hughes \& Hartigan}{1992}]{Hug92} Hughes, J.H., \& Hartigan, P., 1992, AJ 104-2, 680
\bibitem[\protect\citeauthoryear{Ivezic et al. }{1999}]{Ive99} Ivezic, Z., Nenkova, M., \& Elitzur, M., 1999, User Manual for Dusty (University of Kentucky), Internal Report 
\bibitem[\protect\citeauthoryear{Jayawardhana \& Ivanov}{2006}]{Jay06} Jayawardhana, R., \& Ivanov, V.D., 2006, ApJ 647, 167
\bibitem[\protect\citeauthoryear{J{\o}rgensen et al.}{2006}]{Jor06} J{\o}rgensen, J.K., Johnstone, D., van Dishoeck, E.F., \& Doty, S.D., 2006, A\&A 449, 609
\bibitem[\protect\citeauthoryear{J{\o}rgensen et al.}{2007}]{Jor07} J{\o}rgensen, J.K., et al., in preparaton
\bibitem[\protect\citeauthoryear{Knee}{1992}]{Kne92} Knee, L.B.G., 1992, A\&A 259, 283 
\bibitem[\protect\citeauthoryear{Kainulainen et al.}{2005}]{Kai05} Kainulainen, J., Lehtinen, J., \& Harju, J., 2005, A\&A 447, 597
\bibitem[\protect\citeauthoryear{Kenyon \& Hartmann}{1987}]{Ken87} Kenyon, S., \& Hartmann, L., 1987, ApJ 323, 714
\bibitem[\protect\citeauthoryear{Kenyon \& Hartmann}{1995}]{Ken95} Kenyon, S., \& Hartmann, L., 1995, ApJ 101, 117
\bibitem[\protect\citeauthoryear{K\"ohler}{2001}]{Koh01} K\"ohler, R., 2001, AJ 122, 3334 
\bibitem[\protect\citeauthoryear{Lada}{1987}]{Lad87} Lada, C.J., 1987, in IAU Symp. 115, Star Forming Regions, ed. M. Peimbert \& J. Jugaku ( Dordrecht: Reidel), 1
\bibitem[\protect\citeauthoryear{Lada \& Lada}{2003}]{Lad03} Lada, C.J., \& Lada, E.A., 2003, A.R.A\&A 41, 57  (LL03)
\bibitem[\protect\citeauthoryear{Lada et al.}{2006}]{Lad06} Lada, C.J., Muench, A.A., Luhman, K.L., et al. 2006, AJ 131, 1574
\bibitem[\protect\citeauthoryear{Lagrange et al.}{2000}]{Lag00} Lagrange, A.M., Backman, D.E., \& Artymowicz, P., 2000, proceeding of the PPIV conference, pag. 639
\bibitem[\protect\citeauthoryear{Larson et al.}{1998}]{Lar98} Larson, K.A., Whittet, D.C.B., Prusti, T., \& Chiar, J.E., 1998, A\&A 337, 465
\bibitem[\protect\citeauthoryear{Lehtinen, Mattila \& Lemke}{2005}]{Leh05} Lehtinen, K., Mattila, K., \& Lemke, D., 2005, A\&A 437, 159 
\bibitem[\protect\citeauthoryear{Lepin\`{e} \& Duvert}{1994}]{Lep94} Lepin\'{e} J. \& Duvert G. 1994, A\&A 286, 60 
\bibitem[\protect\citeauthoryear{Liseau et al.}{1996}]{Lis96} Liseau, R., Ceccarelli, C., Larsson, B., et al. 1996, A\&A 315, L181
\bibitem[\protect\citeauthoryear{Luhman et al.}{2003}]{Luh03} Luhman, K.L., Stauffer, J.R., Muench, A.A., et al. 2003, AJ 593, 1093
\bibitem[\protect\citeauthoryear{Luhman}{2004}]{Luh04} Luhman, K.L., 2004, AJ 602, 816
\bibitem[\protect\citeauthoryear{Luhman et al.}{2005a}]{Luh05a} Luhman, K.L., D'Alessio, P., Calvet, N., et al. 2005, ApJ 620, 51
\bibitem[\protect\citeauthoryear{Luhman et al.}{2005b}]{Luh05b} Luhman, K.L., Adame, L., D'Alessio, P., et al. 2005, AJ 635, L93
\bibitem[\protect\citeauthoryear{Luhman et al.}{2005c}]{Luh05c} Luhman, K.L., Lada, C.J., Hartmann, L.,  et al. 2005, AJ 631, L69
\bibitem[\protect\citeauthoryear{Lynden-Bell \& Pringle}{1974}]{Lyn74} Lynden-Bell, D., Pringle, J.E. 1974, M.N.R.A.S., 168, 603
\bibitem[\protect\citeauthoryear{Mizuno et al.}{1995}]{Miz95} Mizuno, A., Onishi, T., Yonekura, Y., et al. 1995, AJ 445, L161
\bibitem[\protect\citeauthoryear{Mizuno et al.}{1998}]{Miz98} Mizuno, A., Hayakawa, T., Yamaguchi, N. et al.  1998, ApJ 507, 83
\bibitem[\protect\citeauthoryear{Mizuno et al.}{1999}]{Miz99} Mizuno, A., Hayakawa, T., Tachihara, K., et al. 1999, PASJ 51, 859
\bibitem[\protect\citeauthoryear{Mizuno et al.}{2001}]{Miz01} Mizuno, A., Yamaguchi, R., Tachihara, K., et al. 2001, PASJ 53, 1071
\bibitem[\protect\citeauthoryear{Neufeld et al.}{1998}]{Neu98} Neufeld, D.A., Melnick, G.J., \& Harwit, M., 1998, ApJ 506, L75
\bibitem[\protect\citeauthoryear{Neufeld et al.}{2006}]{Neu06} Neufeld, D.A., Melnick, G.J., Sonnentrucker, P., et al., 2006, ApJ 649, 816
\bibitem[\protect\citeauthoryear{Nisini et al.}{1996}]{Nis96} Nisini, B., Lorenzetti, D., Cohen, M., et al. 1996, A\&A 315, L321
\bibitem[\protect\citeauthoryear{Padgett et al.}{1997}]{Pad97} Padgett, D.L., Strom, S.E., \& Ghez, A., 1997, ApJ 477, 705
\bibitem[\protect\citeauthoryear{Palla \& Stahler}{1999}]{Pal99} Palla, F., \& Stahler, S.W., 1999, ApJ 525, 772
\bibitem[\protect\citeauthoryear{Persi et al.}{2003}]{Per03} Persi, P., Marenzi A.R., G\'{o}mez, M., \& Olofsson, G., 2003, A\&A 399, 995
\bibitem[\protect\citeauthoryear{Porras et al.}{2007}]{Por07} Porras, A., Jorgensen, J., \& Allen, L.E., 2006, ApJ 656, 493
\bibitem[\protect\citeauthoryear{Prusti et al.}{1992}]{Pru92} Prusti, T., Whittet, D.C.B., Assendorp, R., \& Wesselius, P.R., 1992, A\&A 260, 151
\bibitem[\protect\citeauthoryear{Rebull et al.}{2007}]{Reb07} Rebull, L., Stapelfeldt, K.R., Evans II, N.J., et al. 2007, ApJS 171, 447
\bibitem[\protect\citeauthoryear{Reipurth}{1994}]{Rei94} Reipurth, B., 1994,  A general catalogue of Herbig-Haro objects 
\bibitem[\protect\citeauthoryear{Reipurth \& Bally}{2001}]{Rei01} Reipurth, B., \& Bally, J., 2001, A.R.A.\&A 39, 403 
\bibitem[\protect\citeauthoryear{Robitaille et al.}{2006}]{Rob06} Robitaille, T.P., Whitney, B.A., Indebetow, et al. 2006, ApJS 167, 256 (R06)
\bibitem[\protect\citeauthoryear{Robitaille et al.}{2007}]{Rob07} Robitaille, T.P., Whitney, B.A., Indebetow, R., \& Wood, K., 2007, in press (astro-ph/0612690) (R07)
\bibitem[\protect\citeauthoryear{Rowan-Robinson et al.}{2004}]{Row04} Rowan-Robinson, M., Lari, C., Perez-Fournon, I., et al. 2004, MNRAS 351, 1290 
\bibitem[\protect\citeauthoryear{Rydgren}{1980}]{Ryd80} Rydgren, A.E., 1980, AJ 85, 444
\bibitem[\protect\citeauthoryear{Sandell et al.}{1987}]{San87} Sandell, G., Zealey, W.J., Williams, P.M., et al. 1987, A\&A 182, 237
\bibitem[\protect\citeauthoryear{Schwartz}{1977}]{Sch77} Schwartz, R.D., 1977, ApJS 35, 161
\bibitem[\protect\citeauthoryear{Schwartz}{1991}]{Sch91} Schwartz, R.D., 1991, ESO rep. No. 11, p. 93
\bibitem[\protect\citeauthoryear{Spezzi et al.}{2007a}]{Spe07a} Spezzi, L., Alcal\'{a}, J.M., Frasca, A., et al. 2007, A\&A 470, 281 (S07a)
\bibitem[\protect\citeauthoryear{Spezzi et al.}{2007b}]{Spe07b} Spezzi, L., Alcal\'{a}, J.M., Covino, E., et al. 2007, ApJ submitted (S07b)
\bibitem[\protect\citeauthoryear{Steenman \& Thi}{1989}]{Ste89} Steenman, H., \& Thi, P.S., 1989, Ap\&SS 161, 99
\bibitem[\protect\citeauthoryear{Surace et al.}{2004}]{Sur04} Surace, J.A., Shupe, D.L., Fang, F., et al. 2004, Vizier Online Data Catalog 2255, 0
\bibitem[\protect\citeauthoryear{Tachihara et al.}{1996}]{Tac96} Tachihara, K., Dobashi, K., Mizuno, A., et al. 1996, PASJ 48, 489
\bibitem[\protect\citeauthoryear{Teixeira et al.}{2000}]{Tei00} Teixeira, R., Ducourant, C., Sartori, M.J., et al. 2000, A\&A 361, 1143 
\bibitem[\protect\citeauthoryear{Vrba \& Rydgren}{1984}]{Vrb84} Vrba, F.J., \& Rydgren, A.E., 1984, ApJ 283, 123
\bibitem[\protect\citeauthoryear{Vuong et al.}{2001}]{Vuo01} Vuong, M.H., Cambr\'{e}sy, L., \& Epchtein, N., 2001, A\&A 379, 208
\bibitem[\protect\citeauthoryear{Weingartner \& Draine}{2001}]{Wei01} Weingartner, J.C., \& Draine, B.T., 2001, ApJ 548, 296 
\bibitem[\protect\citeauthoryear{Whitney et al.}{2003}]{Whit03} Whitney, B.A., Wood, K., Bjorkman, J.E., \& Cohen, M., 2003, ApJ 598, 1079 
\bibitem[\protect\citeauthoryear{Whittet et al.}{1987}]{Whi87} Whittet, D.C.B., Kirrane, T.M., Kilkenny, D., et al. 1987, MNRAS 224, 497
\bibitem[\protect\citeauthoryear{Whittet et al.}{1991}]{Whi91} Whittet, D.C.B., Laureijs, R.J., \& Zhang, C.Y., 1991, A\&A 251, 524
\bibitem[\protect\citeauthoryear{Whittet et al.}{1994}]{Whi94} Whittet, D.C.B., Gerakines, P.A., Carkner, A.L., et al. 1994, MNRAS 268, 1
\bibitem[\protect\citeauthoryear{Whittet et al.}{1997}]{Whi97} Whittet, D.C.B., Prusti, T., Franco, G.A.P., et al. 1997, A\&A 327, 1194
\bibitem[\protect\citeauthoryear{Young et al.}{2005}]{You05} Young, K.E., Harvey, P.M., Brooke, T.Y., et al. 2005, ApJ 628, 283
\bibitem[\protect\citeauthoryear{Zacharias et al.}{2005}]{Zac05} Zacharias, N., Monet, D.G., Levine, S.E., et al. 2004, AAS 205, 4815  
\bibitem[\protect\citeauthoryear{Zuckerman}{2001}]{Zuc01} Zuckerman, B., 2001, Ann. Rev. A\&A 39, 549


\end{thebibliography}
\end{document}